\documentclass[11pt]{article}
\usepackage[utf8]{inputenc}
\usepackage{geometry}
\usepackage{float}
\usepackage{booktabs}
\usepackage{url}
\usepackage{amsmath}
\usepackage{amsthm}
\usepackage{amssymb}
\usepackage{graphicx}
\usepackage{setspace}
\usepackage[unicode=true,
 bookmarks=false,
 breaklinks=false,pdfborder={0 0 1},backref=section,colorlinks=false]
 {hyperref}
\hypersetup{
 colorlinks,allcolors=red}

\makeatletter

\providecommand{\tabularnewline}{\\}
\theoremstyle{plain}
\newtheorem{thm}{\protect\theoremname}
\theoremstyle{plain}
\newtheorem{prop}[thm]{\protect\propositionname}
\theoremstyle{plain}
\newtheorem*{prop*}{\protect\propositionname}

\usepackage{tikz}
\usepackage{palatino}

\usepackage{titling}
\usepackage{harvard}

\usepackage{amsthm}
\usepackage{rotating}

\@ifundefined{showcaptionsetup}{}{%
 \PassOptionsToPackage{caption=false}{subfig}}
\usepackage{subfig}
\makeatother

\providecommand{\propositionname}{Proposition}
\providecommand{\theoremname}{Theorem}

\begin{document}
\title{Do Investors Hedge Against Green Swans?\\
Option-Implied Risk Aversion to Wildfires\thanks{I would like to thank Serdar Dinc, Matthew E. Kahn, Jens Jackwerth,
Mathieu Naud, Piotr Orłowski, Toan Phan, David Thesmar, Ambika Gandhi
for insightful comments on preliminary versions of this paper. The
standard disclaimers apply.} }
\author{Amine Ouazad\thanks{Associate Professor, Department of Finance and Economics, Rutgers
Business School. 1, Washington Place, Newark, New Jersey, 07102, United
States. \texttt{amine.ouazad@rutgers.edu.}}}
\date{August 2022}
\maketitle
\begin{abstract}
Measuring beliefs about natural disasters is challenging. Deep out-of-the-money
options allow investors to hedge at a range of strikes and time horizons,
thus the 3-dimensional surface of firm-level option prices provides
information on (i)~skewed and fat-tailed beliefs about the impact
of natural disaster risk across space and time dimensions at daily
frequency; and (ii)~information on the covariance of wildfire-exposed
stocks with investors' marginal utility of wealth. Each publicly-traded
company's daily surface of option prices is matched with its network
of establishments and wildfire perimeters over two decades. First,
wildfires affect investors' risk neutral probabilities at short and
long maturities; investors price asymmetric downward tail risk and
a probability of upward jumps. The volatility smile is more pronounced.
Second, comparing risk-neutral and physical distributions reveals
the option-implied risk aversion with respect to wildfire-exposed
stock prices. Investors' marginal utility of wealth is correlated
with wildfire shocks. Option-implied risk aversion identifies the
wildfire-exposed share of portfolios. For risk aversions consistent
with Barro (2012), equity options suggest (i)~investors hold larger
shares of wildfire-exposed stocks than the market portfolio; or (ii)~investors
may have more pessimistic beliefs about wildfires' impacts than what
observed returns suggest, such as pricing low-probability unrealized
downward tail risk. We calibrate options with models featuring both
upward and downward risk. Results are consistent a significant pricing
of downward jumps.
\end{abstract}
\vfill{}

\pagebreak\clearpage{}

\section{Introduction}

Natural disaster risk is receiving increasing attention by both academics
and policymakers preparing for a potential ``Green Swan'' \cite{bolton2020green,rudebusch2021climate}.
What are market participants' beliefs about the wealth impact of natural
disasters? Measuring such beliefs is challenging as they may be (a)~asymmetric,
as firms may lose or gain; (b)~fat-tailed, as natural disasters may
cause large jumps and volatility shifts; (c)~non-stationary, as agents
learn about the probability distribution of disaster risk over time;
(d)~heterogeneous, as firms have different levels of adaptation to
disasters; and (e)~ambiguous as different forecasters provide different
probability distributions. A seminal literature uses surveys to measure
beliefs about assets' risk exposure \cite{bakkensen2017flood}, and
about the evolution of future disaster risk \cite{leiserowitz2010climate,ballew2019climate}.
Yet, eliciting high frequency and spatially heterogeneous beliefs
about climate risk for the United States at both local and national
levels, and for short- and long-run time horizons, remains challenging.

The exchange-traded market for options is a fast-growing global market
with a total volume of 33.3 billion contracts in 2021\footnote{Futures Industry Association, January 19, 2022.}
that allows financial agents to trade hedges for firm-, time-, and
maturity-specific risks. The surface of option prices by maturity
and strike depends on market participants' subjective probability
distribution as well as the marginal utility of consumption or wealth,
thus revealing the set of Arrow-Debreu prices for each level of each
climate-exposed stock \cite{breeden1978prices,ait2003nonparametric,beber2006effect,figlewski2018risk}.
Such surface typically displays a 'smile' or 'smirk' that steepened
significantly after the 1987 market crash \cite{benzoni2011explaining},
as market participants hedge against large downward and upward tail
risks. Teasing out what, in Arrow Debreu state prices, is driven by
the marginal utility of wealth and by beliefs about risks enables
an estimation of investors' aversion to natural disaster risks.

% What this paper does in one paragraph

This paper measures the response of investors' hedging to natural
disaster shocks, compares this response to observed returns, and identifies
investors' option-implied risk aversion with respect to such natural
disasters. It matches the universe of establishments, sales, and employment
of publicly-traded firms between 2000 and 2018 to their daily surface
of option prices by strike and maturity, and to the boundaries of
wildfire perimeters reported by fire departments across the United
States. First, the paper estimates the daily risk neutral distribution
implied\footnote{Option prices are de-americanized using the procedure evaluated by
\citeasnoun{burkovska2018calibration}, which converts American
option prices into European option prices using \possessivecite{cox1979option}
binomial trees to recover implied volatilities. Alternatives involve
solving partial differential equations with a boundary condition for
each firm $\times$day $\times$strike $\times$maturity.} by the surface of option prices\footnote{The paper uses a method close to \possessivecite{ait2003nonparametric}
arbitrage-free prices described in Section~\ref{subsec:Estimation-Technique:-Arbitrage}.
Other important alternatives for estimating the risk neutral distribution
include \citeasnoun{figlewski2009estimating} and \citeasnoun{figlewski2018risk}.} for all day-firms affected by a wildfire, and for an equal-sized
set of control surfaces, at short-, medium-, and long-run time horizons.
This panel approach can non-parametrically estimate the impact of
wildfires on the risk neutral distribution over and above day- and
firm-specific confounders. Second, over the last two decades the paper
estimates a significant impact of exposure to wildfires on the price
in deep out-of-the-money options, controlling for day- and firm-specific
confounders. This increase in prices is over and above what a log
normal distribution of stock prices would imply. Third, the paper
compares such risk neutral probabilities to the physical probabilities,
thus inferring investors' marginal utility for each level of a wildfire-exposed
stock. This yields a risk aversion parameter with respect to each
wildfire-exposed stock. Fourth, the paper develops a continuous time
\citeasnoun{merton1969lifetime} model with a wildfire stock and
the index to tease out risk aversion with respect to the index from
that with respect to the wildfire-exposed stock. Portfolio shares
are typically unobserved, and this paper's novel approach provides,
`in reverse', the weight of wildfire stocks implied by deep out-of-the-money
option prices. Fifth, the paper calibrates investors' marginal utility
at the levels of each wildfire-exposed stock to disentangle what,
in the impact of wildfires on the risk neutral distribution is due
to marginal utility from what is due to change in beliefs.

% Results and detailed walk through

The paper puts forward and implements a fast and efficient way to
identify the impact of natural disasters on investors' risk-neutral
probability distributions for each firm and for each maturity at daily
frequency. Using such daily probability surfaces, the paper estimates
the impact of wildfires on the risk neutral distribution non-parametrically
when affected by a wildfire. This enables an estimation of the impact
of wildfires on beliefs' (a)~asymmetry, (b)~fat tails, and (c)~non-stationarity.
Results suggest that when a firm's network of establishments is exposed
to a wildfire, the risk neutral probability of downward tail events
increases but is heterogeneous across industries, while they have
a small but precisely estimated positive impact on the risk neutral
probability of upward movements. Impacts are observed for short and
long maturities. 

Backing out risk neutral distributions from option prices requires
using arbitrage-free prices. Since such arbitrage opportunities may
be observed, prices are converted into arbitrage-free prices that
do not exhibit possibilities of butterfly, calendar spread, or call
spread arbitrage~\cite{carr2005note}. Yet, wildfires show little
correlation with the difference between arbitrage-free and observed
prices, suggesting that there may be few opportunities to take advantage
of `climate risk' arbitrage opportunities in vanilla option markets.
This also suggests that arbitrage opportunities are unlikely to drive
the paper's main findings.

Investors' hedging against downward and upward tail risks is typically
associated with a steeper implied volatility smile~\cite{bakshi2003stock}.
The surface of implied volatilities is a simple transformation of
the surface of option prices such that, if the firm's stock followed
a geometric diffusion process, the implied volatility would be independent
of strike and maturity. In the short-run, wildfires have a greater
impact on the implied volatility smile for low strikes than for strikes
higher than the forward price: the impact is asymmetric, suggesting
that market participants hedge more against the downside than for
the upside. Over medium and long run time horizons, the lower hedging
for the upside remains: markets price a lower probability of an upside.
Impacts are long-lasting, with a smile that persists after the first
wildfire.

Results on the risk neutral distribution and the volatility smile
are obtained either in a linear panel data regression with day and
firm fixed effects with daily data between 2000 and 2018; they are
also obtained using a more flexible non-parametric approach pooling
all day-firm surfaces after orthogonalization with respect to firm
and day-specific confounders.

Option prices can also enable an identification of the correlation
of investor's wealth with wildfires. Arrow Debreu state prices are
the product of the marginal utility of investors, the physical probability
of each state, and the risk free discount factor. As each wildfire-exposed
stock represents a small share of the S\&P 500, wildfire stocks could
be statistically uncorrelated with investors' wealth, once controlling
for the stock's beta with respect to the index. If movements in wildfire-exposed
stocks were uncorrelated with investors' marginal utility of wealth
or consumption, risk neutral probabilities would be proportional to
the physical probabilities. We compare the Arrow Debreu state prices
for each day-firm to the physical probability distribution of future
returns to estimate investors' risk aversion with respect to the wildfire
stocks. This also identifies the marginal utility of investors at
each level of the wildfire-exposed stock.\footnote{This paper's results suggest that most firms exhibit pricing kernels
with negative slopes consistent with decreasing marginal utilities
of standard microeconomic models: between 87.4\% and 90.5\% of the
firms have a strictly negative slope of the pricing kernel w.r.t.
the log of the forward price. When considering index options, an extensive
literature provides evidence of \emph{pricing kernel puzzles} and
methods to address them, e.g.~\citeasnoun{linn2018pricing}.} The first finding is that the ratio of risk neutral to physical probabilities
significantly differs from a constant for treated stocks. This suggests
that investors' marginal utility is significantly correlated with
wildfire shocks. Results provide a firm-level elasticity of state
prices with respect to the level of the stock.

Such risk aversion with respect to a specific stock can reveal the
weight of the stock in the investor's portfolio. We develop a \citeasnoun{merton1971optimum}
portfolio allocation problem with two risky assets, the index and
a wildfire-exposed stock, to separately identify what, in investors'
demand for equity options, is driven by the beta of the wildfire-exposed
stock w.r.t the index from what is driven by wildfires. Proposition~\ref{proposition:risk_aversion}
of Section~\ref{subsec:physical_vs_risk_neutral_distribution} shows
that the Arrow-Pratt risk aversion with respect to a single stock
is proportional to the relative risk aversion with respect to wealth,
where the proportionality constant is a linear function of the share
of the stock in the investor's portfolio and the beta of the stock
w.r.t. the market. Using the option-implied risk aversion with respect
to the stock, assuming that the investor holds the market portfolio,
and using the estimated betas from CRSP, we can identify the risk
aversion \emph{with respect to wealth} that is consistent with the
option-implied risk aversion \emph{with respect to the wildfire-exposed
stock}.

Assuming the market portfolio, the option-implied relative risk aversion
w.r.t wealth is significantly higher than that suggested by \citeasnoun{barro2006rare},
\citeasnoun{barro2012rare}, or \citeasnoun{mehra1985equity}.
This is consistent with (i)~investors holding shares of wildfire-exposed
stocks larger than what the market portfolio implies, markets may
be segmented as in \citeasnoun{gabaix2007limits}; (ii)~a mismeasurement
of the physical probability distribution, which reflects investors'
beliefs; thus investors' beliefs are not well captured by the realization
of past or future returns; this would include investors' expectations
of rare tail risk, that does not materialize with high enough frequency
in the sample of wildfire-exposed stocks, a phenomenon suggested for
the index by \citeasnoun{barro2012rare}.

The last part of the paper parameterizes the risk neutral distribution
using \possessivecite{merton1976option} jump-diffusion model, and
\possessivecite{kou2004option} double exponential jump model to be
able to observe the pricing of both upward and downward shocks in
the same model. Results of this calibration on the treatment group
(affected by wildfires) and the control group (unaffected) suggest
a small increase in the pricing of upward jumps, but a doubling in
the risk-neutral magnitude of downward jumps.

\citeasnoun{merton1976option} suggests that jumps are a source
of market incompleteness, as riskless delta hedging is not typically
feasible. A set of state-contingent parametric insurance, with a payoff
tied to the occurrence of wildfires for each specific area, could
allow market participants to hedge wildfire risk. The calibration
of the options, by providing the risk neutral distribution, allows
the pricing of such parametric insurance.

This paper contributes to at least three literatures. First, the growing
literature on the impact of natural disaster on option prices. \possessivecite{kruttli2021pricing}
important contribution provides convincing evidence of an increase
in the at-the-money implied volatility of firms affected by a hurricane.
The paper also finds that, for firms hit by hurricanes, the variance
risk premium, i.e. the difference between the implied volatility and
the realized volatility, is a significant predictor of future stock
returns. This paper's model-free approach to the recovery of the risk
neutral distribution from option prices has two benefits that complement
\possessivecite{kruttli2021pricing} key insights. First, this paper
provides the entire probability distribution by stock level and by
time horizon (maturity) and thus can separate symmetric uncertainty
(volatility) from asymmetric expectations of gains and losses. As
such, this paper's approach reveals the response of the moments of
the risk neutral distribution and the downside and upside hedging
of investors: this enables an identification of downward and upward
tail risks separately, as investors perceive the possibility of firms'
recovery in the short-, medium-, and long-run. This enables the identification
of jump risk separately from stochastic volatility, as the calibration
of option prices as in \citeasnoun{duffie2000transform} enables
the identification of the jump diffusion processes consistent with
the surface of option prices. An emerging literature studies the skewness
risk premium \cite{orlowski2020nature}. Second, this paper's approach
has the benefit of identifying investors' option-implied risk aversion,
and thus reveals the importance of wildfire shocks for investors'
portfolio. Results suggest that wildfire shocks matter more than if
investors held the market portfolio.

This paper also contributes to the literature on the estimation of
beliefs. An important literature uses surveys to measure beliefs about
the specific risk exposure of assets \cite{bakkensen2017flood}, and
about the evolution of future climate risk \cite{leiserowitz2010climate,ballew2019climate}.
The economics literature has shown that stated beliefs have predictive
power for individual behavior \cite{manski2018survey}, predict portfolio
allocation \cite{giglio2021five} and shape macroeconomic outcomes
\cite{fuster2010natural}. This paper suggests that option prices
provide a low-cost, high-frequency, and spatially heterogeneous way
of measuring beliefs in natural disaster risk at multiple time horizons.
The absence of a significant correlation between arbitrage opportunities
and wildfire risk suggests that this recovers well-behaved impacts
on the risk neutral probability distribution.

This paper contributes to the literature on risk aversion and asset
pricing. In the mortgage-backed securities (MBS) market, \citeasnoun{gabaix2007limits}
suggests that prepayment risk, which is diversified in the aggregate,
is priced in MBS yields. They also find that the sign for the correlation
between income and yields is the opposite of what would be expected.
Their conclusion is that markets are segmented. This paper is also
in the flavor of \citeasnoun{froot1999pricing}, which finds that
the price of catastrophe reinsurance is unusually high, lending support
for a segmented trading of such assets; in this paper, we find that
the pricing of deep out of the money puts is significantly higher
than the pricing when investors hold a diversified portfolio.

The response of option prices to transitory shocks is akin to \possessivecite{dessaint2017managers}
finding that the managers of firms located next to hurricane-hit areas
tend to express concerns about hurricanes in their 10-Ks/10-Qs, and
increase cash holdings, even when actual risk remains unchanged. This
paper suggests that such concerns translate into the pricing of downward
tail risk in out-of-the-money options, but also in upward tail risk
pricing. It is also closely related to \citeasnoun{orlik2014understanding}
on Black Swans: when people form expectations like Bayesian econometricians,
and when they estimate tail risk probabilities dynamically, small
changes in estimated skewness cause large changes that ``whip around''
the probabilities of unobserved tail events.

Finally, this paper contributes to the literature on firms' adaptation
to natural disaster risk \cite{kahn2020climate}. This paper provides
a method for identifying the option-implied price of wildfire insurance,
a set of state-contingent and spatially-located Arrow Debreu assets
whose price is the risk neutral price of downward tail risk. This
paper's wildfire data suggests a significant upward trend in the probability
that a Zip code is in a wildfire perimeter. Results suggest long-run
impacts of wildfires on the implied volatility smile, but such long-run
impacts are priced in after the first exposure to a wildfire rather
than cumulatively per wildfire. Hence option prices may provide an
adaptation signal~\cite{kahn2021adapting}. Out of the money put
and call options are part of the set of hedging devices that enable
investors to weather natural disaster risks. And, in reverse, investors'
demand for such options reveals their expectations of firms' ability
to adapt to natural disasters. In this sense, this paper's causal
estimates are the reduced-form estimates of firms' adaptation responses.

The paper is structured as follows. Section \ref{sec:datasets} describes
the main data set: the daily set of option implied volatilities and
prices by strike and maturity, matched to each publicly-traded firm's
geocoded network of establishments and wildfire perimeters. Section
\ref{sec:wildfires_and_rnd} describes the relationship between option
prices and their implied risk neutral distribution. It estimates the
firm-level risk neutral distribution at daily frequency between 2000
and 2018 for listed firms using the convexity of option prices. It
estimates the impact of wildfires on the risk neutral moments, the
lower and upper tails, as well as the central part of the probability
distribution. Section \ref{sec:panel_data_evidence} estimates the
model-free impact of wildfires on the volatility smile, using daily
longitudinal panel data controlling for day and firm-level unobservables.
Section \ref{sec:marginal_utility} then estimates the relationship
between the risk neutral distribution and the physical distribution
for wildfire-exposed stocks, thus identifying a correlation between
investors' marginal utility and the value of wildfire-exposed stocks.
This yields an option-implied risk aversion. Section \ref{sec:Calibrating-Risk-Neutral}
calibrates the surface of option prices to tease out whether wildfires
cause in upward and downward jump intensity or magnitude. Section
\ref{sec:conclusion} concludes.

\section{Data Sets}

\label{sec:datasets}

Estimating the impact of natural disasters on the surface of option
prices, on Arrow-Debreu prices, and on investors' beliefs requires
four data sets. First, the universe of establishments of publicly-traded
companies. Second, the spatial footprint of natural disasters at daily
frequency. Third, option prices by maturity and strike at daily frequency.
Fourth, stock prices and dividends to build physical probability distributions
that can be compared to the risk-neutral distribution of Arrow-Debreu
prices.

\subsection{The Geography of Publicly-Traded Companies' Establishments}

The first source of data is Dun and Bradstreet's NETS for the 50 states
and the District of Columbia between 2000 and 2018. NETS provides
the geographic location of establishments, with their employment,
sales, and 8-digit NAICS code at annual frequency. \citeasnoun{barnatchez2017assessment}
compares NETS data to the County Business Patterns (CBP), the Nonemployer
Statistics, and the Quarterly Census of Employment and Wages (QCEW)
and finds that the data set covers three quarters of U.S. private
sector employment. The study finds that the county-level correlation
of NETS employment and CBP employment is typically above 0.99. We
follow the convention of \citeasnoun{echeverri2019chasing} in omitting
the last two years of data, which can be subject to revisions. Other
studies using NETS and comparing it to administrative data counts
include \citeasnoun{neumark2011small}. Recent work using NETS includes
\citeasnoun{song2019firming}.

A key feature of NETS is its inclusion of the ticker symbol when the
establishment is the headquarters of a publicly-traded firm. It also
links a parent firm with its subsidiaries. Hence the data enable the
estimation of each publicly-traded firm's geographic footprint. We
compute the number of establishments, the employment, and the sales
of each publicly traded firm in each 5-digit Zip code. As NETS is
longitudinal, the data also enable the estimation of the relocation
of a firm's establishments over time in response to climate shocks.
Thus this paper's estimates use the firm's latest measured geographic
footprint and exposure to wildfires, which may respond to expectations
or experiences of climate risk.

\subsection{Wildfire Risk}

\label{subsec:wildfire_data}

Wildfire perimeters and dates are provided by the Development and
Application team of Wildland Fire Management Research at the National
Interagency Fire Center. Such center coordinates the actions of local
firefighting agencies and records the polygon of each fire's perimeter,
the date of the fire, the number of acres burnt, the reporting agency;
for a total of 43,114 wildfire perimeters on 829 separate dates, with
an average of 726 perimeters per year, 67 million acres in total,
and an average of 3,771 acres per perimeter. For each 5-digit Zip
Code (Census ZCTAs) we record the area of the Zip code intersecting
the wildfire perimeter in the US National Atlas Equal Area coordinate
reference system. Wildfires affected 1,964 5-digit Zip codes. The
share of a publicly-traded firm's establishments, sales, and employment
hit by a wildfire depends on the share of its establishments in each
of the US's 33,120 ZCTAs. We use the boundaries of 2014 ZCTAs. The
log number of establishments, employees, and sales hit by a wildfire
according to this definition are presented on Appendix Figure~\ref{fig:wildfire_exposure}.
The figures suggest a substantial increase in the number of establishments,
employees, and sales hit by a wildfire over time.

\subsection{Option Prices and Implied Volatility}

\label{subsec:optionmetrics}

Our option data comes from OptionMetrics' IvyDB US, which covers all
US-listed equity options. We consider data on the same time period
as the Dun and Bradstreet data, i.e. from January 2000 till December
2018. The data include the strike, maturity, call/put indicator, best
bid/ask prices and we use the midpoint as the option price. We estimate
the moneyness of the option using ratio of the strike over the forward
price. Implied volatility is obtained by OptionMetrics using binomial
trees, which account for the possibility of early exercise. Option
prices are de-Americanized using the methodology described by \citeasnoun{burkovska2018calibration},
which converts binomial-tree implied volatilities into European option
prices.\footnote{\citeasnoun{burkovska2018calibration} provides numerical bounds
for the absolute statistical biases in prices potentially introduced
by this approach, ranging between $10^{-6}$ and $10^{-23}$.} In total, this comprises 2.36 billion equity option $\times$ day
$\times$ maturity $\times$ strike observations. To work with a more
manageable data set amenable to a panel regression with multi-way
fixed effects, we (i)~include all option prices of treated (wildfire-exposed)
firms, and (ii)~a random sample of 500 untreated firms on a random
5\% of trading days. We end up with a sample of 16,842,463 option
prices.

%\subsection{Stock Prices} \label{subsec:crsp}

%Observing stock prices and dividends enables a potential estimation of the physical distribution of future stock prices, which can then be compared to the risk neutral distribution. This is done in Section~\ref{sec:marginal_utility}. Our data set of stock returns comes from the Center for Research in Security Prices (CRSP).

\subsection{Descriptive Statistics}

\begin{figure}
\caption{Wildfires and the Nationwide Distribution of Publicly Listed Firms'
Employment}
\label{fig:map_employment_wildfires}

\emph{\footnotesize{}This map presents the geographic distribution
of employment of publicly listed firms between 2000 and 2018. Black
points indicate the location of the headquarters of public firms.
Colors indicate the log number of employees of such public firms per
squared mile in each ZIP code.}{\footnotesize\par}

\emph{\footnotesize{}}\subfloat[{\footnotesize{}Geographic Distribution of the Employment of Establishments
of Publicly-Listed Companies}]{\emph{\footnotesize{}}{\footnotesize\par}
\centering{}{\footnotesize{}\includegraphics[clip,scale=0.6]{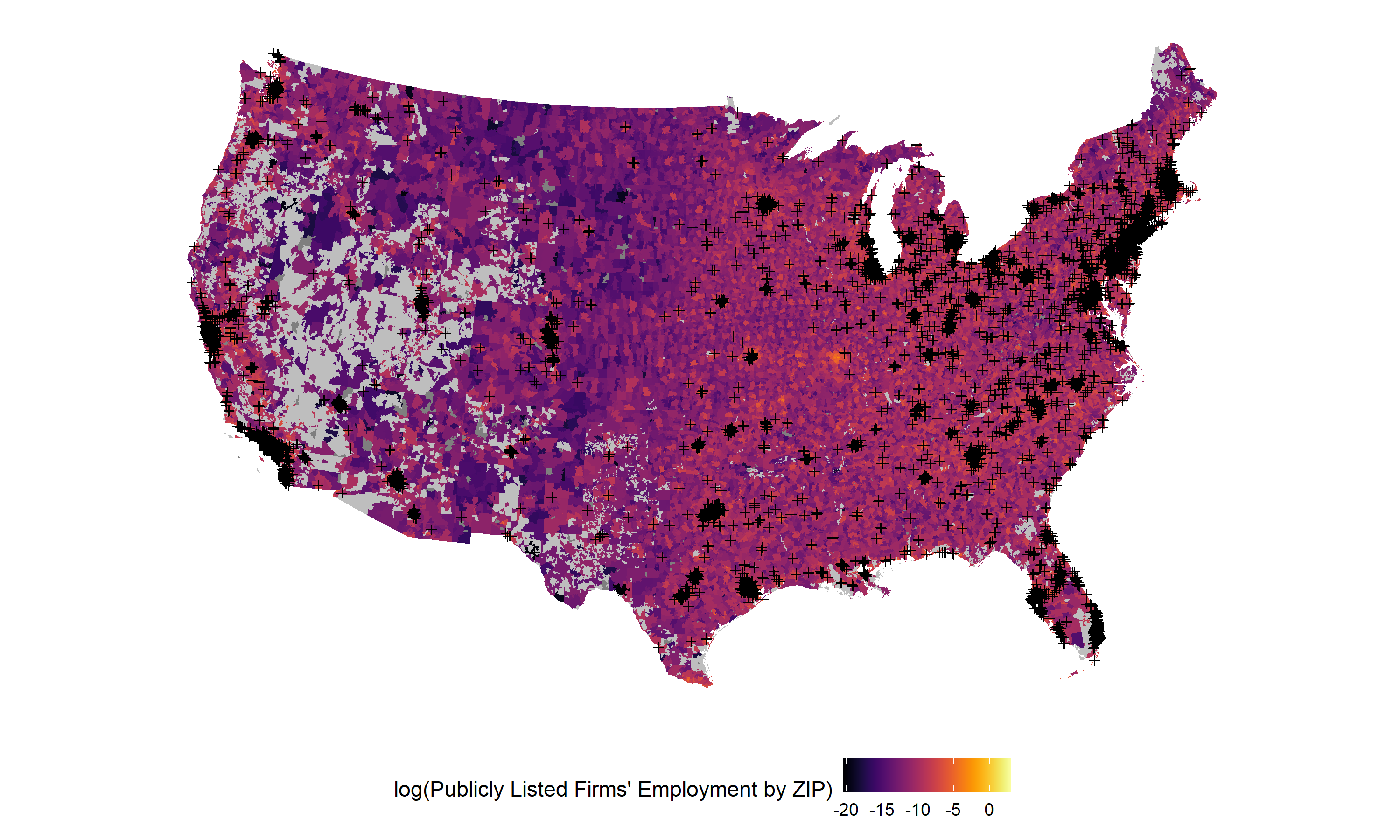}}{\footnotesize\par}}{\footnotesize\par}

\emph{\footnotesize{}This map presents the number of years with at
least one wildfire for each 5-digit ZIP code, between 2000 and 2018.
Fire perimeters from the National Interagency Fire Center. ZIP code
boundaries from the Census' Zip Code Tabulation Areas. In the paper's
regressions, fires are recorded at daily frequency.}{\footnotesize\par}
\centering{}{\footnotesize{}}\subfloat[{\footnotesize{}Geographic Distribution of 2000--2018 Wildfires }]{{\footnotesize{}}{\footnotesize\par}
\centering{}{\footnotesize{}\includegraphics[clip,scale=0.6]{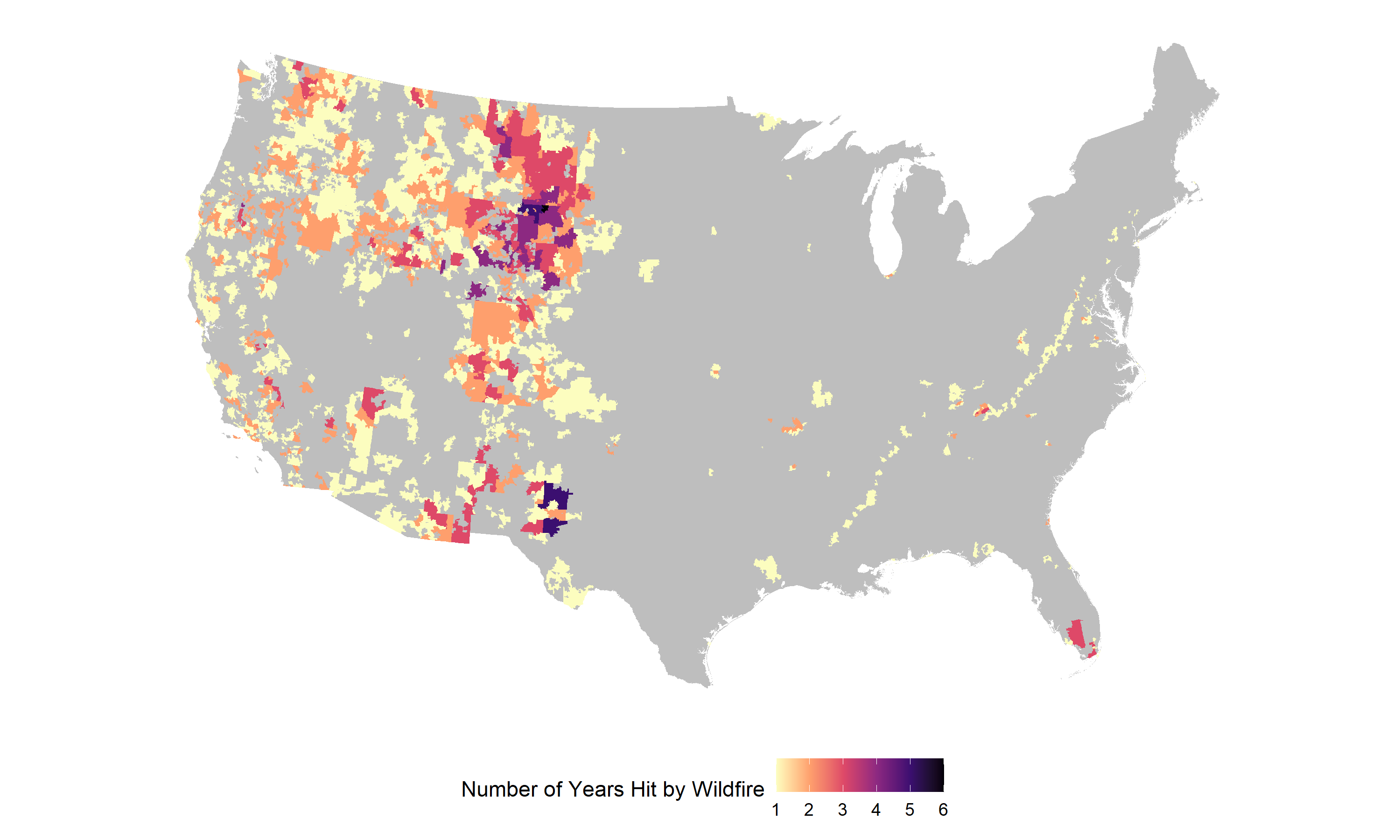} }{\footnotesize\par}}{\footnotesize\par}
\end{figure}

Figure~\ref{fig:map_employment_wildfires}(a) presents the spatial
distribution of publicly-listed companies' employment. For each listed
company, the map estimates the number of employees in each 5-digit
Zip code on average between 2000 and 2018, represented by colors from
black to yellow. The black dots are for the headquarters. The map
suggests that NETS data covers the coterminous United States, and
that the employment captured in this paper extends beyond the location
of the headquarters. Figure~\ref{fig:map_employment_wildfires}(b)
presents the spatial distribution of wildfires between 2000 and 2018.
The colors correspond to the number of years that a 5-digit Zip code
was hit by a wildfire. The map suggests that wildfires hit most of
the forests of the Western part of the United States, forests of the
Pacific Northwest and Southwest, the Intermountain area, the Northern
region, the Rocky Mountain region, and the Southwestern region. They
also affected the George Washington and Jefferson forests in the Appalachians,
the Ozark and Ouachita forests in Arkansas, Southernwestern Florida,
but also the Sam Houston forest in Texas, the Tallageda forest in
Alabama.

\section{Option-Implied Risk Neutral Probabilities of Natural Disasters}

\label{sec:wildfires_and_rnd}

\subsection{Identifying Probability Distributions from the Surface of Option
Prices}

The surface of option prices by maturity and by strike reveals market
participants' set of Arrow Debreu prices at each level of the future
stock price and for each maturity. This appears when considering the
price of a European option, which can be exercised at maturity, before
considering the case of American options.\footnote{The de-Americanization approach used in this paper is described in
Section~\ref{subsec:optionmetrics} and assessed by \citeasnoun{burkovska2018calibration}
using Monte Carlo simulations.} At time $t$, the payoff of a European call option with maturity
$T$ and with strike $K$ is the expectation of the payoff at $T$:
\begin{equation}
C(K,T,t)=e^{-r(T-t)}\mathbb{E}_{t}^{\mathbb{Q}}\left[\max(S_{T}-K,0)\right],\label{eq:price_european_option}
\end{equation}
where the expectation is taken at $t$ with respect to the risk neutral
distribution, whose pdf is denoted $f^{*}(S_{T})$, consistent with
the notations of \citeasnoun{ait1998nonparametric}. This risk neutral
distribution $f^{*}$ typically differs from the physical distribution
of the stock price, noted $f$. We investigate what we learn from
their differences in Section~\ref{subsec:physical_vs_risk_neutral_distribution}.
The price of the option is thus: 
\begin{equation}
C(K,T,t)=e^{-r(T-t)}\int_{K}^{\infty}(s-K)f^{*}(s)ds,\label{eq:price_european_option_integral}
\end{equation}
where $r$ is the risk-free rate.

When the strike $K$ increases marginally, it has two impacts on the
call price. First, the lower bound of integration changes, and second
the payoff $S_{T}-K$ declines. \citeasnoun{breeden1978prices}
shows that this leads to a simple relationship between the derivatives
of the option price with respect to the strike and the risk neutral
distribution: 
\begin{equation}
\left.\frac{\partial^{2}C}{\partial K^{2}}\right|_{K=S_{T}}=e^{-r(T-t)}f^{*}(S_{T}),\qquad\textrm{and}\qquad\left.\frac{\partial C}{\partial K}\right|_{K=S_{T}}=-e^{-r(T-t)}\left[1-F^{*}(S_{T})\right],\label{eq:litzenberger}
\end{equation}
where $F^{*}$ is the cumulative risk neutral distribution function.
Hence this suggests that the convexity of option prices at the strike
$K=S_{T}$ identifies the price of an Arrow-Debreu asset that pays
off \$1 when $S=S_{T}$ at maturity $T$. This key result, implemented
using arbitrage-free approaches of \citeasnoun{ait2003nonparametric}
enables an estimation of the Arrow-Debreu prices of states in which
a climate shock hits.

This approach provides the risk neutral distribution at each time
$t$, for each stock price $S_{T}$, and for each maturity $T$. It
enables an estimation that depends on investors' time horizon, an
estimation of belief updating, and recovers a rich probability distribution
parameterized by exposure to wildfires, allowing a comparison $f^{*}(S_{t}\vert\text{Disaster}=1)-f^{*}(S_{t}\vert\text{Disaster}=0)=\Delta f^{*}(S_{t})$
at each future level of the stock for a range of maturities.

\subsection{Estimation Technique: Arbitrage Free Prices and the Convexity of
Option Prices\label{subsec:Estimation-Technique:-Arbitrage}}

\possessivecite{breeden1978prices} approach described in equation
(\ref{eq:litzenberger}) suggests that the convexity of option prices
reveals the risk neutral distribution for each maturity. Practical
implementation requires arbitrage-free option prices.

Observed option prices may display arbitrage opportunities~\cite{carr2005note},
and a naive approach consisting of taking the second-difference of
observed prices may not yield the risk neutral distribution. First,
call prices may not be convex, which may be a source of butterfly
arbitrage \cite{carr2005note}. Second, call prices may not be decreasing
in the strike, which may lead to a call spread arbitrage opportunity.
Third, observed call prices may not be higher than $S-Ke^{-r(T-t)}$,
leading to potential calendar spread arbitrage.

We address this concern by estimating arbitrage-free option prices
using a method close to the approach of \citeasnoun{ait2003nonparametric}.
We first de-Americanize option prices using the method described in
Section~\ref{subsec:optionmetrics}. We then estimate the closest
prices, in the least-squares sense, that satisfy the three no-arbitrage
conditions above. This is done by a constrained linear least squares
similar to \citeasnoun{ait2003nonparametric} but allowing here
for an unequal grid of strikes whenever such grid is observed. Such
unequal strike steps are common when considering options on individual
equities. This first step yields a vector of arbitrage-free call prices
for each observed strike.

Finally, we estimate the second derivative with a local polynomial
regression of order 4 with a Gaussian kernel. The second-derivative
of option prices is the coefficient of the second order term in the
strike. In contrast with \citeasnoun{malz2014simple}, we allow
the probabilities to be nonzero at the boundaries of the support of
traded strikes and assess the impact of wildfires on the support of
traded strikes in a second step.\footnote{Another choice here is to extrapolate the risk neutral distribution
outside the support of observed strikes \cite{figlewski2009estimating,birru2012anatomy},
which requires distributional assumptions. This paper does not make
specific distributional assumptions about probability distributions
and rather estimates the impact of wildfires on the support.} Both steps (arbitrage free and second derivative) rely on a constrained
least squares approach with an interior point optimization algorithm.\footnote{Both steps use Matlab's \texttt{lsqlin} procedure.}We
use a grid of 100 equally spaced strikes $K\in\left[\underline{K}_{jtT},\overline{K}_{jtT}\right]$,
with the same range as the observed strikes $K_{jtTk}$ for $k=1,2,\dots,$.
$\left[\underline{K}_{jtT},\overline{K}_{jtT}\right]$ is the support.
Formally, the risk neutral distribution at point $K$ is the coefficient
$\beta_{2}$ in the optimization program: 

\begin{equation}
\begin{aligned}\min_{\beta_{p},p=0,\dots,4}\quad & \sum_{i=1}^{N}\left[C_{i}-\sum_{p=0}^{4}\beta_{p}(k_{i}-K)^{p}\right]K_{h}(k_{i}-K)\end{aligned}
\label{eq:rnd_estimation}
\end{equation}
with $\beta_{2}\geq0$. The function $K_{h}(x)=K(x/h)/h$ is a Gaussian
kernel, where the bandwidth $h$ was selected as in \citeasnoun{ait2003nonparametric}.

Important alternative methods include \citeasnoun{jackwerth1996recovering}
and \citeasnoun{figlewski2009estimating}. Recent reviews include
\citeasnoun{figlewski2018risk} and \citeasnoun{cuesdeanu2018pricing}.

\subsection{Wildfires and Arbitrage Opportunities }

To see whether wildfires cause the emergence or the increase in arbitrage
opportunities, we compute the difference between the arbitrage-free
prices of \citeasnoun{ait2003nonparametric} and the observed prices.
We regress such difference on the treatment indicator variable at
different maturities. Results are presented on Appendix Table~\ref{tab:wildfires_arbitrage},
with three different approaches for the dependent variable. Column
(1) uses the dollar difference, column (2) the absolute value of the
dollar difference, and column (3) the log of the absolute value of
the difference. The table presents results for 5 different ranges
of maturities. Results are similar when pooling all maturities. Results
suggest that most coefficients on the treatment indicator are not
significant, and often of small magnitude. These results also suggest
that there may not be `climate risk' arbitrage opportunities based
on the butterfly, the calendar, and call spreads.

\subsection{PG\&E's Risk Neutral Distribution during the October 2017 Wildfires }

The Northern California wildfires broke a record as the most destructive
wildfires in California's history, with an estimated 44 deaths, 250,000
acres hit \cite{mass2019northern}, 100,000 evacuated and 8,900 structures
destroyed.\footnote{California Statewide Fire Summary, Monday, October 30, 2017}.
The California Department of Forestry and Fire Protection released
a report on June 8, 2018 stating that:\footnote{Michael Mohler, "CAL FIRE Investigators Determine Causes of 12 Wildfires
in Mendocino, Humboldt, Butte, Sonoma, Lake, and Napa Counties",
June 8, 2018.} 
\begin{quote}
After extensive and thorough investigations, CAL FIRE investigators
have determined that 12 Northern California wildfires in the October
2017 Fire Siege were caused by electric power and distribution lines,
conductors and the failure of power poles. 
\end{quote}
The PG\&E Fire Victim Trust filed a lawsuit \footnote{Justic John Trotter (Rer.), Trustee of the PG\&E Fire Victim Trust
v. Lewis Chew et al., Supreme Court of California for the County of
San Francisco. \url{https://www.cpmlegal.com/media/news/15076_2021-02-24\%20Amend_Complaint__PGE_LIT_.pdf}} on February 24, 2021 arguing that "The 2017 North Bay Fires Could
Have Been Prevented If PG\&E Had Implemented A De-Energization Program"
and that "PG\&E Should Have Cut Off Power Because It Had Failed to
Maintain Vegetation in Violation of Applicable Regulations." While
it took significant time for courts and CAL FIRE to describe their
assessment of responsibilities, the evidence below suggests that option
markets reacted quickly and provided investors with hedges against
tail risk and rising, potentially stochastic, volatility.

\begin{figure}
\caption{The Risk Neutral Distribution of PG\&E during the October 2017 Wildfires}
\label{fig:pge_rnd_surface}

\emph{These four figures display the risk neutral distribution obtained
using the convexity of option prices with respect to the strike, as
in \citeasnoun{breeden1978prices}. The practical implementation
of this recovery is based on arbitrage-free option prices and a local
polynomial regression, a method similar to \citeasnoun{ait2003nonparametric}
described in Section~\ref{subsec:Estimation-Technique:-Arbitrage}.}

\medskip{}

\begin{center}
\begin{centering}
\subfloat[September 22, 2017]{\includegraphics[scale=0.5]{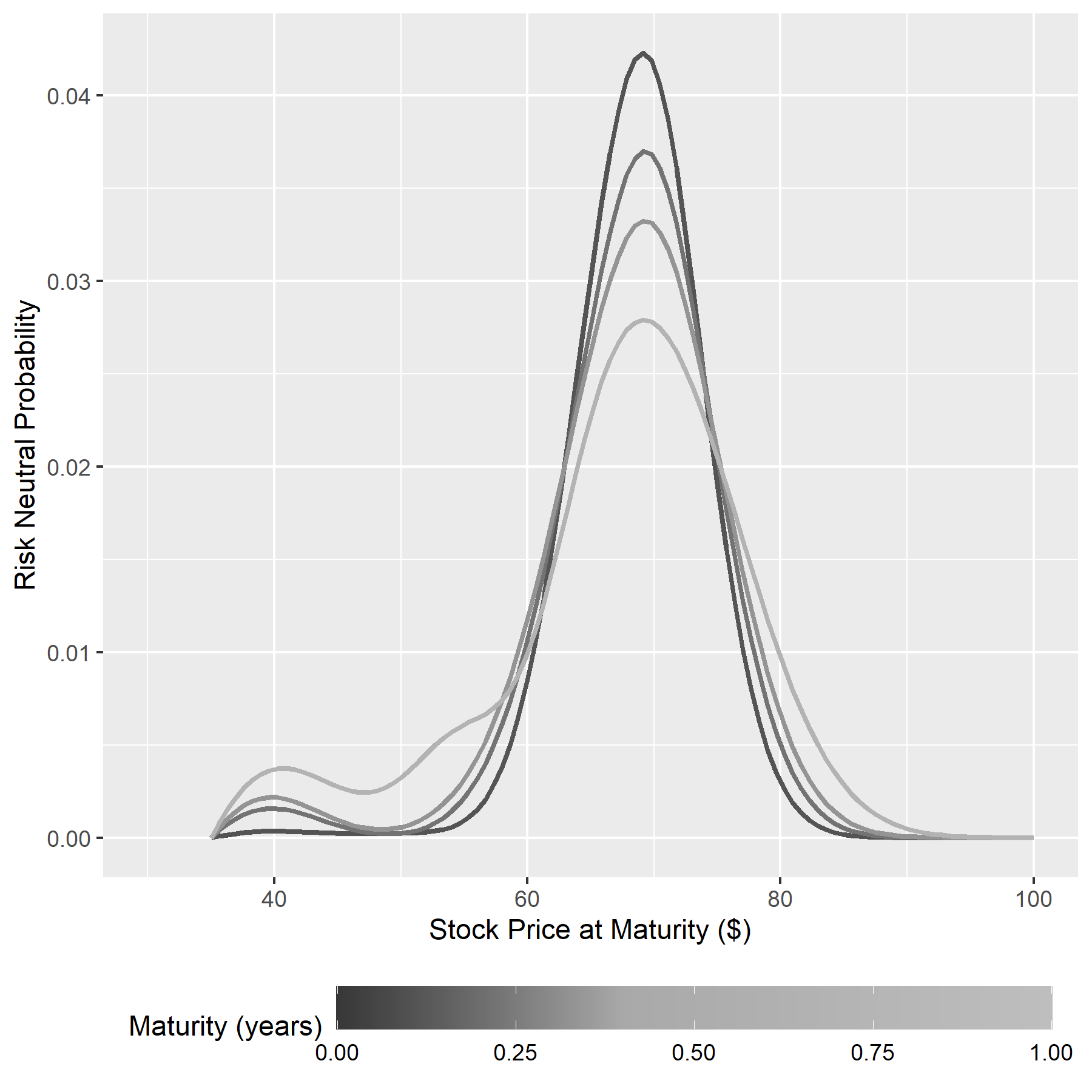} 

}~~~~~~~~~~ \subfloat[September 27, 2017]{\includegraphics[scale=0.5]{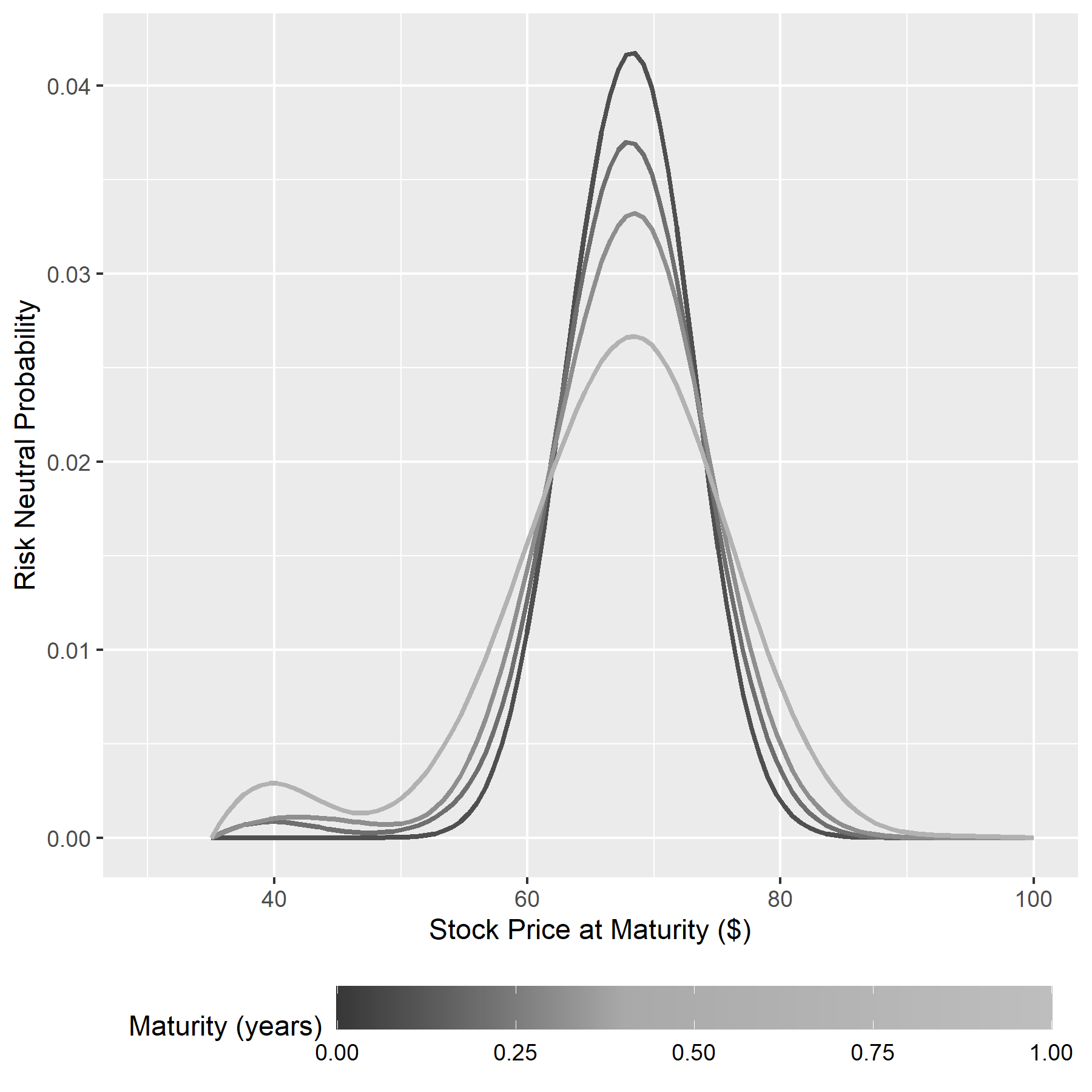} 

}
\par\end{centering}
\begin{centering}
\subfloat[October 12, 2017]{\includegraphics[scale=0.5]{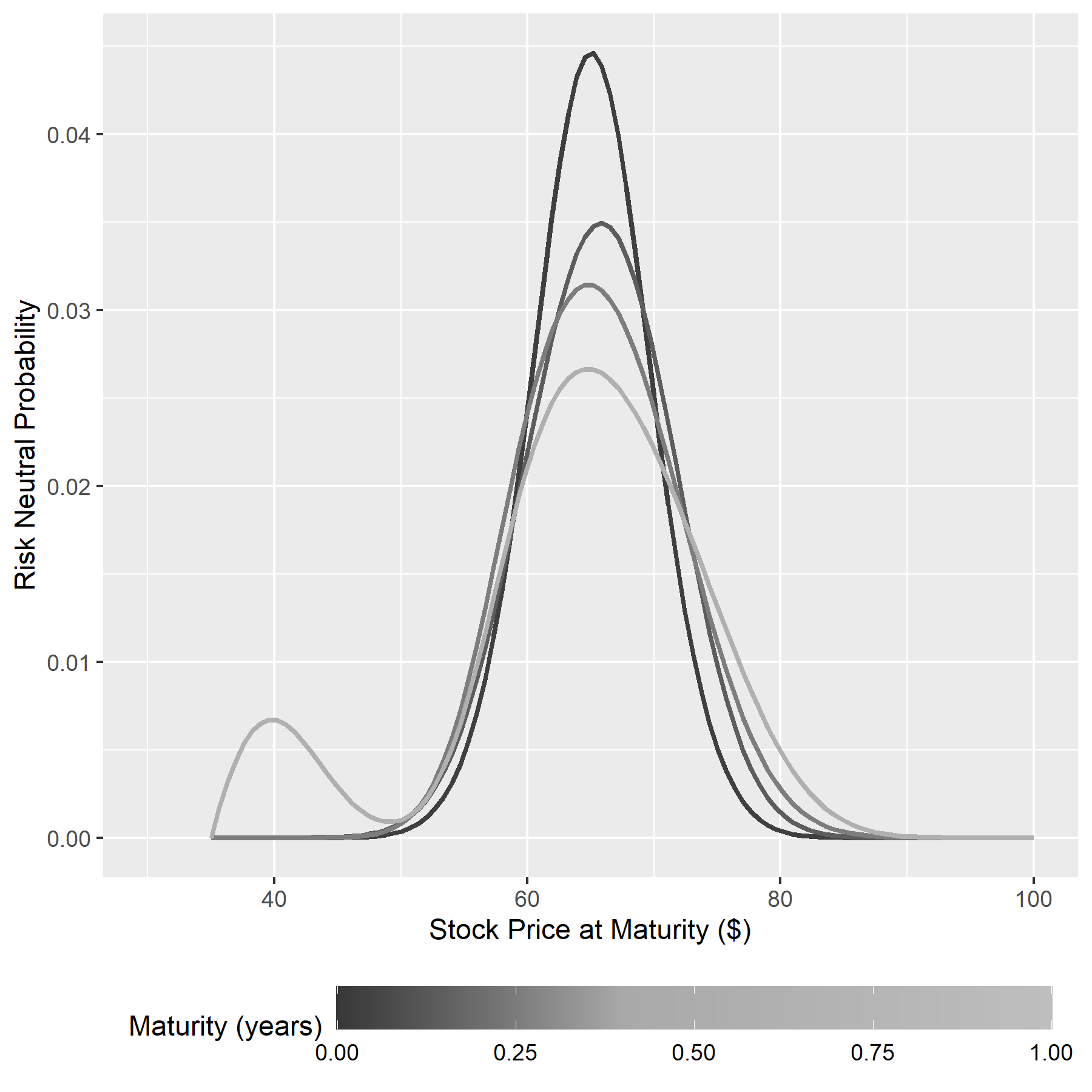} 

}~~~~~~~~~~ \subfloat[November 7, 2017]{\includegraphics[scale=0.5]{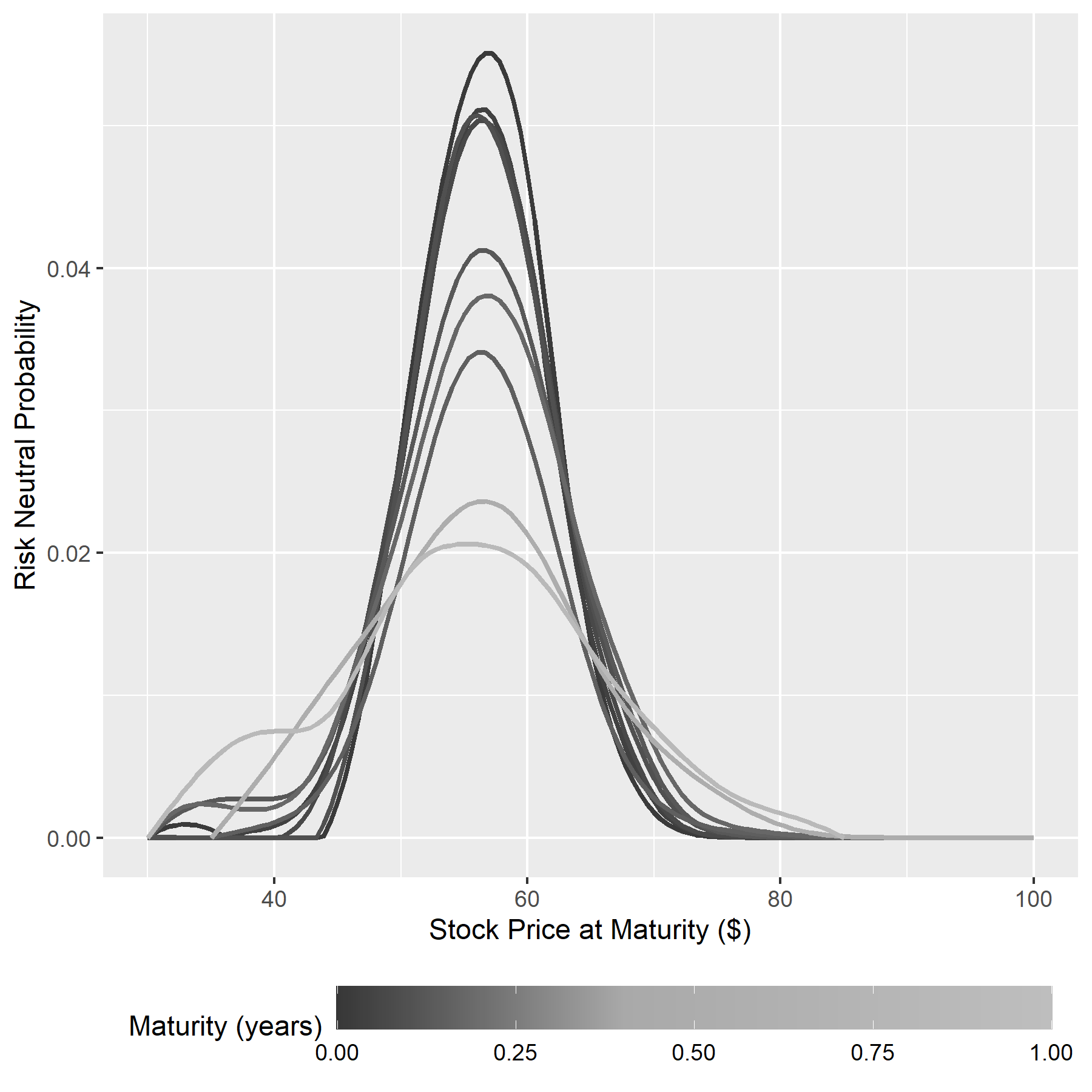} 

}
\par\end{centering}
\end{center}
\end{figure}

Results are presented on Figure~\ref{fig:pge_rnd_surface} for October
2, October 12, November 7 and November 30. The colors correspond to
the traded maturities of the options. The maturity is in fraction
of a year. The RND distribution suggests a clear asymmetry in investors'
risk neutral expectations. Hedges against tail risk appear as soon
as October 2, and become more evident during the wildfire (figure
(b)) and after the wildfire. In the aftermath of the wildfire, hedges
for downside risk appear for both shorter-dated and for longer-dated
options. The risk neutral distribution displays both negative skewness
and negative kurtosis that become more pronounced over time. % provide moments here.

\subsection{Panel Data Impacts on the Risk Neutral Distribution: Identification
Challenges}

\label{subsec:identification_concerns}

Estimating the causal impact of wildfires on listed firms' out-of-the-money
option prices, risk neutral probabilities, and implied volatility
is challenging for at least four reasons. First, firms affected by
wildfires may already have stock prices that are more likely to exhibit
average skewness and kurtosis, asymmetric jumps and stochastic volatility,
independently of wildfires. Second, wildfires may occur on days where
the market is overall more likely to experience skewness and kurtosis
in returns, jumps and stochastic volatility. Third, wildfires may
coincide with other adverse or positive events for the firm. Fourth,
option prices may exhibit arbitrage opportunities, including non-convexities.

The causal impact of wildfires on out-of-the-money option prices may
also be heterogeneous across firms depending on their adaptation or
expected ability to adapt to wildfire shocks. Firms that are expected
to be resilient to wildfires may experience a smaller or no increase
in the price of out-of-the-money puts. Firms that are expected to
gain or be affected only shortly may experience increases in the price
of out-of-the-money calls.

\subsection{Panel Data Estimates: Wildfires and Arrow Debreu State Prices\label{subsec:Panel-Data-Estimates:}}

We then turn to the estimation of the impact of wildfires on the risk
neutral distribution or Arrow Debreu state prices. The approach described
for PG\&E is applied systematically to stocks of the treatment group,
and, for computational reasons, to a random sample of the control
group of the same size as the treatment group. 

As in the previous sections, we wish to identify the causal impact
of wildfires over and above the confounding impact of shocks specific
to the day of wildfires and shocks specific to firms. The following
regression estimates the impact of wildfires controlling for non-time-varying
firm-level and day-specific confounders: 
\begin{equation}
f_{j,t,T,m}^{*}=\sum_{\ell=1}^{M}\nu_{\ell}\mathbf{1}(m=\ell)+\mathbf{1}(\textrm{Treated}_{j,t})\sum_{\ell=1}^{M}\delta_{\ell}\mathbf{1}(m=\ell)+\textrm{Firm}\times\text{Moneyness}_{jm}+\textrm{Date}_{t}+\varepsilon_{j,t,T,k}\label{eq:specification_rnd}
\end{equation}
where $f_{j,t,T,m}^{*}$ is the risk neutral distribution obtained
using the \citeasnoun{ait2003nonparametric} approach for firm $j$
on day $t$ at maturity $T$ for moneyness levels on a grid of $M$
equally-spaced moneynesses on the observed support. $\mathbf{1}(\textrm{Treated}_{j,t})$
an indicator variable equal to 1 either when (a)~firm $j$ is exposed
to a wildfire on day $t$, (b)~firm $j$ has been exposed to a wildfire
any time before $t$, (c)~firm $j$ has been exposed to its last
recorded wildfire any time before $t$. $\delta_{\ell}$ is the set
of coefficients of interest. $\nu_{\ell}$ captures the baseline distribution.
Standard errors are double-clustered as in \citeasnoun{cameron2011robust}
at the firm and date levels.

This specification is flexible as it allows an estimation of the impact
of wildfire on each part of the risk neutral distribution separately.
It is also demanding in terms of inference and statistical power when
the grid of moneyness levels $m=1,\dots,M$ is fine-grained.

Another approach is to estimate a treatment effect at each moneyness
level, weighing observations depending on how far (in a kernel weighting
sense) from the moneyness level of interest. We are interested in
measuring the treatment effect for moneyness level $m\in\{1,2,\dots,M\}$:
\begin{equation}
\text{TE}_{m}=E(f_{j,t,T,m}^{*}\vert\text{Treated}_{j,t}=1)-E(f_{j,t,T,m}^{*}\vert\text{Treated}_{j,t}=0),\label{eq:treatment_effect}
\end{equation}
and each expectation can be approximated using local polynomial regression.
This yields an estimator for the treatment effect based on a two-way
fixed effects regression:
\begin{equation}
f_{j,t,T,m}^{*}=\delta_{m}\mathbf{1}(\textrm{Treated}_{j,t})+\textrm{Firm}_{j}+\textrm{Date}_{t}+\varepsilon_{j,t,T,m},\label{eq:specification_rnd-2}
\end{equation}
weighted by a smooth function of the distance between the moneyness
of interest and the moneyness of the observation. This smooth function
is a Gaussian kernel here. We estimate the treatment effects with
different bandwidths of the kernel, to test for robustness. We also
control for firm and date fixed effects as before, and cluster standard
errors as before.

The sample has 630 tickers, 387 treated at least one day, 243 firms
in the control group, 45.25 million observations across 539 days pre-
and post-wildfires, a median maturity of 51 days. Appendix Figure~\ref{fig:wildfire_exposure}
displays the evolution of treated firms over time.

Different supports of strikes across firms, time, and maturity could
be a potential issue in the analysis. For instance, risk neutral probabilities
may not be changing but rather be traded more often. Two additional
regressions alleviate concerns about selection into the sample: 1)
a regression focusing only on firms with moneyness levels that include
at least the segment $[0.5,1.5]$ yields similar results, and 2) a
\emph{selection} \emph{regression }of the probability of being quoted
on the same set of covariates shows that options close to at the money
are more likely to enter the sample that options deep out of the money,
which is the opposite of what would be expected if tails were more
likely to be traded. 

\begin{figure}

\caption{Impact of Wildfires on the Risk Neutral Distribution}
\label{fig:impact_rnd_fixed_effects}

\emph{These figures present the regression of the risk neutral distribution
implied by Out of The Money (OTM) puts and calls, on the following
covariates: (1)~indicators for the 100 moneyness bins, (2)~interactions
between the treatment -- exposure to a wildfire -- and the 100 moneyness
bins, (3)~firm$\times$moneyness and day fixed effects. The points
correspond to (2), the interaction between the treatment and the 15
moneyness bins. Standard errors at 95\% are double-clustered at the
firm and day levels.}

\begin{center}

\subfloat[Treatment Effect on Risk Neutral Probabilities, Panel Data with Firm$\times$Moneyness
and Day Fixed Effects, 100 Moneyness Bins]{

\includegraphics[scale=0.4]{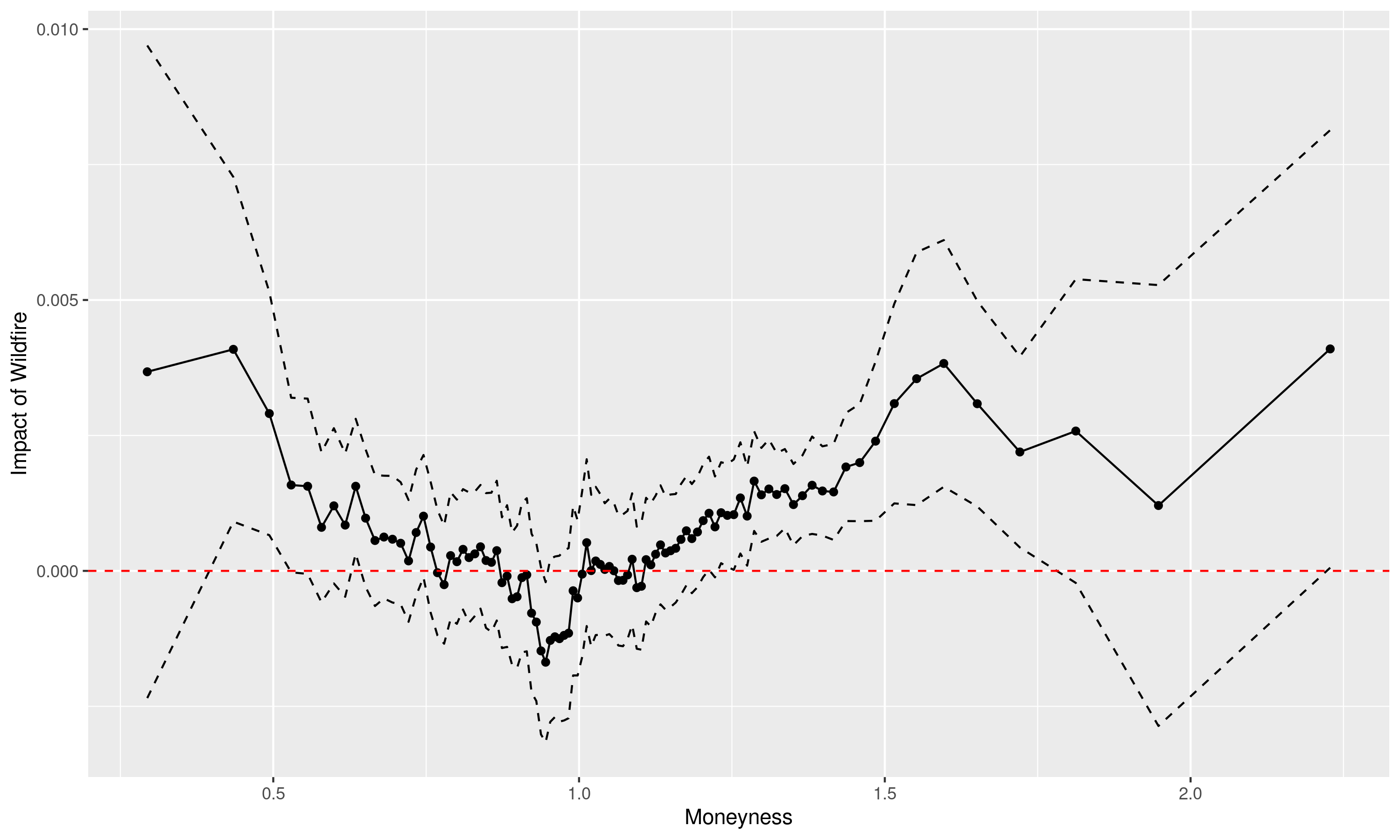}}

\subfloat[Treatment Effect on Risk Neutral Probabilities, Local Polynomial Regression,
Bandwidth=20]{

\includegraphics[scale=0.4]{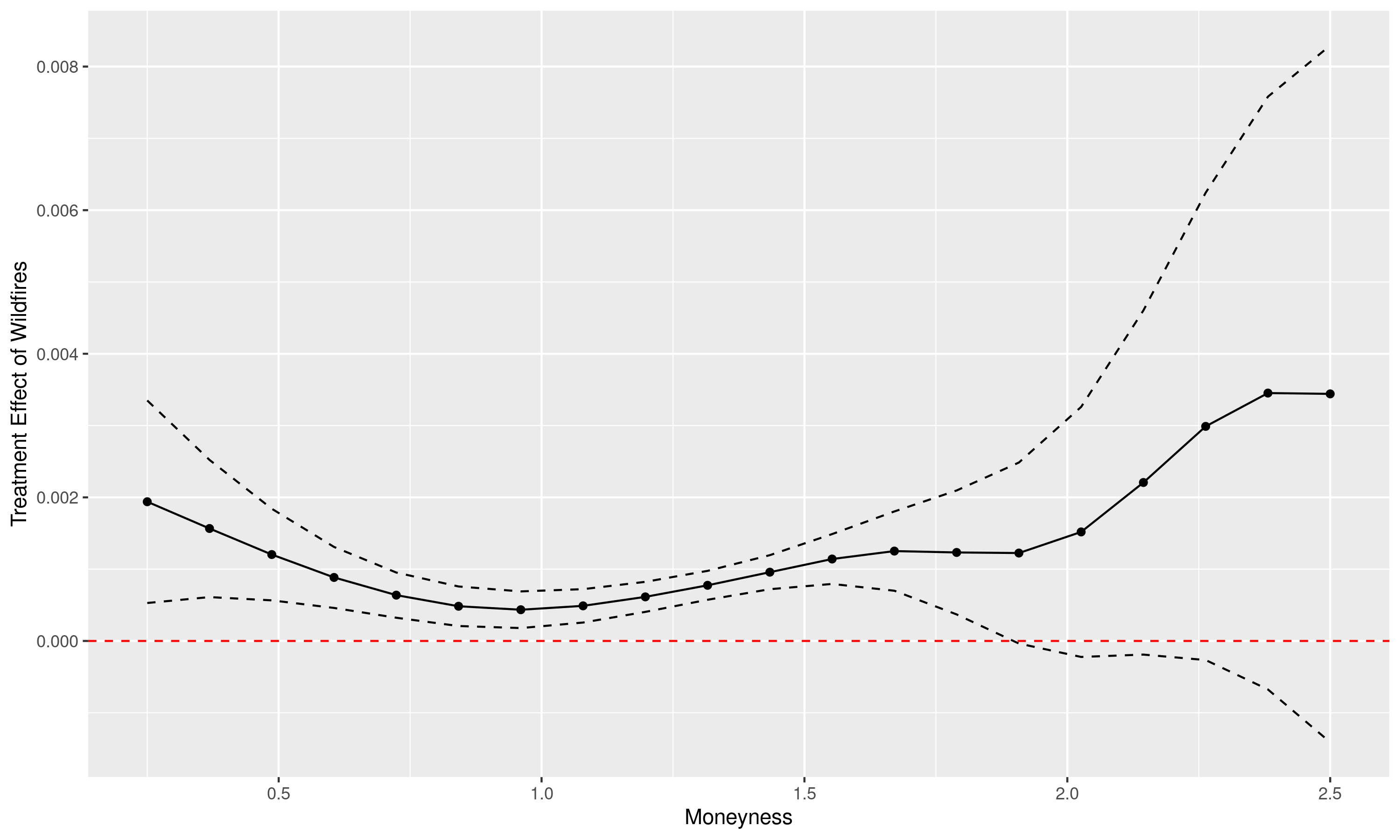}}

\subfloat[Treatment Effect on Risk Neutral Probabilities, Local Polynomial Regression,
Bandwidth=5]{

\includegraphics[scale=0.4]{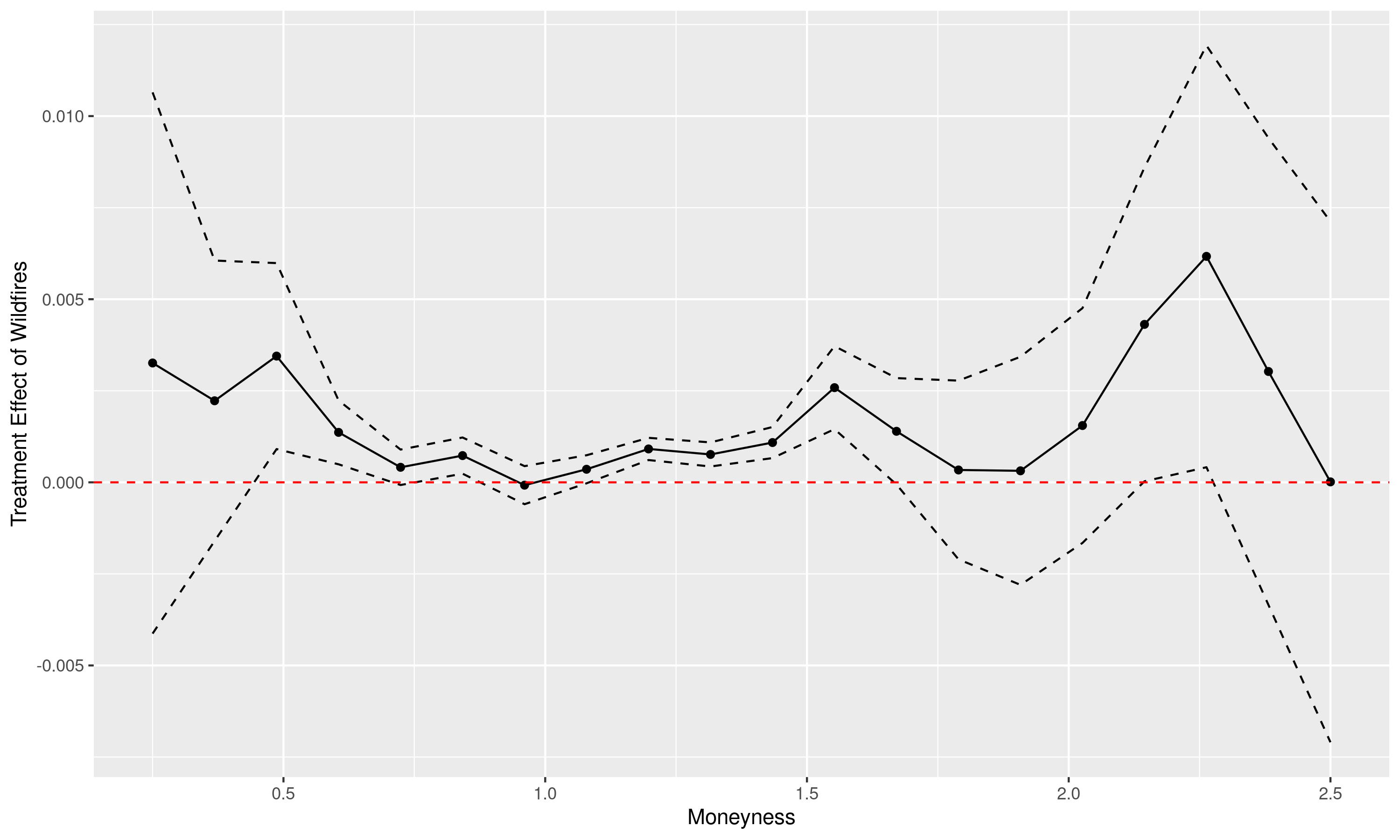}}

\end{center}
\end{figure}

Results are presented on Figure~\ref{fig:impact_rnd_fixed_effects}.
The figure and the table present estimates of the coefficients $\delta_{\ell}$
in specification (\ref{eq:specification_rnd}). The segment of moneyness
levels from 0.1 to 1.8 is divided into 30 quantiles of equal numbers
of observations. The first column presents results for out-of-the-money
puts, and the second column presents results for out-of-the-money
calls.

Results are presented on Figure~\ref{fig:impact_rnd_fixed_effects}.
Panel (a) presents the two-way fixed effect approach of specification~\eqref{eq:specification_rnd}
with 100 moneyness bins and double-clustered standard errors. Panels
(b) and (c) present the local polynomial approach~\eqref{eq:specification_rnd-2}
with two different bandwidths. Both approaches suggest a positive
impact of wildfires on the risk neutral probabilities at the tails. 

On panel (a), in the upper tail, the impact is significant at 95\%
for moneyness levels above 1.208 and below 1.762. In the lower tail,
the impact is significant for moneyness levels between 0.39 and 0.5132,
as well as between 0.54 and 0.57, and 0.61 and 0.64. Thus the impact
is more precisely estimated in the upper tail, suggesting expectations
of an upside.

On panels (b) and (c), the standard errors of the local polynomial
approach are less conservative. One benefit of panel (a) is that it
accounts for the joint correlation of residuals across moneyness levels.
Panels (b) and (c) are smoother as they use the entire panel for estimation
of each treatment effect. They suggest a higher risk neutral probability
at the tails, suggesting that wildfires lead to fat tails, or leptokurtic
distributions.

\begin{figure}
\caption{By Industry -- Impact of Wildfires on the Risk Neutral Distribution}
\label{fig:impact_rnd_fixed_effects-1}

\emph{These figures present the regression of the risk neutral distribution
implied by Out of The Money (OTM) puts and calls, for three separate
NAICS industries. Each point is the estimate of the impact of wildfires
on the risk neutral probabilities, controlling for firm$\times$moneyness
and day fixed effects. }

\begin{center}

\subfloat[NAICS 4 - Trade, Transportation and Warehousing]{

\includegraphics[scale=0.4]{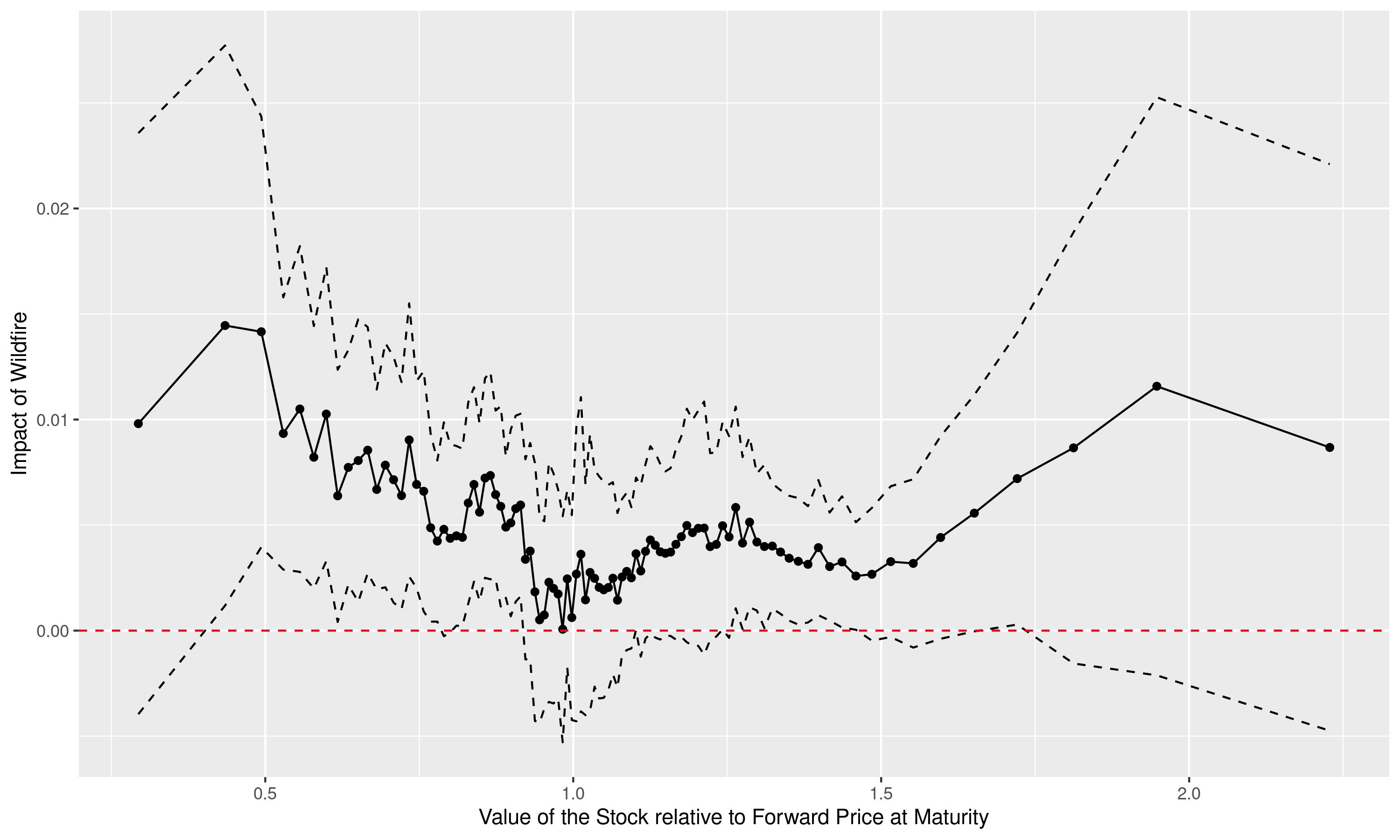}}

\subfloat[NAICS 5 -- Finance, Insurance, Real Estate]{

\includegraphics[scale=0.4]{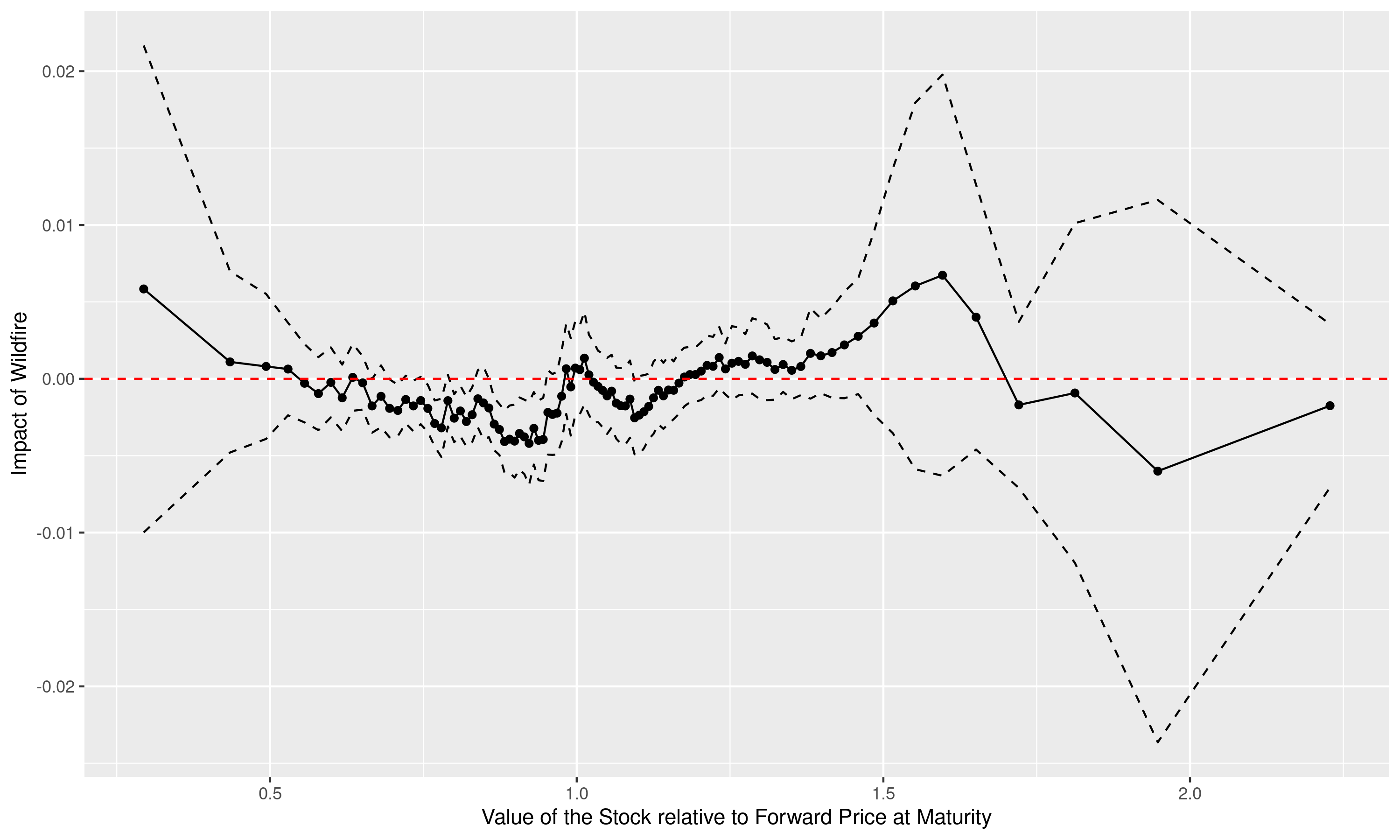}}

\subfloat[NAICS 7 -- Entertainment, Recreation, Accomodation, Food Services]{

\includegraphics[scale=0.4]{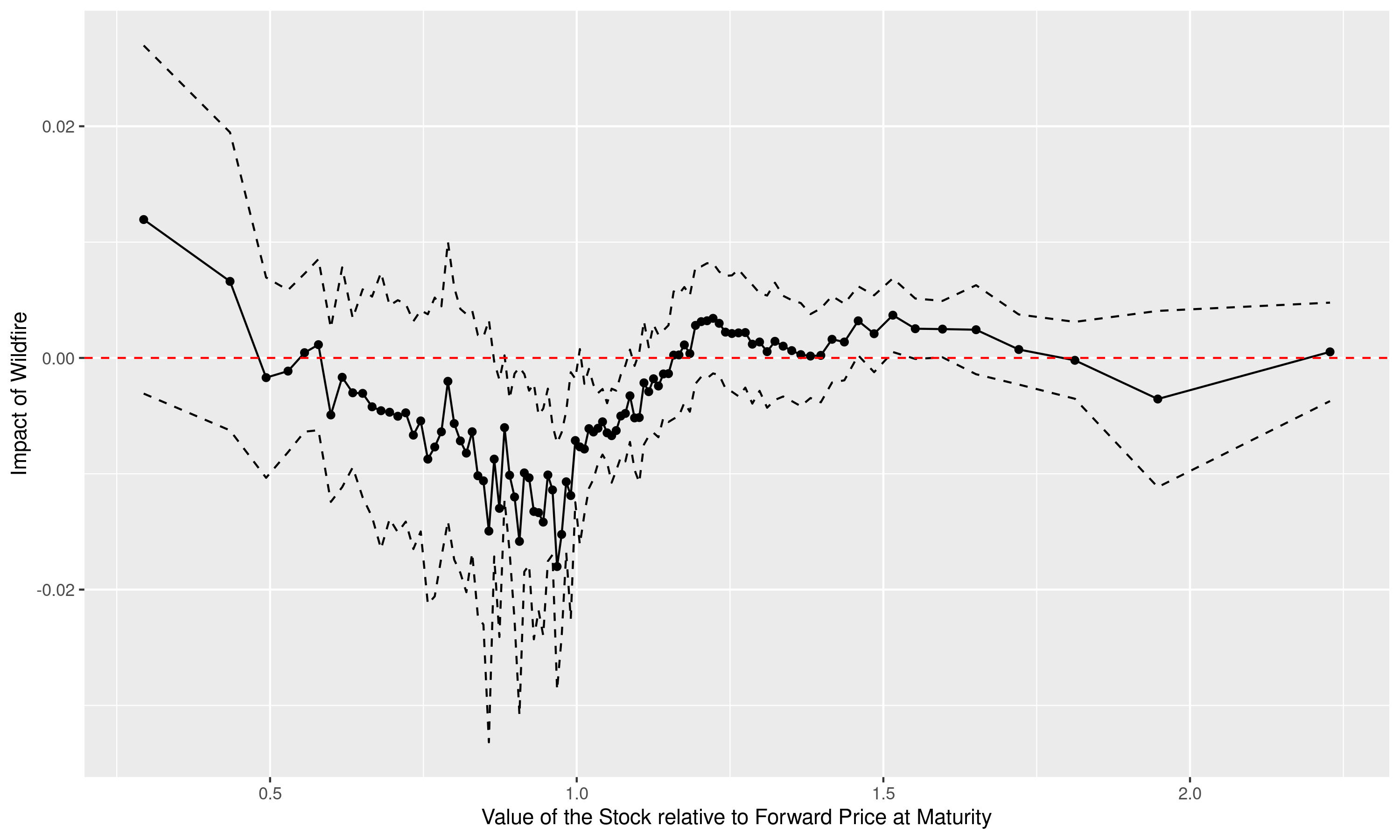}}

\end{center}

\emph{Dotted lines are clustered standard errors at 95\%. Black dots
are the coefficients of a 2-way fixed effect regression with firm$\times$moneyness
and day fixed effects.}
\end{figure}

This is clearly visible on Figure~\ref{fig:impact_rnd_fixed_effects-1},
which estimates the treatment effect separately for each 1-digit NAICS
industry. It keeps all observations of the control, but only retains
those observations of the treatment group that belong to a specific
1-digit NAICS. Limiting the number of observations could lead to more
imprecise estimates, but the results of panel (a), Figure~\ref{fig:impact_rnd_fixed_effects-1},
suggest a strongly significant impact on the risk neutral probability
for low moneyness states in the trade, transportation and warehousing
industry. Interestingly here the effect is asymmetric, with higher
expectations of a downside than an upside. Effects are significant
on both sides, but the point estimates are more than twice as large
on the downside. Panel (b) shows symmetric impact in the NAICS 5 --
Finance, Insurance, Real Estate, with a lower probability of almost
at-the-money stock prices. This is also visible on Panel (c), for
NAICS 7 -- Entertainment, Recreation, Accomodation, Food Services,
where there is a lower probability of being at the forward price at
maturity.

\begin{figure}
\caption{By Maturity: Impact of Wildfires on The Risk Neutral Distribution,
Panel Data Evidence}
\label{fig:rnd_panel-1}

\emph{These graphs present the non-parametric impact of a wildfire
on the risk neutral distribution, estimated using the longitudinal
panel of risk neutral probabilities, controlling for day and firm$\times$moneyness
fixed effects. Risk neutral probabilities (the dependent variable)
are implied by each firm's daily surface of call prices. Risk neutral
probabilities are derived using \possessivecite{breeden1978prices}
approach, implemented using \possessivecite{ait2003nonparametric}
method.}

\begin{center}

\subfloat[Treatment Effect, Maturity Below the Median]{

\includegraphics[scale=0.6]{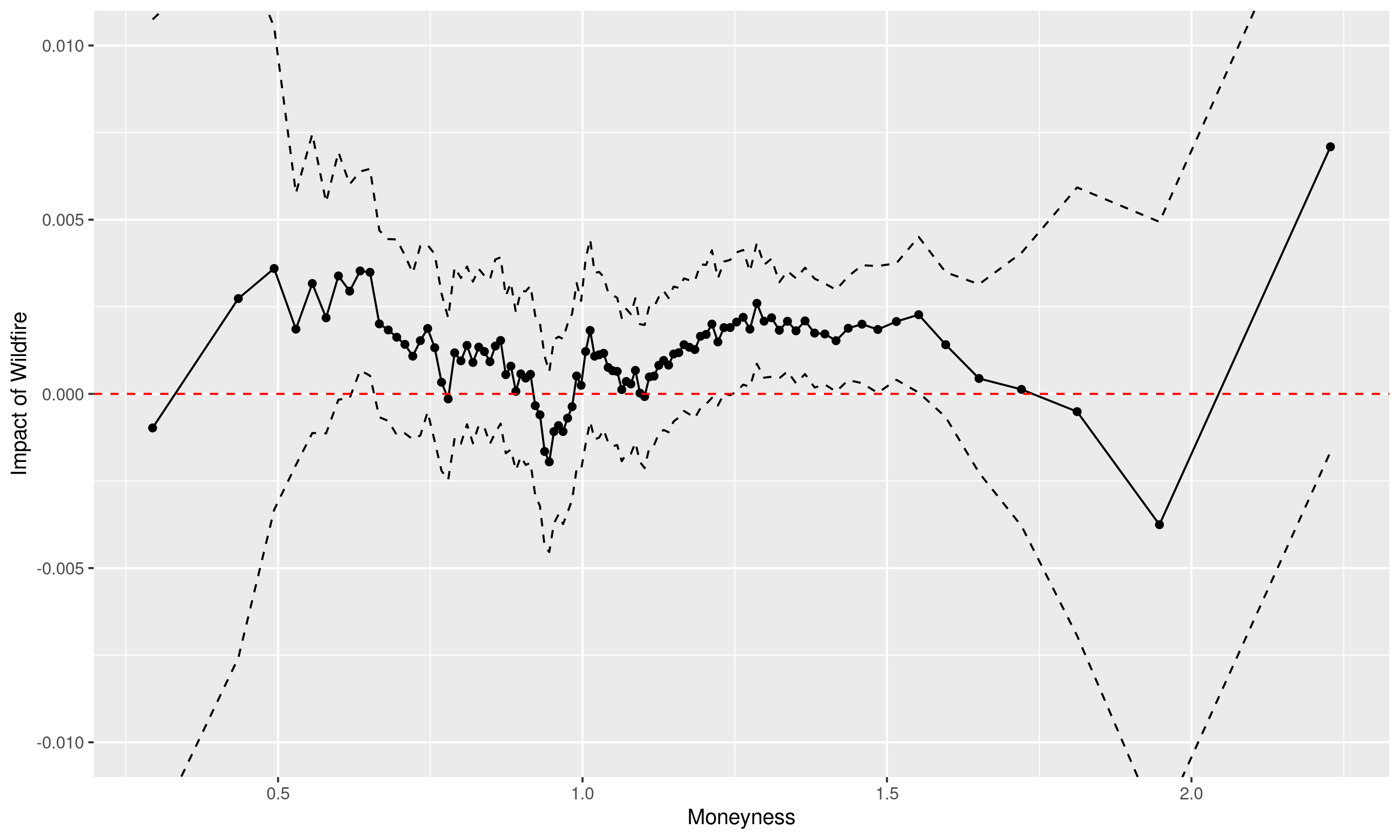}}

\subfloat[Treatment Effect, Maturity Above the Median]{

\includegraphics[scale=0.6]{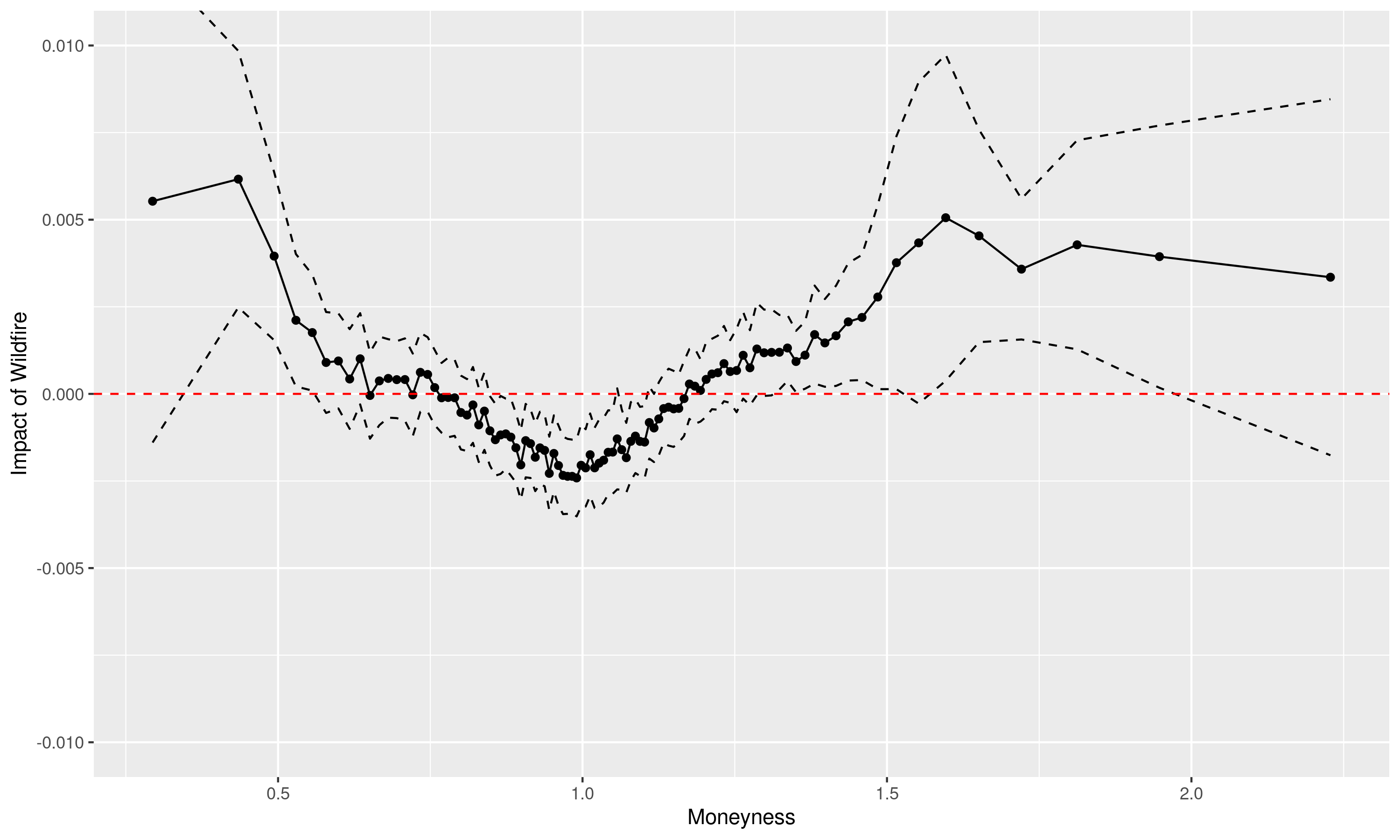}}

\end{center}

\emph{Dotted lines are clustered standard errors at 95\%. Black dots
are the coefficients of a 2-way fixed effect regression with firm$\times$moneyness
and day fixed effects.}
\end{figure}

Figure~\ref{fig:rnd_panel-1} performs an analysis by splitting the
sample in two, for maturity levels below the median of 51 days, and
above the median. Results are significant for both samples, but the
impact on the downside is mostly visible for higher maturity levels,
suggesting that investors hedge on the downside.

Overall, in a specification saturated with firm $\times$ moneyness
and day fixed effects and conservative double-clustered standard errors
with risk neutral probabilities obtained using \citeasnoun{breeden1978prices},
results suggest that wildfires have significant impacts on the risk
neutral probability distribution at the tails. While the effects seem
to price the upside more precisely than the downside on average for
the entire sample, the market hedges against the downside in the trade,
transportation, warehousing industry.

\section{Natural Disaster Risk and the Volatility Smile}

\label{sec:panel_data_evidence}The volatility is a deviation from
the standard Black and Scholes approach whose first appearance in
the aftermath of the 1987 Black Monday is well-documented~\cite{ait1998nonparametric,derman2016volatility}.
\citeasnoun{birru2012anatomy} suggests that such implied volatility
smile became more pronounced during the 2008 crisis. \citeasnoun{yan2011jump}
suggests that the slope of the smile is a significant predictor of
jump risk and future stock returns. 

Whether the physical risk exposure to wildfires leads to a \emph{steepening
}of such smile is an empirical question.

This section provides evidence of a steepening volatility smile and
skew for firms exposed to wildfires by a systematic panel data approach
on the daily implied volatility surface of listed equity options between
2000 and 2018. Estimating the impact of wildfires has an advantage
over the use of the risk neutral distribution, as it requires less
processing and relies solely on the binomial tree approach readily
available in OptionMetrics~\cite{cox1979option}. The smile or skew
focuses on the tails, which provides information on the pricing of
large deviations, complementary to the volatility information at the
money \cite{kruttli2021pricing}.

\subsection{The Risk Neutral Distribution and the Implied Volatility Smile}

We model below the relationship between the implied volatility surface
and the risk neutral probability distribution.

\possessivecite{black1973pricing} approach indeed parameterizes
underlying stock price dynamics using a single parameter, the volatility
$\sigma$, when this stock follows a diffusion process, $\frac{dS_{t}}{S_{t}}=rdt+\sigma dB_{t}$
where the risk-neutral trend of the stock is the risk-free rate, and
$B_{t}$ is a Brownian motion. In this world, option prices at different
strikes and maturities are redundant, as the price of one option,
typically the at-the-money (ATM) call option, provides the price of
options of all maturities and strikes.

The implied volatility $IV$ is the volatility that makes the observed
price at maturity $T$ and strike $K$ consistent with the Black and
Scholes price. While stocks do not follow geometric Brownian motions,
the market price of an option is typically quoted through its Black-Scholes
implied volatility. If $C$ is the observed option price, and $BS(\sigma,K,T,S_{t},r)$
is the Black and Scholes price of an option with volatility $\sigma$,
the implied volatility is: 
\begin{equation}
C=BS(IV(C,K;T,S_{t},r),K;T,S_{t},r),
\end{equation}
where $r$ is the risk-free rate. $IV$ is independent of the strike
$K$ and maturity $T$ if the stock follows a geometric Brownian motion.

A more pronounced volatility smile is typically associated with a
corresponding shift in the underlying risk-neutral distribution, or
Arrow Debreu state prices~\cite{bakshi2003stock}. However \citeasnoun{bakshi2003stock}
documents that the moments of the risk neutral distribution are typically
different from the distribution of implied volatilities. This is visible
by noticing that the first and second derivatives of the implied volatility
with respect to the strike is a weighted average of the risk neutral
cdf and the risk neutral pdf, with non-linear weights. In general,
applying the chain rule:

\begin{equation}
\frac{dIV}{dK}(C,K)=\frac{\partial IV}{\partial C}\frac{\partial C}{\partial K}+\frac{\partial IV}{\partial K}=-e^{-r(T-t)}\left[1-F^{*}(S_{T})\right]\frac{\partial IV}{\partial C}+\frac{\partial IV}{\partial K}\label{eq:relation_dIV_cdf}
\end{equation}
where the implied volatility function is a deterministic function
of the strike $K$and the price $C$ of the call, the inverse of the
Black and Scholes pricing function $BS$ w.r.t. to the volatility. 

Thus the slope $dIV/dK$ in implied volatility along the range of
out-of-the-money strikes is related to a shift in the cumulative distribution
function, weighted by $\frac{\partial IV}{\partial C}$ and with a
constant $\frac{\partial IV}{\partial K}$. An increasing slope of
the implied volatility surface at the tails is related higher values
of the cumulative distribution function.\footnote{The convexity of the implied volatility smile is weighted average
of the cdf and the pdf. 
\begin{equation}
\frac{d^{2}IV}{dK^{2}}=-e^{-r(T-t)}f^{*}(S_{T})\frac{\partial IV}{\partial C}-e^{-r(T-t)}\left[1-F^{*}(S_{T})\right]\frac{\partial^{2}IV}{\partial C^{2}}+\frac{\partial^{2}IV}{\partial K^{2}}\label{eq:relationship_IV_pdf}
\end{equation}
Thus, in general the relationship between the implied volatility surface
$IV(C,K)$ and the risk neutral probability distribution $f^{*}(S_{T})$
is non-linear.} 

\subsection{PG\&E's Volatility Smile During the October 2017 Wildfires\label{subsec:PG=000026E's-Volatility-Smile}}

\begin{figure}
\caption{The Volatility Surface of PG\&E Before and During the October 2017
Northern California wildfires}
\label{fig:pge_iv_surface} \emph{The implied volatility surface is
flat when the underlying stock follows a Geometric Brownian Motion.
Here, in contrast, data from PG\&E before, during, and after the October
2017 wildfires displays a steepening implied volatility smile. Results
are described in Section~\ref{subsec:PG=000026E's-Volatility-Smile}.}

\medskip{}

\begin{center}
\begin{centering}
\subfloat[September 22, 2017]{\includegraphics[scale=0.45]{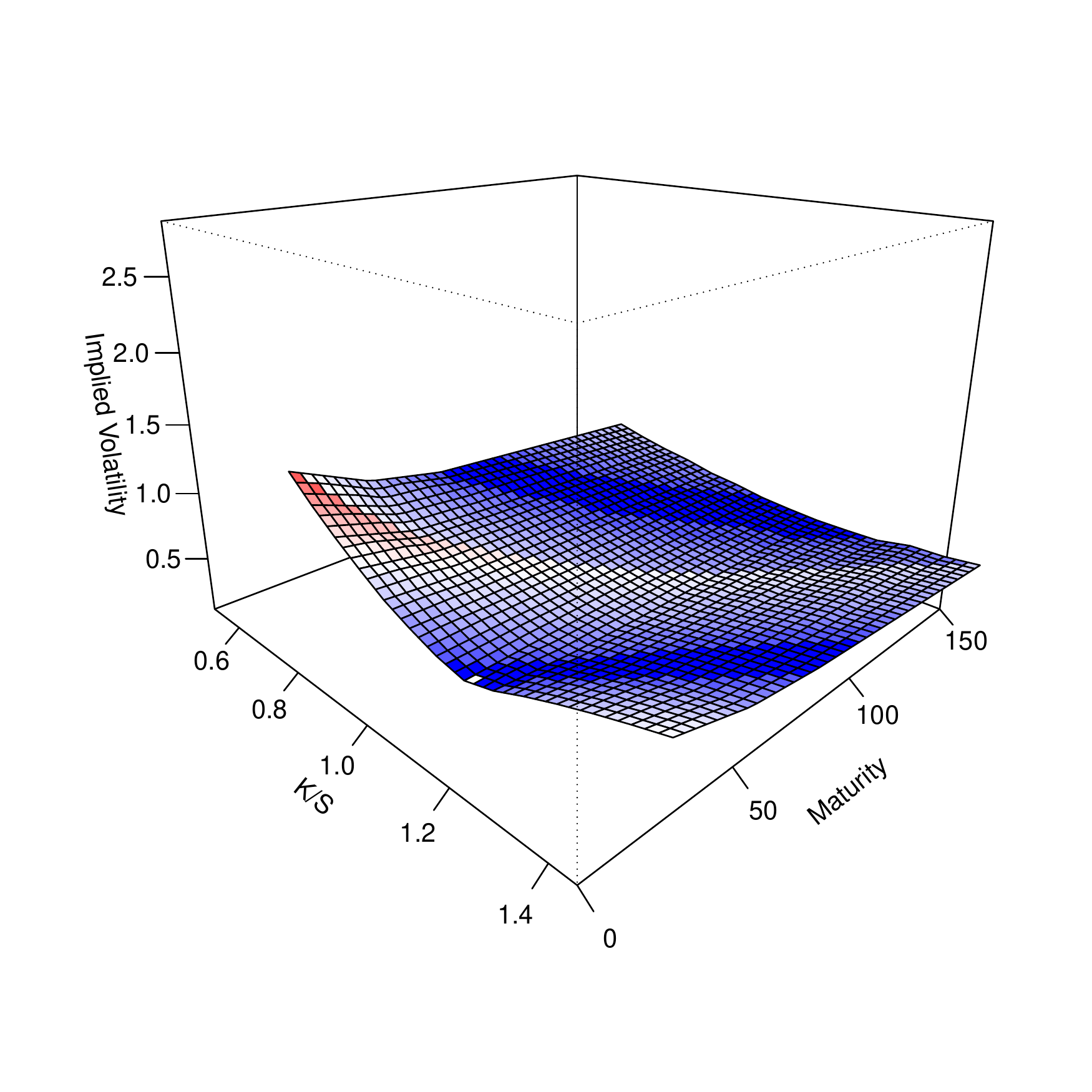}

}\subfloat[September 27, 2017]{\includegraphics[scale=0.45]{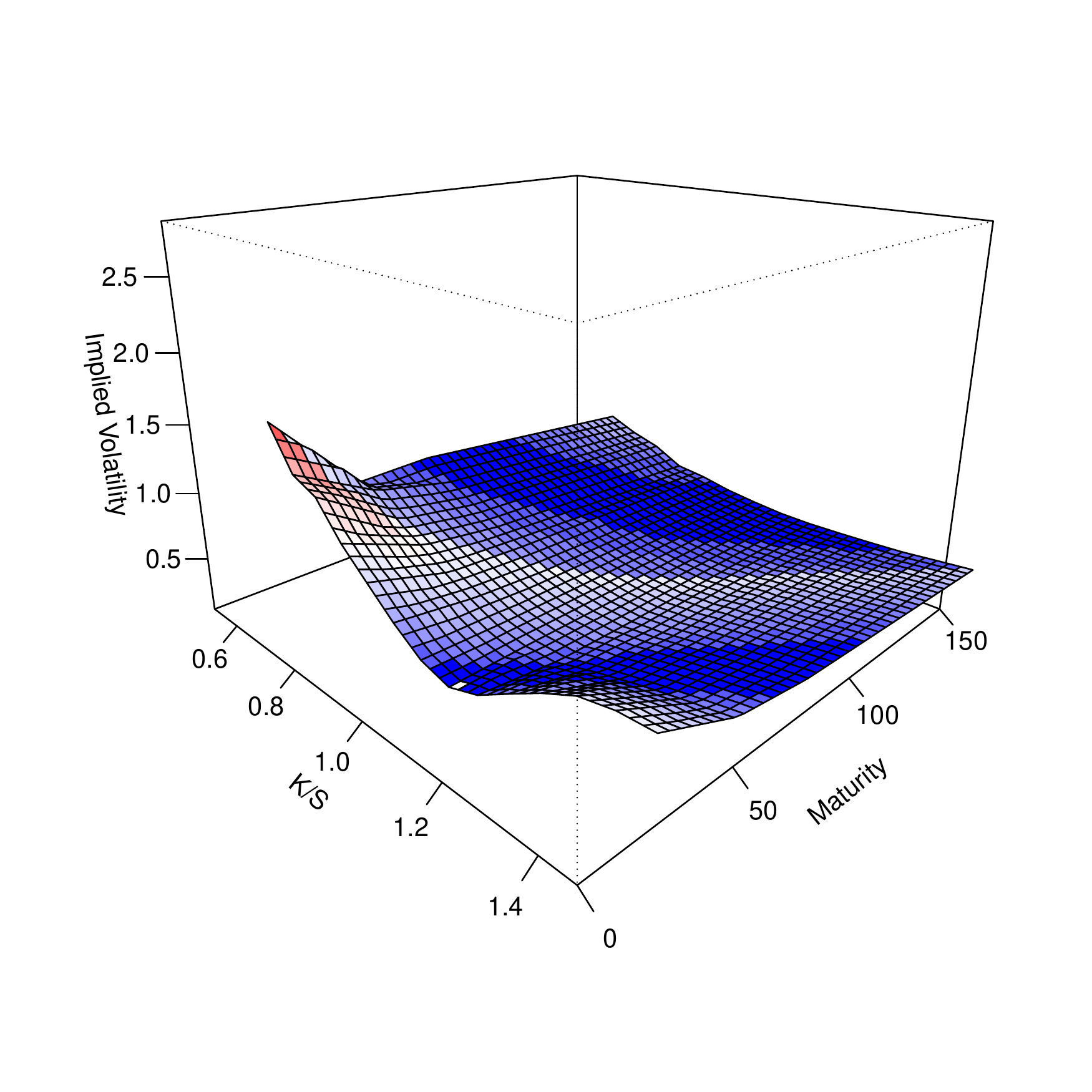}

}
\par\end{centering}
\begin{centering}
\subfloat[October 12, 2017]{\includegraphics[scale=0.45]{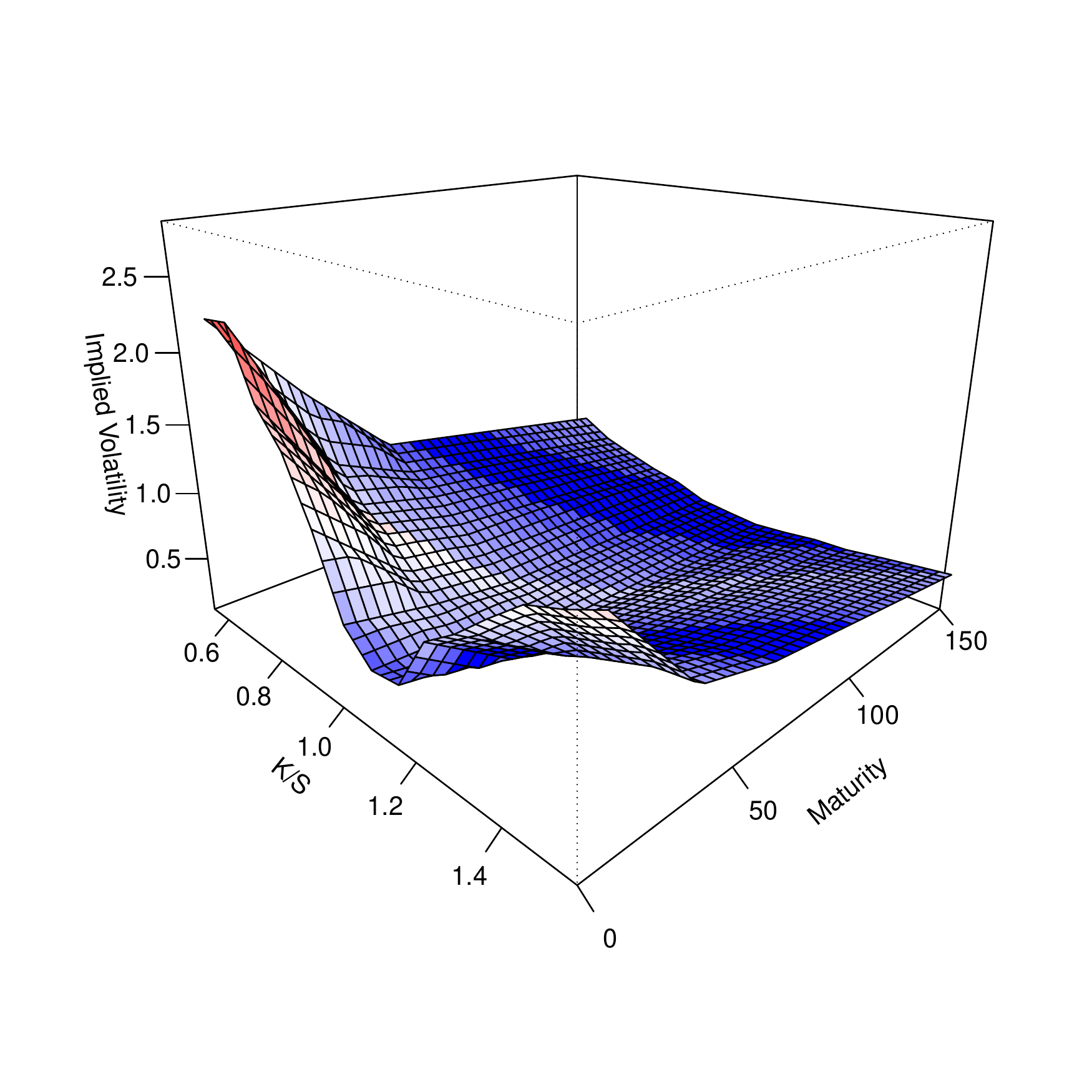}

}\subfloat[November 7, 2017]{\includegraphics[scale=0.45]{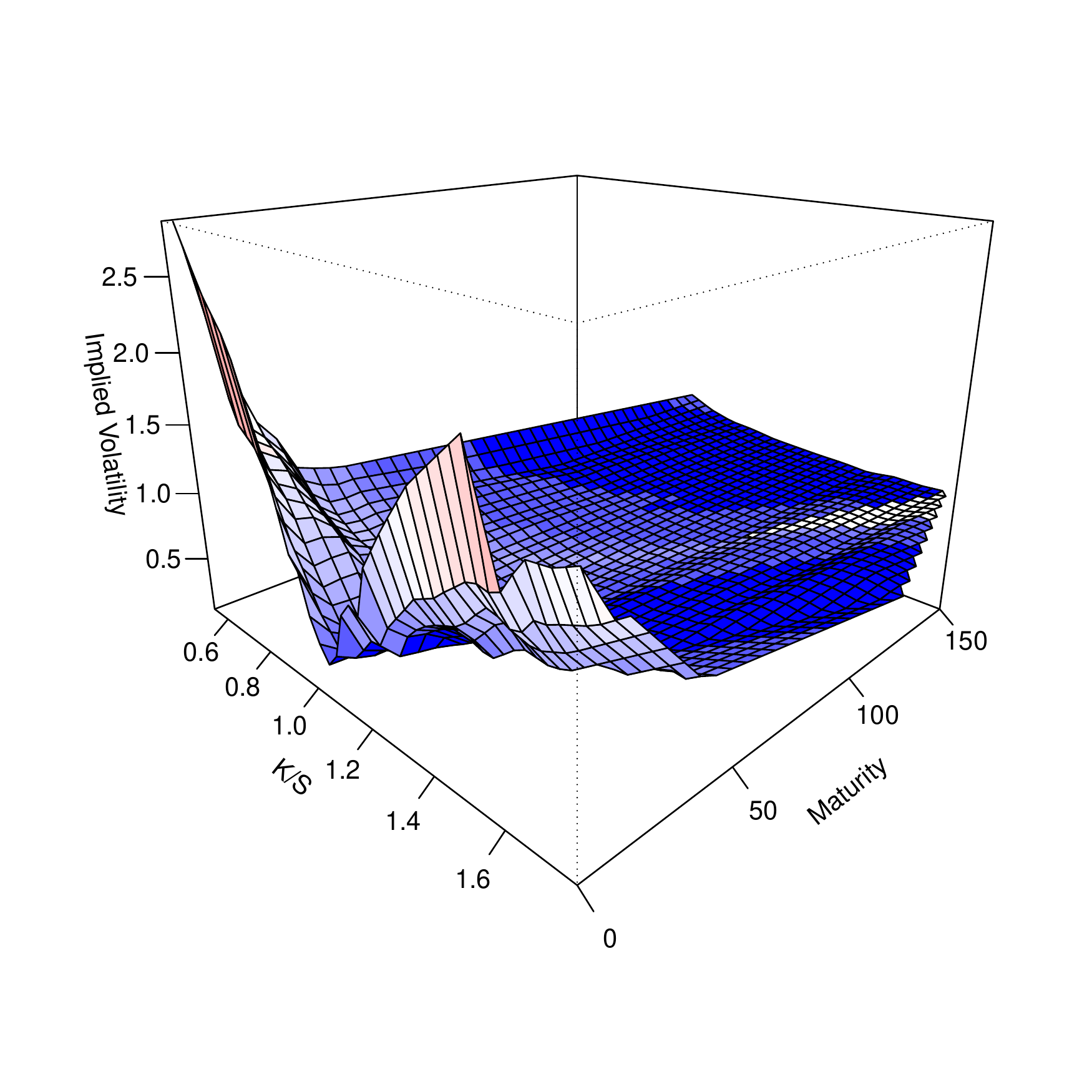}

}
\par\end{centering}
\end{center}

\emph{Implied volatility from OptionMetrics, obtained using \possessivecite{cox1979option}
binomial-tree approach for American options.}
\end{figure}

Figure \ref{fig:pge_iv_surface} presents the evolution of the implied
volatility smile for the PG\&E stock (ticker: PCG) during the October
2017 wildfires. Maturity is in days, and $K/S$ is the ratio of the
strike over the forward price. Axis scales are constant across figures.
The figures consider out-of-the-money calls and puts only. Hence,
when $K/S>1$, the data is that of calls, while $K/S<1$ is the implied
volatility data for puts.

The wildfires started on the evening of 8 October 2017, and lasted
until October 31, 2017. Figures 1(a) and 1(b) are thus estimated prior
to the beginning of the wildfires; Figure 1(c) is the set of implied
volatilities during the wildfire and 1(d) is drawn after the wildfires.
The graphs suggest two phenomena. First, the trading and pricing of
shorted-dated options. On September 22, 2017, options with maturities
of 28, 56, 84, and 175 are quoted. On October 12, options with maturities
of 8, 36, 64, 155 days are traded. On November 7, there are 9 maturities
traded: 3, 10, 17, 24, 31, 38, 45, 129, 220 days. This thicker market
is also visible in the number of traded strikes: from 14 strikes until
October 12, to 46 smiles on November 7. The minimum traded strike
also decreases, from \$35 to \$30. The second phenomenon is the increase
in implied volatilities that is larger for out-of-the-money options
than for at-the-money options: on September 22, the ATM implied volatility
ranges between 0.13 and 0.148 from short-dated to long-dated options.
The smile is already visible on September 22, with implied volatilities
for deep out of the money ranging between 0.96 and 0.42. The smile
becomes substantially more pronounced during the wildfires, with deep
OTM IVs for puts ranging between 2.19 (short-dated) and 0.45, while
ATM calls and puts have IVs ranging between 0.28 and 0.19. This smile
becomes significantly more pronounced in November. Hence, while ATM
implied volatilities increase during the wildfire, most of the volatility
is at the tails, for deep out-of-the-money calls and puts.

While the PG\&E smile suggests that information is contained in deep
out of the money options, it also suggests controlling for sources
of endogeneity, such as pre-existing trends in deep OTM implied volatility
and for market-level volatility. These endogeneity concerns and the
identification strategy conditioning for unobservables are addressed
in Section~\ref{subsec:identification_concerns}.

\subsection{The Out-of-the-Money Volatility Smile: Linear Fixed Effect Panel
Regression}

A `model-free' linear specification addresses the first three identification
concerns. The effect of exposure to a wildfire on the volatility surface
by strike and maturity is estimated controlling for firm-specific
and time-specific unobservables with a specification similar to \citeasnoun{dumas1998implied}.
\begin{eqnarray}
IV(K/S,T)_{i,t} & = & \alpha+\beta(K/S)_{it}+\gamma\sqrt{T}+\delta\sqrt{T}\times(K/S)_{it}\nonumber \\
 &  & +Treated_{i,t}~(\alpha^{\tau}+\beta^{\tau}(K/S)_{it}+\gamma^{\tau}\sqrt{T}+\delta^{\tau}\sqrt{T}\times(K/S)_{it})\nonumber \\
 &  & +Firm_{i}+Date_{t}+\varepsilon_{it}\label{eq:linear_specification}
\end{eqnarray}
A separate regression is estimated for calls and for puts. $IV(K/S,T)_{i,t}$
is the implied volatility of option $i$ on day $t$ obtained using
binomial trees. It is regressed on the moneyness of the option $K/S$,
i.e the ratio of the strike over the forward price. The coefficient
$\beta$ on this first covariate captures the volatility skew. $\sqrt{T}$
is the square root of maturity. $\delta$ measures a phenomenon observed
in the case of PG\&E on Figure~\ref{fig:pge_iv_surface}: it measures
how the volatility skew depends on the maturity of the option. $\textrm{Treated}_{it}=0,1$
is an indicator variable for the treatment status. 

The coefficients of interests are $\beta^{\tau}$, $\gamma^{\tau}$,
and $\delta^{\tau}$. $\beta^{\tau}$ measures the impact of wildfires
on volatility skew. $\gamma^{\tau}$ measures the impact of the wildfire
on the term structure of implied volatilities. $\delta^{\tau}$ captures
the impact of wildfires on the variation in the skew along maturities.

$\textrm{Firm}_{i}$ captures non-time-varying firm-specific differences
in implied volatilities. $\textrm{Date}_{t}$ controls for increases
or declines in implied volatility specific to each day. Standard errors
are double-clustered at the firm and date levels, as in \citeasnoun{cameron2011robust}.

A firm is treated if either 10\% of its establishments, its employment
or its sales are in a ZIP code affected by a wildfire on day $t$.
Table~\ref{tab:by_industry}, panels (a) and (b) presents industry-level
statistics on the share of treated firm $\times$ date observations.
74.4\% of 6-digit NAICS industries have at least one treated firm
$\times$ date observation, the median industry has 1.2\% of treated
observations, and the industry with the largest share has 8\% of treated
observations. The top 3 is Home Health Care Services, Ammunition Manufacturing,
Petroleum and Petroleum Productions Merchant Wholesalers. These panels
suggest that the share of treated firm $\times$ date observations
is quite homogeneous across industries, and that this paper's results
are therefore unlikely to be due to a single or a few industries. 

\subsection{The Out-of-the-Money Volatility Smile: Non-Parametric Analysis}

\label{subsec:non-parametric-panel-estimates}

While the linear panel fixed effect provides a model-free set of estimators
that are arguably independent of the specific parametric assumptions
of option pricing models, the functional form of the implied volatility
is better captured by a non-parametric approach relying on flexible
function forms. Option pricing that feature stochastic volatility
\cite{heston1993closed}, jumps \citeasnoun{kou2004option}, or
combine both stochastic volatility and jumps \citeasnoun{duffie2000transform}
predict more general functional forms for the implied volatility.
Two models are identified in Section~\ref{sec:Calibrating-Risk-Neutral}.

An alternative approach regresses the implied volatility of the set
of options on a flexible functional form that differs between the
control and the treatment groups. 
\begin{eqnarray}
IV(K/S,T)_{i,t} & = & g(K/S,T)+\mathbf{1}(Treated_{i,t})\Delta g(K/S,T)+Firm_{i}+Date_{t}+\varepsilon_{it},\label{eq:nonparametric}
\end{eqnarray}
where $g(K/S,T)$ is a two-dimensional function of moneyness and maturity
for the control group. $\Delta g(K/S,T)$ is the impact of the treatment
on the functional form. As before, the regression includes firm and
day fixed effects. These functional forms are estimated using a Frisch-Waugh-Lovell
approach by first orthogonalizing the implied volatilities with respect
to the firm and day fixed effects. This yields $IV^{\perp}(K/S,T)_{it}$.
The functional forms $g$ and $g+\Delta g$ are then estimated using
a local polynomial regression \cite{cleveland1979robust} of order
2 of $IV^{\perp}(K/S,T)_{it}$ on the surface of $(K/S,T)$.

\subsection{Linear Panel Data Results}

Estimates of the impact of a wildfire on the volatility skew are presented
in Table~\ref{tab:panel_data_evidence}. The upper panel is for puts,
and the lower panel for calls. Column (1) is an OLS regression with
double-clustered standard errors. Column (2) includes firm fixed effects.
Column (3) includes day fixed effects. Column (4) is a two-way fixed
effect regression with double-clustered standard errors.

Baseline coefficients suggest a volatility skew for both puts ($\beta<0$)
and for calls ($\beta>0$), inconsistent with the Black and Scholes
geometric Brownian motion. Coefficients also suggest that implied
volatility is a decreasing function of the square root of maturity,
for both puts and calls ($\gamma<0$). The skew is less pronounced
for longer-dated options ($\delta$) in the control group.

In the treatment group, results suggest that the volatility skew is
significantly steeper for both puts and for calls. In the regression
including both fixed effects, the coefficient $\beta^{\tau}$ is significant
at 5\% for both puts and calls. For puts, its magnitude is approximately
$0.159/0.413=38.5\%$ of the baseline skew. For calls, its magnitude
is $0.155/0.048=3.2$ times the magnitude of the baseline skew. For
calls, the effect is statistically significant in columns (1)--(3).
Thus the impact on puts is more precisely estimated than on calls,
even though the magnitude of the impact on calls is larger. This suggests
an asymmetry and a heterogeneity in the impact of wildfires: wildfires
may lead to significant increases in the prices of out-of-the-money
puts, which insure investors against downside risk. And more heterogeneity
in the impact on the price of out-of-the-money calls, which allow
investor to sell the upside.

Table~\ref{tab:by_industry}, panel (c), presents results estimated
separately for each 3-digit NAICS industry. For puts, results suggest
that the steepening of the volatility smile of out-of-the-money puts
is present in a majority of industries (66\%), and that 57\% of industries
exhibit a steepening significant at 99\%. For calls, 26\% of industries
exhibit a steepening of the volatility smile, suggesting an asymmetry
between downward risk (observable with puts) and upward risk (observable
with calls). For calls, 24.3\% of industries exhibit an effect significant
at 99\%.

\subsection{Non-Parametric Treatment Effect Along the Volatility Surface}

\begin{figure}
\caption{The Impact of Wildfires on the Volatility Smile: Non-Parametric Approach}
\label{fig:non_parametric}

\emph{These four figures present the estimation of the non-parametric
specification~(\ref{eq:nonparametric}), using both in the money
and out of the money options. Implied volatilities are orthogonalized
with respect to firm and day fixed effects. A local polynomial regression
then regresses orthogonalized implied volatilities on strike and maturity
in the control and the treatment groups. Figure (a) presents the IV
surface in the control group, denoted $g$. Figure (b) presents the
treatment effect, denoted $\Delta g$. Hence the IV surface in the
treatment group is $g+\Delta g$. Figure (c) presents a cross section
for calls of 3-day maturity. Figure (d) presents the cross section
for calls of 11-day maturity. Cross sections of puts display similar
results. Cross sections of option IVs with longer maturities display
similar results as for the 11-day maturity.}

\begin{center}

\subfloat[IV Surface, Control Group $g$]{\includegraphics[scale=0.45]{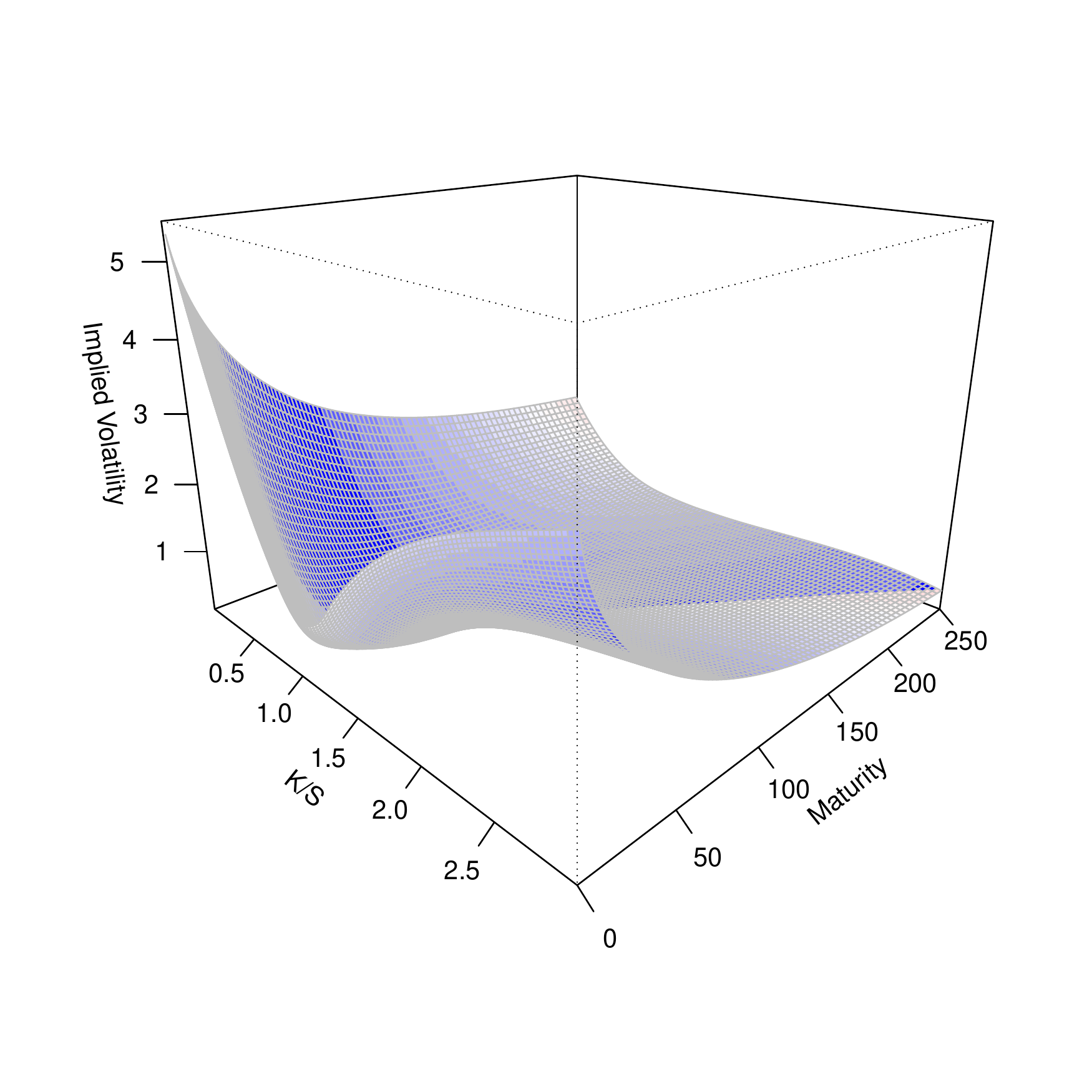}

}\subfloat[IV Surface, Treatment Effect, $\Delta g$]{\includegraphics[scale=0.45]{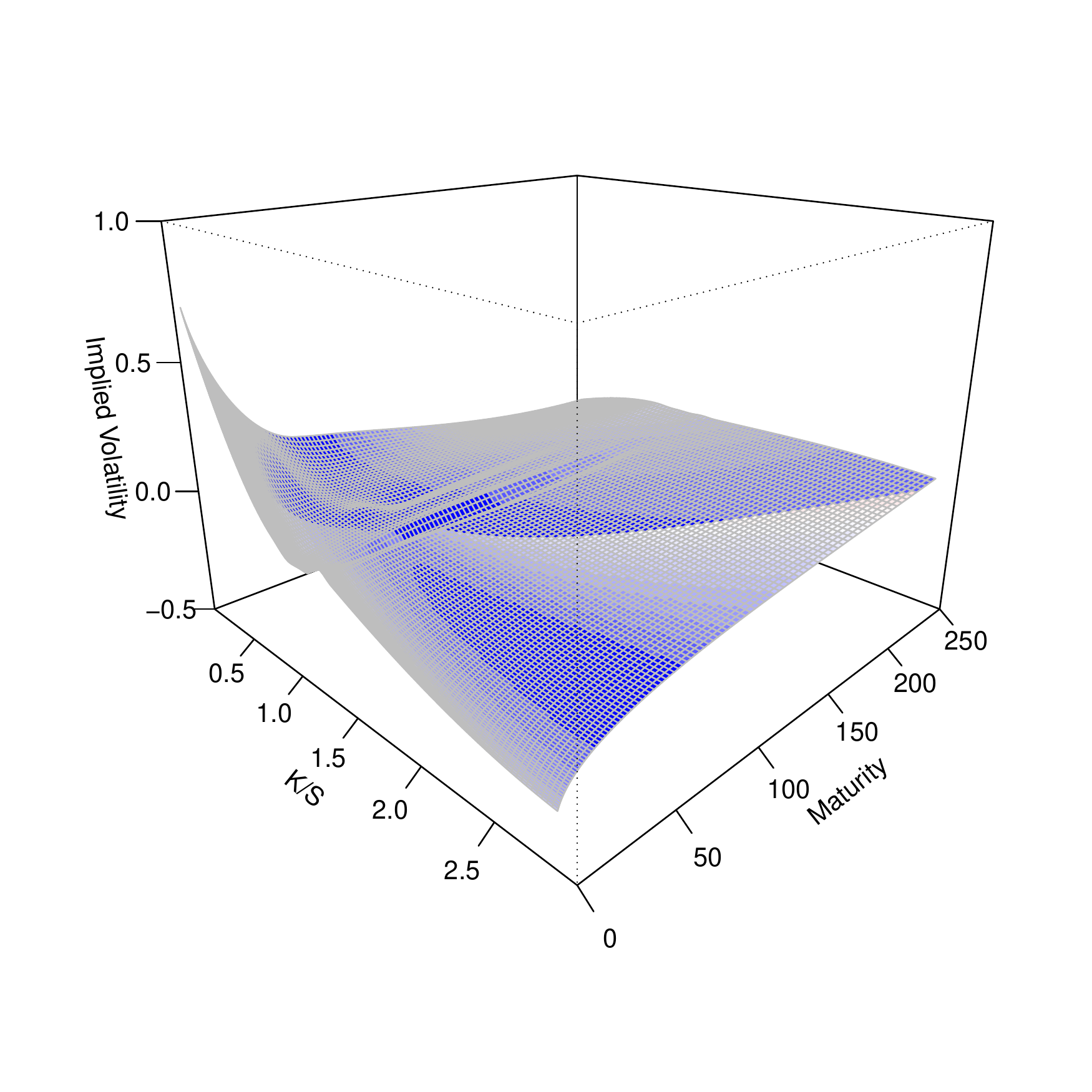}

}
\begin{centering}
\subfloat[IV Cross Section, Calls, 3-Day Maturity]{\includegraphics[scale=0.4]{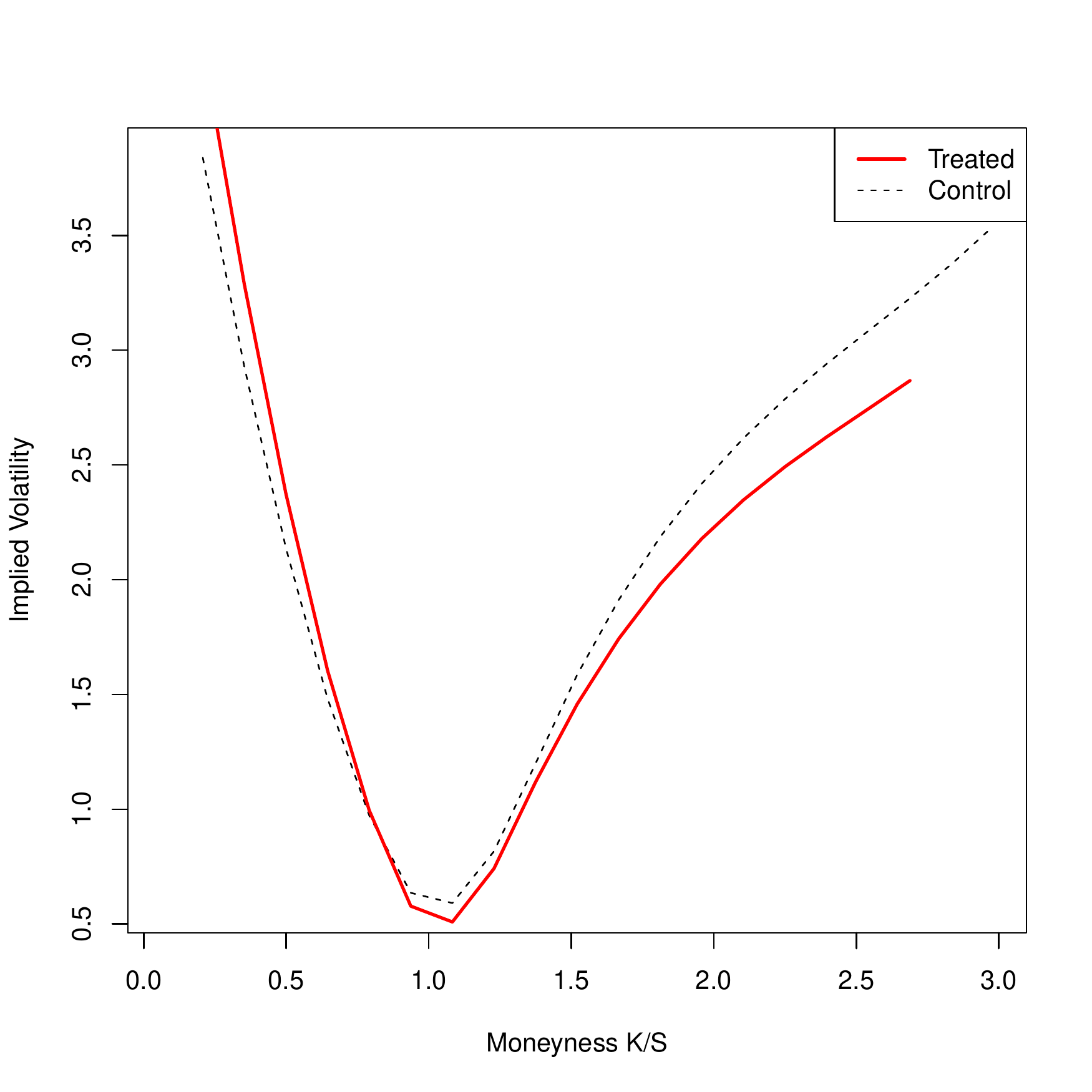}

}\subfloat[IV Cross Section, Calls, 11-Day Maturity]{\includegraphics[scale=0.4]{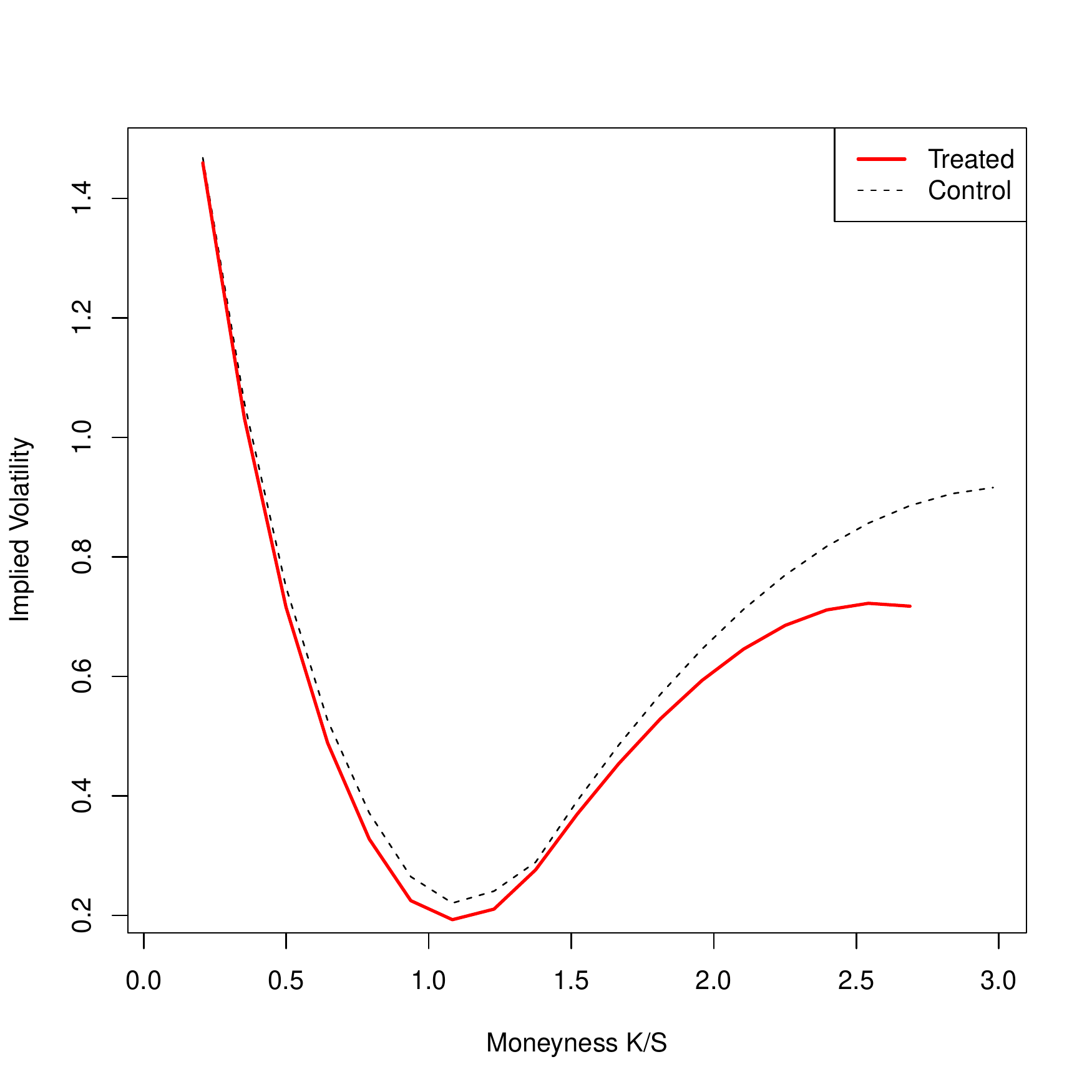}

}
\par\end{centering}
\end{center}
\end{figure}

The non-parametric enables an estimation of the treatment effect that
is specific to each strike and maturity. As such, it also reassures
us that the linear specification used previously is not driving results.
The non-parametric results are presented on Figure~\ref{fig:non_parametric}.
Figure (a) presents the shape of the IV surface in the control group,
while (b) presents the shape of the treatment effect $\Delta g$,
as defined in specification~(\ref{eq:nonparametric}). Figures (c)
and (d) present cross sections at two different maturities. This analysis
is performed only on calls, both in-the-money and out-of-the-money.

Figure (a) suggests that the average IV surface in the control displays
a smile: as documented in past literature, option IVs are higher at
low and high moneyness levels, and the smile is steeper for shorter-dated
options. Figure (b) suggests that the IV of wildfire-exposed stocks
is higher for moneyness less than 1, and higher for moneyness levels
higher than 1. This is consistent with investors pricing in a higher
risk-neutral probability of downward shocks and a lower probability
of an upward shock. Figures (c) and (d) show that the pricing of the
downward shock happens for short-dated options. For longer maturities,
investors are less likely to price upward shocks, but there is no
difference in the pricing of downward shocks.

\subsection{Long-Run: Linear Panel Data Approach}

The previous approach focuses on effects on the days of the wildfire.
This section estimates the permanent impact of wildfire exposure on
the skews of calls and puts. Wildfires may have a permanent impact
on puts and calls if wildfires affect expected stock price dynamics
or the dividend flow.

To do so, we estimate two sets of regressions, where the Treatment
variable is replaced by either an `After the first wildfire' indicator
variable, or an `After the last wildfire' indicator variables. Other
covariates are kept as in the baseline regression.

The results are presented on Table~\ref{tab:permanent_effects}.
All fours specifications include day and firm fixed effects, and double-cluster
standard errors by firm and by day. Columns (1) and (3) are for puts,
and columns (3) and (4) are for calls. Columns (1) and (2) are for
the permanent impact after the first wildfire, and columns (3) and
(4) for the permanent impact after the last wildfire. The results
suggest that the permanent effects are on calls. This is consistent
with the previous non-parametric evidence that suggested that long-dated
options priced a lower probability of upward stock price movements.
This is consistent with this evidence, that call options respond permanently
to wildfire exposure. The impact on the slope of calls' implied volatility
is $-0.152$ after the first wildfire: a call with strike at 1.5 times
the forward price of the stock has a $0.228$ lower implied volatility
in the treatment group than in the control group. The impact ($-0.132$)
is similar after the last wildfire: a call with strike at 1.5 times
the forward price of the stock has a $0.198$ lower implied volatility
in the treatment group than in the control group. The impact is stronger
for short-dated options than for long-dated options: there is no permanent
impact on call options with a 136-day maturity ($(0.152/0.013)^{2}$)
after the first wildfire, and with a 144-day maturity after the last
wildfire.

\section{Wildfires and Investors' Marginal Utility}

\label{sec:marginal_utility}

A key question is whether the results presented in the previous section
are small, diversifiable shocks on a single stock in the market portfolio
or whether they are correlated with investors' wealth and its marginal
utility. This would occur if, for instance, investors hold higher
shares of wildfire-exposed stocks than what the market portfolio would
suggest.

By comparing the risk neutral distribution to the physical distribution
of wildfire-exposed stock returns, we can identify investors' marginal
utility at each level of the wildfire-exposed stock. This allows an
estimation of the correlation between the investors' marginal utility
of wealth and the value of a wildfire stock. A simple proposition,
derived from a Merton portfolio model with a wildfire stock and the
index, provides a relationship between the Arrow Pratt risk aversion
with respect to a wildfire-exposed stock and the risk aversion with
respect to wealth. 

Results suggest a risk aversion w.r.t. wealth that is arguably inconsistent
with a representative agent investing in the market portfolio. Multiple
hypothesis are consistent with this result. Wildfire stocks could
be held by heterogeneous investors who hold significantly more shares
exposed to natural disaster risk than what the market portfolio suggests.
Investors may be hedging against rare tail risk, that is not observed
in realized returns, and thus not captured in the physical distribution
measured in this paper. This latter explanation is consistent with
the option hedging on the index. 

\subsection{Pricing Kernel and Investors' Marginal Utility of Wealth}

\label{subsec:physical_vs_risk_neutral_distribution}

This section (i)~estimates state prices and the pricing kernel for
all wildfire-exposed stocks, (ii)~shows that the relationship between
the state prices of wildfire-exposed stocks and the value of such
stocks depends on the weight of wildfire-exposed stocks in the portfolio,
the stocks' beta, and investors' risk aversion; thus (iii)~we provide
an empirical method to estimate a stock-specific Arrow-Pratt aversion
and a statistical test of this null hypothesis of independence. 

As a testable benchmark, consider a continuous-time investment problem
over the time period $t\in\left[0,T\right]$ where the investor maximizes
the expected utility of terminal wealth. At each time $t$, the investor
can invest a share $q_{t}^{w}$ in the potentially wildfire-exposed
stock $S_{t}^{w}$ and a share $q_{t}$ of wealth in a diversified
portfolio with price $S_{t}$: 
\begin{eqnarray}
\frac{dS_{t}}{S_{t}} & = & \mu(S_{t},t)dt+\sigma(S_{t},t)dB_{t}\\
\frac{dS_{t}^{w}}{S_{t}^{w}} & = & \alpha(S_{t}^{w},t)dt+\beta\frac{dS_{t}}{S_{t}}+\sigma^{w}(S_{t}^{w},t)dB_{t}^{w},
\end{eqnarray}
where $dB_{t},dB_{t}^{w}$ are two independent Brownian motions, and
$\mu,\alpha$ (resp. $\sigma,\sigma^{w}$) are functions from $\mathbb{R}\times\mathbb{R}^{+}$
to $\mathbb{R}$ (resp., to $\mathbb{R}^{+}$). The portfolio $S_{t}$
follows a diffusion process whose trend $\mu(S_{t},t)$ and volatility
$\sigma(S_{t},t)$ are general functional forms that depend on both
the price and time. The wildfire exposed stock follows a similar diffusion
process with general functional forms but with a beta on the portfolio
$S_{t}$.\footnote{We introduce later jumps where the magnitude of the jump $J(S_{t},t)$
and the intensity $\lambda(S_{t},t)$ depend on the stock price and
time.}

The investor picks $q_{t}$ and $q_{t}^{w}$ to maximize the utility
of terminal wealth: 
\begin{eqnarray}
 &  & \max_{q_{s},q_{s}^{w}\vert t\leq s\leq T}E\left[U(W_{T})\right]\nonumber \\
s.t. &  & dW_{s}=\left[r+q_{s}\left(\mu(S_{s},s)-r\right)+q_{s}^{w}\left(\alpha(S_{s}^{w},s)+\beta\mu(S_{s},s)-r\right)\right]W_{s}ds\nonumber \\
 &  & \qquad\qquad+(q_{s}+\beta q_{s}^{w})\sigma(S_{s},s)W_{s}dB_{s}+q_{s}^{w}\sigma^{w}(S_{s}^{w},s)W_{s}dB_{s}^{w}\label{model:investor}
\end{eqnarray}

If $V(W,S,S^{w},t)$ is the investor's value function, the Euler equation
becomes: 
\begin{equation}
\frac{\partial V(W_{s},S_{s},S_{s}^{w},s)}{\partial W}=e^{-r(s-t)}\frac{\partial V(W_{t},S_{t},S_{t}^{w},t)}{\partial W}\zeta_{s}
\end{equation}
where $\zeta_{s}$ is the pricing kernel, a random variable at time
$s$ that depends both on the value $S_{t}$ of the diversified portfolio
and the value $S_{t}^{w}$ of the wildfire-exposed stock.

The risk neutral density of the wildfire stock is related to the physical
density $f$ by the expectation of the pricing kernel condition on
the value of the wildfire stock: 
\begin{equation}
f_{t}^{*}(S_{T}^{w})=e^{-r(T-t)}f_{t}(S_{T}^{w})E(\zeta_{T}\vert S_{T}^{w},S_{t}^{w})\label{eq:relationship_risk_neutral_distribution_physical_distribution}
\end{equation}
This has multiple implications. First, the ratio of the risk neutral
distribution to the physical density identifies the stochastic discount
factor $e^{-r(T-t)}E(\zeta_{T}\vert S_{T}^{w},S_{t}^{w})$. The risk
neutral distribution for wildfire-exposed stocks was estimated in
the previous section. Second, the pricing kernel is the marginal utility
of wealth, up to a multiplicative constant. As marginal utility is
decreasing in wealth, the estimated $\zeta_{T}$ should be decreasing
in wealth.\footnote{An extensive literature describes and explains apparent paradoxes
in the non-monotonic shape of the pricing kernel. See for instance
\citeasnoun{beare2016empirical}, \citeasnoun{linn2018pricing},
\citeasnoun{cuesdeanu2018pricing}.} Third, the beta of the wildfire-exposed stock matters, as the expectation
of the pricing kernel is taken conditional on the wildfire-exposed
stock, and thus depends on the joint distribution of the value $S_{T}$
diversified portfolio and the value $S_{T}^{w}$ of the wildfire-exposed
stock. Intuitively, if the beta is large, declines in $S_{T}^{w}$
are correlated with declines in $S_{T}$, and thus a high Arrow Debreu
state price for low values of $S_{T}^{w}$ may be an indication of
the $\beta$.\footnote{The panel data regressions of Sections \ref{sec:panel_data_evidence}
(Wildfires and the Volatility Smile) and \ref{sec:wildfires_and_rnd}
control for a day fixed effect, and thus for the value $S_{T}$ of
the index. }

Since the pricing kernel is equal to the marginal utility up to a
multiplicative constant, its relationship with wealth identifies risk
aversion. \citeasnoun{rosenberg2002empirical} suggests the Arrow-Pratt
measure \cite{arrow1964role,pratt1964risk}. \footnote{An important alternative is to use absolute risk aversion, as in \citeasnoun{jackwerth2000recovering}.
Both approaches yield a similar set of stylized facts.} 
\begin{equation}
\gamma=-W_{T}\frac{\zeta_{T}'(W_{T})}{\zeta_{T}(W_{T})}\label{eq:gamma_risk_aversion}
\end{equation}
where $\zeta'_{T}$ is the derivative of the state price w.r.t. the
value of the wildfire-exposed stock. When terminal utility is CRRA,
this is equal to the constant relative risk aversion.

Here, as suggested by \citeasnoun{rosenberg2002empirical}, we can
obtain the Arrow-Pratt measure projected on to the value of a wildfire-exposed
stock to estimate the sensitivity of investors' marginal utility of
wealth to the value of such stock. 
\begin{equation}
\gamma^{w}=-S_{T}^{w}\frac{\zeta'_{T}(S_{T}^{w})}{\zeta_{T}(S_{T}^{w})}\label{eq:elasticity_wrt_stock}
\end{equation}
which is the elasticity of the pricing kernel w.r.t. the value of
the wildfire-exposed stock. This quantity can be estimated for all
of this paper's treated firms following the method outlined in the
next subsections.

(\ref{eq:elasticity_wrt_stock}) can be estimated and it provides
key information on the sensitivity of wealth to changes in wildfire-exposed
stocks and on investors' risk aversion. 
\begin{prop}
\textbf{Sensitivity of State Prices $\zeta_{T}$ with Respect to the
Wildfire-Exposed Stock}\label{proposition:risk_aversion} Assume the
stocks' parameters $\mu,\alpha,\beta,\sigma,\sigma^{w}$ are non-time-varying,
the investor has CRRA preferences $u(w)=\frac{w^{1-\gamma}}{1-\gamma}$
where $\gamma$ is risk aversion with respect to wealth. Then a closed
form ties the sensitivity~$\gamma^{w}$ of state prices w.r.t. to
the wildfire stock at maturity to the investor's risk aversion $\gamma$
with respect to wealth and the share $q^{w}$ of the wealth invested
in the wildfire-exposed stock: 
\begin{equation}
\gamma^{w}=(q^{w}+\rho\frac{\sigma}{\sigma_{w}}q)\gamma\label{eq:relationship_arrow_pratt_risk_aversion}
\end{equation}
where $q$ is the share of wealth invested in the diversified portfolio;
$\rho$ is the correlation between the returns on the wildfire stock
and the diversified portfolio, $\rho=\beta\sigma/\sqrt{\beta^{2}\sigma^{2}+(\sigma^{w})^{2}}$;
$\sigma$ and $\sigma^{w}$ the volatilities of the stock and the
wildfire-exposed stock respectively. 
\end{prop}

The proof is presented in Appendix Section~\ref{sec:Proof-of-Proposition}.
The intuition of the proof of Proposition 1 is simple. The proof first
expresses the marginal utility of terminal wealth as a function of
both $\log S_{T}^{w}$ and $\log S_{T}$. $(\log S_{T}^{w},\log S_{T})$
is a joint bivariate normal distribution whose correlation depends
on the beta of the wildfire stock and the variances. Since $\gamma^{w}$
depends on the conditional expectation of a bivariate normal, the
classic bivariate normal expectation formula applies, conditional
on the value of $S_{T}^{w}$, which gives the coefficient $q^{w}+\rho\frac{\sigma}{\sigma_{w}}q$.
Details are provided on page~\pageref{sec:Proof-of-Proposition}. 

Monte Carlo simulations with more complex stochastic processes for
$S_{T}^{w}$ suggest the result holds for Brownian motions with jumps
and stochastic volatility.

Investors' demand for insurance against low values of $S_{T}^{w}$
may thus be due to a high beta, a high share invested in a wildfire-exposed
stock, or a high risk aversion. We use these results empirically below
in sections~\ref{subsec:empirics_pricing_kernel_PGE} and \ref{subsec:empirics_pricing_kernel_panel}
to estimate risk aversion $\gamma$.

\subsection{Estimating and Forecasting the Physical Distribution of $S_{T}^{w}$}

\subsubsection*{Estimation of a Firm-Specific GARCH-Wildfire Process}

The pricing kernel $E(\zeta_{T}\vert S_{T}^{w})$ w.r.t. to the wildfire-exposed
stock price $S_{T}^{w}$ is the ratio of the risk neutral and the
physical distribution. First, we use the previous section's daily
and stock-specific risk neutral distributions $f^{*}(S_{T}^{w})$
for each stock of a treated firm. Second, the so-called physical distribution
$f(S_{T}^{w})$ of the future level of the stock price $S_{T}^{w}$
is estimated using a time series GARCH-Wildfire model that accounts
for jumps in returns and in volatility caused by wildfires. Such physical
distribution reflects investors' beliefs and thus our approach requires
using different assumptions for investors' expectations: myopic (based
on the previous history of returns), foresight (based on the post-wildfire
series of returns), and stationary (assuming that the GARCH model
has similar autoregressive parameters before and after the wildfire).

The GARCH-Wildfire model allows returns to have time-varying volatility
and wildfire jumps:
\begin{align}
\log(S_{jt}/S_{jt-1}) & =\alpha_{j}+\beta_{j}R_{t}+\sigma_{jt}\varepsilon_{jt}+\delta_{j}W_{jt-1}\nonumber \\
\sigma_{jt}^{2} & =\omega_{j}+\zeta_{j}\sigma_{jt-1}^{2}+\xi_{j}\varepsilon_{jt-1}^{2}+\rho_{j}\varsigma_{t}+\gamma_{j}W_{jt-1}\label{eq:garch_wildfire}
\end{align}
where notations are similar to the important \citeasnoun{barone2008garch},
with a number of additions. 

First, $W_{jt}=1$ when a wildfire hit the firm either in the current
period or in previous periods (in this case we include lags $W_{jt-k}$).
Wildfires cause jumps in the stock's return, denoted $\delta_{j}$
and in volatility, denoted $\gamma_{j}$. 

Second, the model accounts for the beta of the stock w.r.t. to the
index, which is potentially correlated with the occurence of the wildfire.
This controls for shifts in the index return in the estimation of
$\delta_{j}$. The market return is modeled as $R_{jt}=\mu+\varsigma_{t}\epsilon_{jt}$
with an autoregressive structure for the variance $\varsigma_{t}^{2}=\omega+\zeta\varsigma_{t-1}^{2}+\xi\epsilon_{t-1}^{2}$. 

Third, the model allows the volatility of the stock to depend on market
volatility. 

When indexed by $j$, parameters $(\alpha_{j},\beta_{j},\delta_{j},\omega_{j},\zeta_{j},\xi_{j},\rho_{j},\gamma_{j})$
are estimated separatelty for each stock.

\subsubsection*{Forward-Looking Physical Distribution of Wildfire-Exposed Stocks
at Time $t+k$}

The physical distribution $f(S_{jt+k}^{w})$ of a wildfire-exposed
stock $j$ at $t+k$ given the history of observations of the stock
and the index at $t$ is modelled as the distribution of the GARCH-Wildfire
stock with estimated parameters $(\widehat{\alpha_{j}},\widehat{\beta_{j}},\widehat{\delta_{j}},\widehat{\omega_{j}},\widehat{\zeta_{j}},\widehat{\xi_{j}},\widehat{\rho_{j}},\widehat{\gamma_{j}})$
conditional on the initial volatility and prior wildfire exposure.
The physical distribution of market returns~$R_{t+k}$ is also estimated
using the parameters of the estimated GARCH $(p,q)$ process. Forward-looking
wildfire probabilities are estimated using the history of wildfire
probabilities for each firm. We discuss further in the paper how the
risk neutral distribution can be used to \emph{identify }the forward-looking
wildfire probability.

\subsection{Panel Data Estimation of Firm-Level Arrow-Pratt Risk Aversions}

When investors have CRRA preferences, there is a linear relationship
between the log risk neutral distribution $\log f_{jt}^{*}$ for stock
$j$, the log of the physical distribution $\log f_{jt}$ and the
risk free return, on the log of the forward stock price. For each
stock $j$, 
\begin{equation}
\log f_{jt}^{*}(S_{j,T}^{w})-\log f_{jt}(S_{j,T}^{w})+r_{t,T}(T-t)=-\gamma_{j}^{w}\log S_{j,T}^{w}\label{eq:log_fstar_and_log_f}
\end{equation}
Hence firm-specific Arrow-Pratt risk aversion is obtained by regressing
the left-hand side on the right-hand side for a grid of forward prices
$S_{T,t,k}^{j}$, for equally spaced forward prices ranging from the
lower bound of the support of option strikes to the upper bound of
such support: 
\begin{equation}
\log\widehat{f_{jt}^{*}(S_{j,T,k}^{w})}-\log\widehat{f_{jt}(S_{j,T,k}^{w})}+\widehat{r_{t,T}}(T-t)=-\widehat{\gamma_{j}^{w}}\log S_{j,T,k}^{w}+\text{Firm}_{j}+\textrm{Date}_{t}+\text{Maturity}_{T}+\varepsilon_{j,t,T,k}^{w}\label{eq:risk_aversion_regression}
\end{equation}
where we can control for firm, date, and maturity fixed effects. For
maturity we control for 10 bins of maturities, and the choice of the
number of maturity bins does not affect the estimation. We restrict
the estimation to the range of forward prices $S_{j,T,k}^{w}$ on
the support of both the risk neutral and physical distributions as
Euler condition~\ref{eq:relationship_risk_neutral_distribution_physical_distribution}
implies that their supports should coincide. $r_{t,T}$ is the risk-free
rate, obtained by calibrating the LIBOR time series at the option's
maturity using the Intercontinental Exchange (ICE) time series. Ordinary
least squares regression \ref{eq:risk_aversion_regression} performed
for each firm $j$ separately provides a firm-specific risk aversion
index $\widehat{\gamma_{j}^{w}}$ and an associated standard error
$SE_{j}$.

\bigskip{}

In regression \ref{eq:risk_aversion_regression}, the risk neutral
distribution, the physical distribution and the risk-free rate are
estimates rather than the true values. \citeasnoun{hausman2001mismeasured}
suggests that such left-hand side measurement error leads to larger
standard errors, smaller $t$ statistics, and more imprecise estimators
of $\hat{\gamma}$, but does not affect the unbiased nature of the
OLS estimator of $\gamma_{j}^{w}$.

\subsection{Empirical Results: PG\&E}

\label{subsec:empirics_pricing_kernel_PGE}

\begin{figure}
\caption{Identifying Investors' Marginal Utility of Wildfire Shocks: State
Prices from the Ratio of the Risk Neutral Distribution vs. the Physical
Distribution, the PG\&E Case}
\label{fig:rnd_physical}

\emph{In the two upper panels, the bold line is the risk neutral distribution
$f^{*}$ on September 22nd (before the fires) and October 12th (during
the fires), as obtained using the method of Section~\ref{subsec:physical_vs_risk_neutral_distribution}.
The dotted lines is the physical distribution $f$ assuming backward-looking
expectations. Three approaches for building the physical distribution
are described in Section \ref{subsec:physical_vs_risk_neutral_distribution}.
The two bottom panels present the pricing kernel, obtained by taking
the ratio $f^{*}/f$ of the RND and the physical distribution.}

\bigskip{}

\centering\subfloat[September 22, 2017]{\includegraphics[clip,scale=0.45]{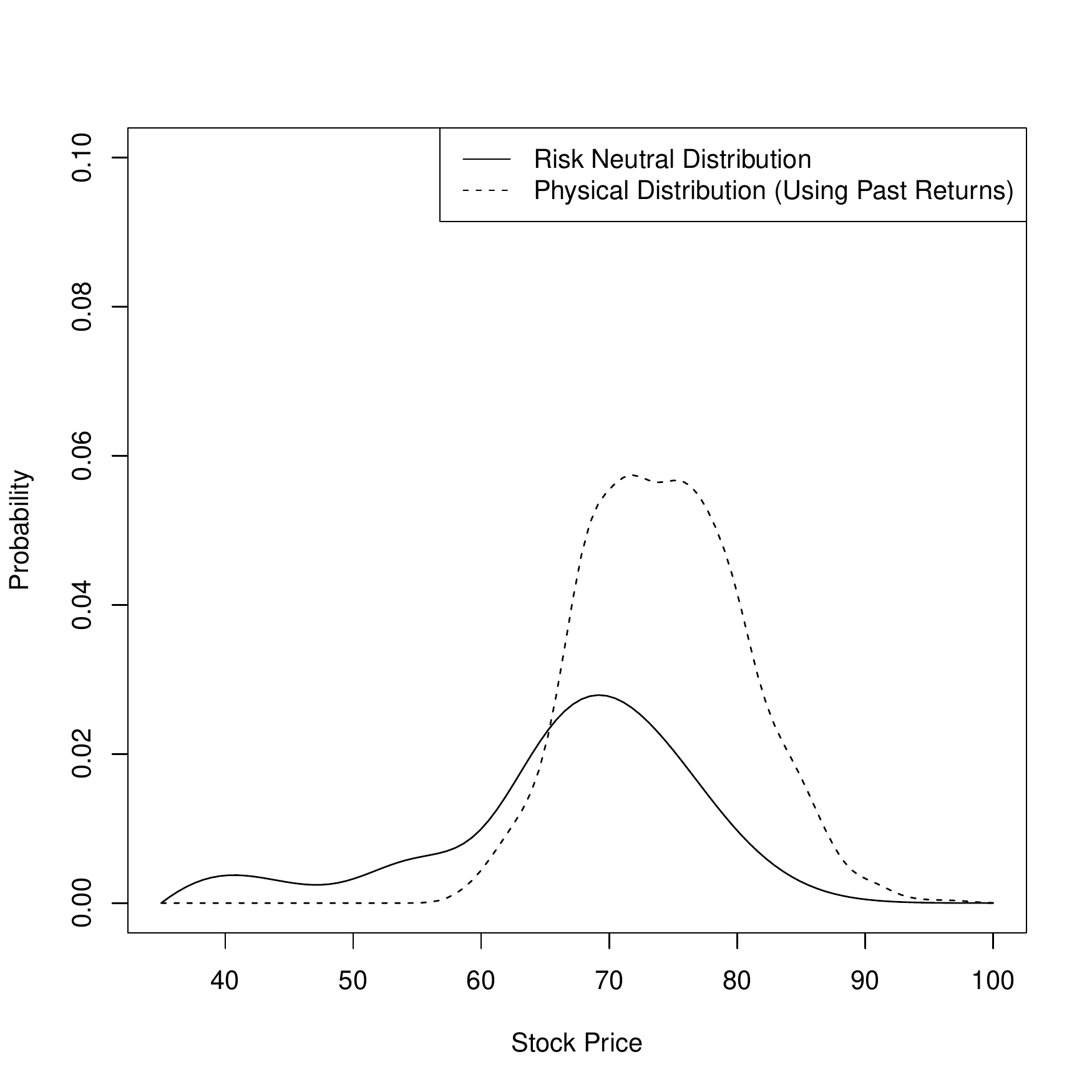} 

}\subfloat[October 12, 2017]{\includegraphics[clip,scale=0.45]{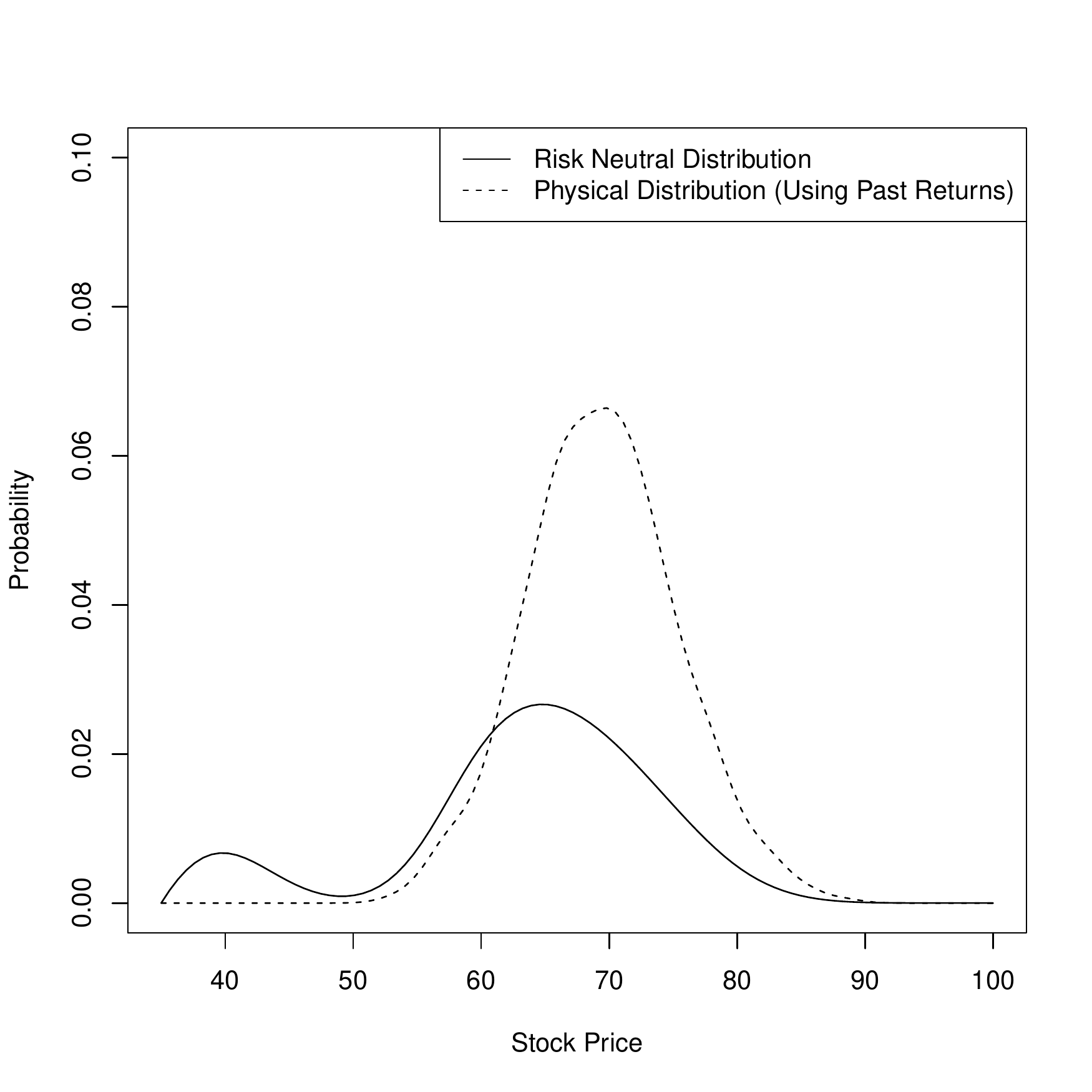} 

}

\bigskip{}

\subfloat[Pricing Kernel, Backward-Looking Expectations]{\includegraphics[clip,scale=0.45]{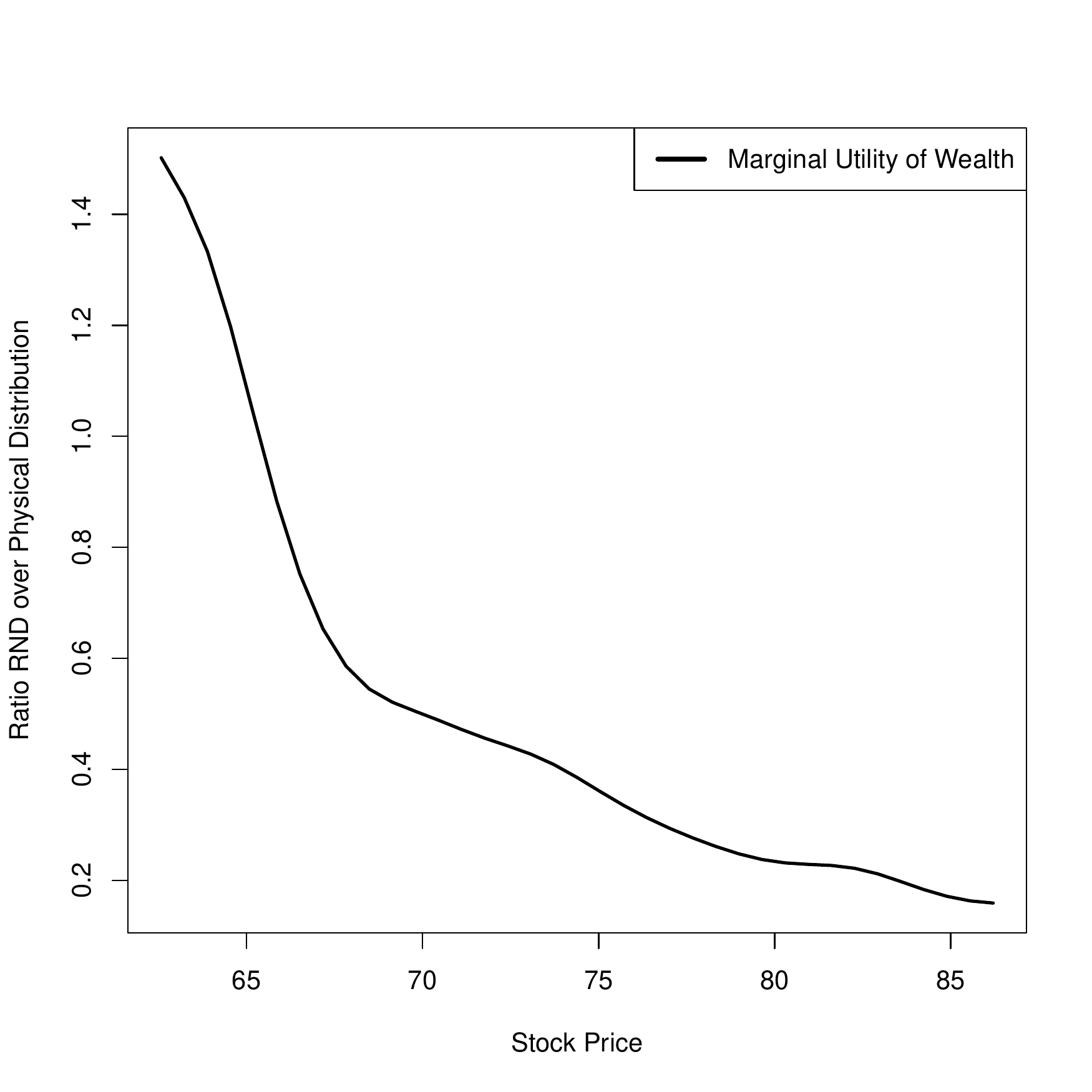} 

}\subfloat[Pricing Kernel, Backward-Looking Expectations]{\includegraphics[clip,scale=0.45]{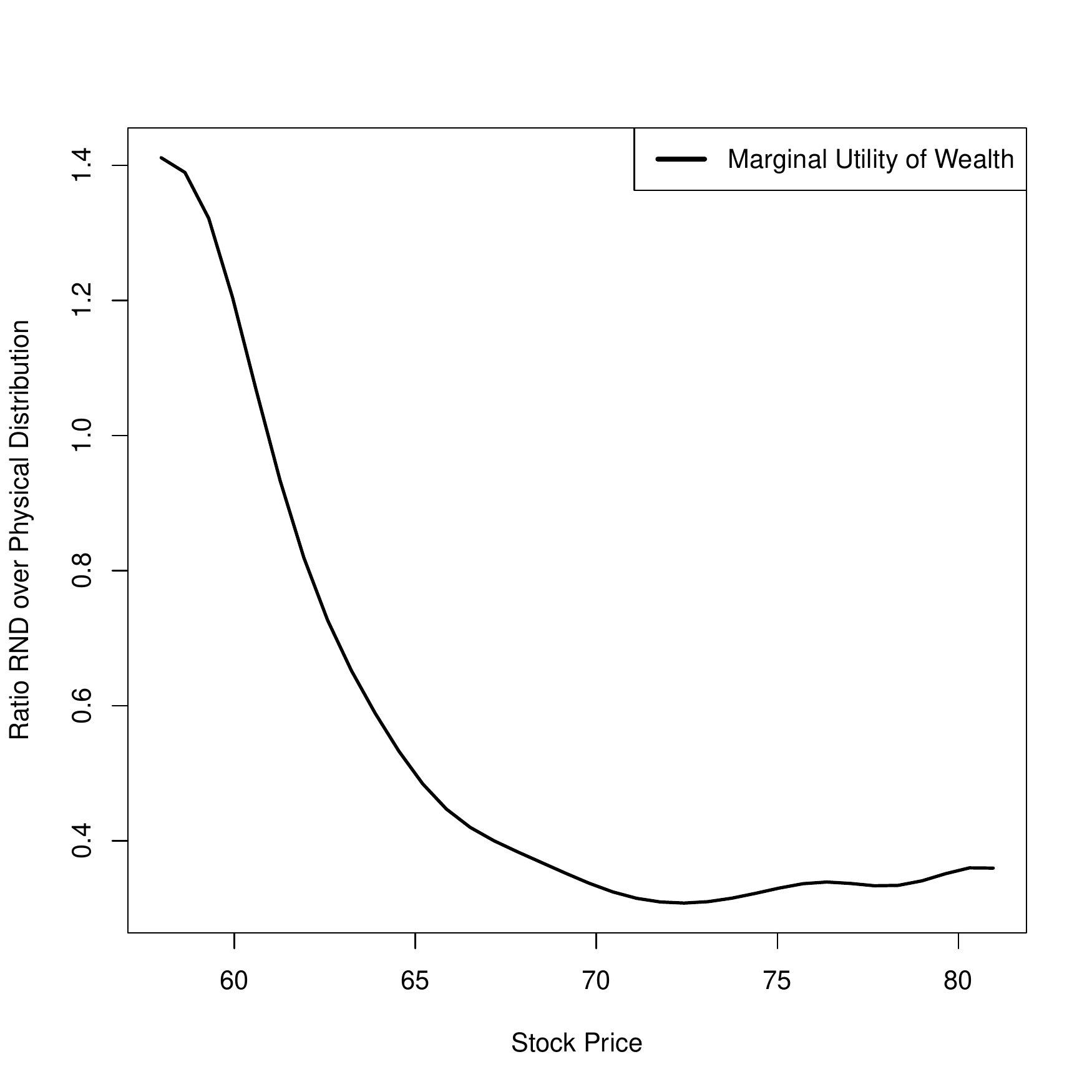}

}

$\gamma^{w}=-S_{T}^{w}\frac{\zeta'(S_{T}^{w})}{\zeta(S_{T}^{w})}\simeq6.19$
\qquad{}\qquad{}\qquad{}\qquad{}\qquad{}$\gamma^{w}=-S_{T}^{w}\frac{\zeta'(S_{T}^{w})}{\zeta(S_{T}^{w})}\simeq4.56$
\end{figure}

Figure~\ref{fig:rnd_physical} gives a visual intuition of the test
and its results for PG\&E. It presents the risk neutral distribution
estimated using option prices and the physical distribution estimated
using past returns. The lower panels present the ratio of the two
probability distributions. The ratio is taken for stock prices in
the support of the physical distribution of future prices. As before,
the RND is obtained using arbitrage-free prices and a local polynomial
regression using the \citeasnoun{breeden1978prices} butterfly spread
method.

The upper panels of Figure~\ref{fig:rnd_physical} suggest that the
risk neutral distribution is above the physical distribution for low
values of the stock, consistent with higher values of Arrow-Debreu
state-contingent assets for lower values of the stock. Similarly,
the panels suggest that, for higher stock prices, the physical probabilities
are higher the risk neutral probabilities, consistent with lower values
of Arrow-Debreu state-contingent prices for higher wildfire-exposed
stock prices.

The lower panels of Figure~\ref{fig:rnd_physical} take the ratio
of the two distributions, to present an estimate of the pricing kernel.
Up to a constant equal to the risk-free discount factor, this is the
marginal utility of the investor in model~(\ref{model:investor})
if the physical distribution is correctly specified. An extant literature
estimates this pricing kernel for the S\&P500 index~\cite{chernov2000study,figlewski2009estimating,birru2012anatomy,figlewski2018risk,reinke2020risk}
and highlights the potential \emph{pricing kernel puzzles }that emerge.
We address this systematically for all of the stocks in our sample
in the next section.

An estimation of the elasticity $\gamma^{w}$ of state prices w.r.t.
to the PG\&E stock for these two dates, September 22, 2017, and October
12, 2017, provides estimates of $\widehat{\gamma^{w}}=6.19$ and $\widehat{\gamma^{w}}=4.56$.
Proposition~\ref{proposition:risk_aversion} can be used to identify
risk aversion from such elasticity of state prices w.r.t PG\&E. Given
that the PG\&E stock represents 0.157\% of the S\&P 500 on October
1, 2017, the beta of the stock is $\widehat{\beta}=0.78$, this would
suggest investors' risk aversion equal to $\widehat{\gamma}=\widehat{\gamma^{w}}/(q^{w}+\widehat{\rho}\frac{\hat{\sigma}}{\widehat{\sigma^{w}}}q)\simeq242.8$
on September 22, 2017 and 178.8 on October 12, 2017. These are far
higher than conventional measures of relative risk aversion \cite{mehra1985equity,meyer2005relative,guiso2018time}.

\subsection{Empirical Results: Panel Data}

\label{subsec:empirics_pricing_kernel_panel}

Firm-specific risk aversion $\gamma_{j}$ is estimated stock by stock
for each wildfire event of the treatment group. For each stock, and
for each day that a listed firm is exposed to a wildfire, the risk
neutral distribution is estimated using \possessivecite{ait2003nonparametric}
approach of applying the \citeasnoun{breeden1978prices} approach
using a local polynomial regression on arbitrage-free prices. The
physical distribution is estimated using the GARCH-Wildfire model,
estimated separately for each firm. The day- and maturity-specific
risk-free rate is obtained using the LIBOR. The physical distribution
is estimated on the same grid as the risk neutral distribution. This
yields 1,481,076 observations for 361 firms across 233 days.

\begin{figure}
\caption{Distribution of the Elasticity $\gamma_{j}^{w}$ of State Prices w.r.t.
Wildfire-Exposed Stocks}
\label{fig:risk_aversions_panel}

\emph{\footnotesize{}This figure presents the distribution of estimated
firm-level elasticity of state prices (investors' marginal utility)
parameters $\gamma_{j}^{w}$ with respect to the value of wildfire-exposed
stocks $S_{T}^{w}$. This is done using regression~\eqref{eq:log_fstar_and_log_f},
i.e.by regressing $\log f^{*}(S_{j,T}^{w})-\text{log}f(S_{j,T}^{w})-r_{t,T}(T-t)$
on $\log$ moneyness for each firm separately. $f^{*}$ is the risk
neutral distribution obtained using option prices and the \citeasnoun{breeden1978prices}
approach; $f$ is the physical distribution, obtained by using the
estimated GARCH-Wildfire process separately for each firm and forecasting
$S_{jt}^{w}$ using the time series of pre-wildfire returns. $r_{t,T}$
is the risk-free interest rate, calibrated using ICE's LIBOR. $T$
is the maturity of the option.}\bigskip{}

\emph{\footnotesize{}For most firms, investors' marginal utility declines
when the value of the wildfire stock $S_{T}^{w}$ is higher, consistent
with consumer theory, $\gamma^{w}>0$ . Estimates left of the red
dotted line exhibit the pricing kernel puzzle, but 87\% of firms do
not exhibit such a puzzle.}{\footnotesize\par}

\bigskip{}

\begin{centering}
{\footnotesize{}(i)~Distribution of $\hat{\gamma}_{j}^{w}$}{\footnotesize\par}
\par\end{centering}
\begin{centering}
{\footnotesize{}\includegraphics[scale=0.75]{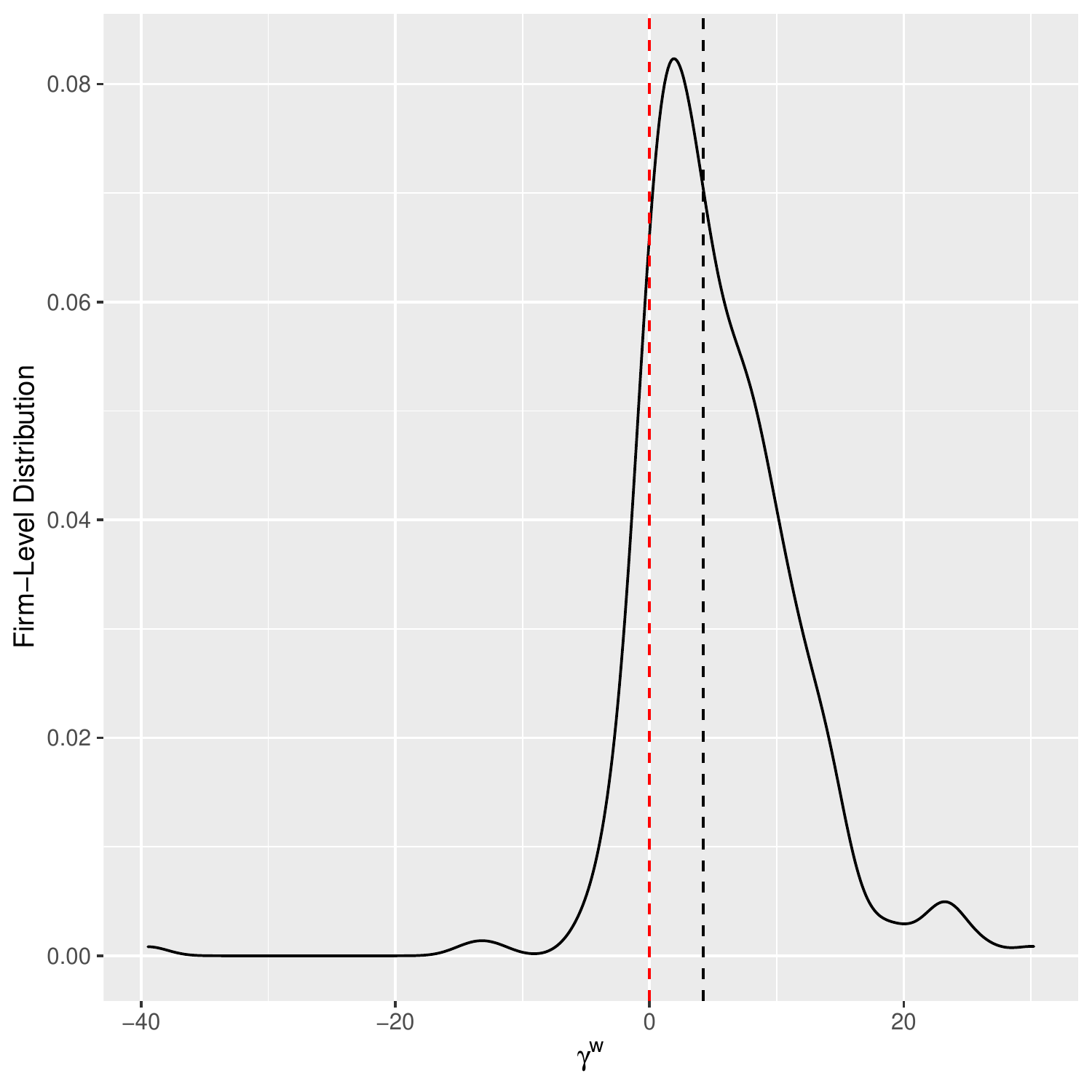}}{\footnotesize\par}
\par\end{centering}
\begin{centering}
{\footnotesize{}\bigskip{}
}{\footnotesize\par}
\par\end{centering}
\begin{centering}
{\footnotesize{}(ii)~Descriptive Statistics}{\footnotesize\par}
\par\end{centering}
\begin{centering}
{\footnotesize{}\bigskip{}
}{\footnotesize\par}
\par\end{centering}
\begin{centering}
{\footnotesize{}}%
\begin{tabular}{lcccccccc}
\toprule 
{\footnotesize{}Point Estimate (firm-level) } & {\footnotesize{}Mean } & {\footnotesize{}S.D. } & {\footnotesize{}P25 } & {\footnotesize{}P50 } & {\footnotesize{}P75 } & {\footnotesize{}$1(\gamma)<0$ } & {\footnotesize{}J } & {\footnotesize{}N }\tabularnewline
\midrule 
{\footnotesize{}$\widehat{\gamma_{j}^{w}}$ } & {\footnotesize{}5.29 } & {\footnotesize{}6.38 } & {\footnotesize{}1.34 } & {\footnotesize{}4.37 } & {\footnotesize{}8.56 } & {\footnotesize{}0.13 } & {\footnotesize{}311 } & {\footnotesize{}1,479,976 }\tabularnewline
\midrule 
{\footnotesize{}t-Statistic} & {\footnotesize{}Mean } & {\footnotesize{}S.D. } & {\footnotesize{}P25 } & {\footnotesize{}P50 } & {\footnotesize{}P75 } & {\footnotesize{}$1(t)<-1.96$ } & {\footnotesize{}J } & {\footnotesize{}N }\tabularnewline
\midrule 
{\footnotesize{}$\widehat{\gamma_{j}}/\widehat{SE_{j}}$ } & {\footnotesize{}11.07 } & {\footnotesize{}10.88 } & {\footnotesize{}3.40 } & {\footnotesize{}10.38 } & {\footnotesize{}17.56 } & {\footnotesize{}0.08 } & {\footnotesize{}311 } & {\footnotesize{}1,479,976 }\tabularnewline
\bottomrule
\end{tabular}\bigskip{}
\par\end{centering}
\emph{\footnotesize{}$J$: number of distinct wildfire-exposed firms
with moneyness ranging between 0.5 and 1.5 times the forward price.
$N$: number of day $\times$ strike $\times$ firm observations,
wildfire-exposed firms. }{\footnotesize{}$1(\gamma)<0$ : exhibits
a pricing kernel puzzle (point estimate). $1(t)<-1.96$: the pricing
kernel puzzle is statistically significant.}{\footnotesize\par}
\end{figure}

Results are presented on Figure~\ref{fig:risk_aversions_panel}.
Panel (i) presents the distribution of firm-level risk aversions.
Panel (ii) suggests that the average relative risk aversion is 0.292.
The 75th percentile is between 0.497 and 0.571. The lower part of
panel (ii) presents the t statistics, of an average of 23.0, 20.2,
and 31.4 respectively.

On panel (i), observations left of the vertical line are for negative
risk aversions, for which a pricing kernel puzzle is present. An extensive
literature describes the presence of non-monotonic parts of the pricing
kernel for specific subsets of the range of forward prices \cite{beare2016empirical,cuesdeanu2018pricing,linn2018pricing,figlewski2018risk}.
The evidence here is on the overall monotonicity of the pricing kernel.
Panel (ii) suggests that 87\% of firms do not exhibit a pricing kernel
puzzle (1-0.13=0.87). This pricing kernel puzzle is significant at
95\% for only 8\% of firms (second row of estimates). This suggests
that a substantial majority of pricing kernels are downward sloping.
Any upward sloping segment of the state prices would lead to downward
biased estimates of $\hat{\gamma}$. Our estimates should thus be
interpreted as lower bounds on the elasticity of state prices w.r.t.
the wildfire-exposed stock.

We then turn to the result of proposition~\ref{proposition:risk_aversion}
to estimate investors' risk aversion w.r.t. wealth. The proposition
states that, for a geometric brownian motion with wildfire jumps,
the $\gamma^{w}=\gamma(q^{w}+\rho\frac{\sigma}{\sigma^{w}}q)$. Such
calculation is performed for each firm exposed to a wildfire separately.
$q^{w}$ is estimated assuming the investor holds the S\&P 500. The
share of stocks in the investor's portfolio is obtained using the
Survey of Consumer Finances (SCF). The correlation $\rho$ is a function
of each stock's $\beta$, the variance of returns, and the variance
of market returns. The table below displays the risk aversion parameters
thus estimated using each firm-level data, using the forecast physical
distribution $f(S_{jT}^{w})$. If the model (including beliefs) are
correctly specified, estimating relative risk aversion using different
stocks $j$ should yield similar estimates $\hat{\gamma}$. Hence
below, and again in the case of a well-specified model, $\gamma_{j}$
should only differ across $j$s due to differences in the share of
the stock in the market portfolio and differences in the $\beta_{j}$
and volatility $\sigma_{j}^{w}$ of the stock.
\begin{center}
\begin{table}[H]
\caption{Option-Implied Relative Risk Aversion with respect to Wealth Assuming
the Market Portfolio is Held}

\begin{center}

\begin{tabular}{lcccc}
\toprule 
Parameter  & Mean  & P25  & Median  & P75 \tabularnewline
\midrule 
$\hat{\gamma}=\frac{\gamma^{b}}{q^{w}+\rho\frac{\sigma}{\sigma^{w}}q}$  & 114.41 & 38.56 & 91.44 & 181.79\tabularnewline
\bottomrule
\end{tabular}

\end{center}

\bigskip

\emph{Risk aversion w.r.t wealth under the assumption that: the share
$q^{w}$ of the asset in the investor's portfolio is equal to its
share in the S\&P 500 times the share of stocks in the investor's
portfolio; correlations and variances using the observed returns of
the stock and the index; share of stocks $q$ in the investor's portfolio}
\emph{using the Survey of Consumer Finances.}
\end{table}
\par\end{center}

Results suggest a median risk aversion parameter of 38.56 (P25) to
181.79 (P75), with a median of 91.44 and mean of 114.41. The distribution
is as expected right-skewed given the demand for hedges for stocks
with little impact of wildfires on their physical distribution. \citeasnoun{barro2006rare}
states that the usual view of the literature is that $\gamma$ should
be between 2 and 5. In the wildfire-exposed sample, using firm-level
$\gamma^{w}$s , 88\% of the estimates suggest a risk aversion w.r.t
wealth above 5. This is the case even as our estimates $\widehat{\gamma_{j}}$
of $\gamma_{j}$ are lower bounds due to the use of estimates $\widehat{f^{*}}$
and $\widehat{f}$ of $f^{*}$ and $f$.

These higher than expected levels of risk aversion could be consistent
by at least two mechanisms:
\begin{enumerate}
\item Investors holding larger amounts $q_{j}^{w}$ of wildfire-exposed
stocks than the market portfolio. 
\item A difference between the observed physical distribution of returns
(which was estimated using a GARCH-Wildfire model) and that used by
investors to price options. Investors hold beliefs, and this probability
distribution $f^{b}(S_{T}^{w})$ may significantly differ from the
$f(S_{T}^{w})$ as estimated using the observed time series. Investors
may hedge against large and rare downward jumps, or they may hedge
against stochastic volatility. Options on wildfire-exposed stocks
may price disasters, the micro counterpart of the market-level approach
of \citeasnoun{barro2006rare}.
\end{enumerate}
This second mechanism is a version of what \citeasnoun{cuesdeanu2018pricing}
coins the Peso problem, when observed historical returns do not include
an event, but subjective distributions incorporate investors' fears.

We implement an empirical approach for these two points below.

\section{Discussion}

\subsection{Identification of Option-Implied Portfolio Shares Exposed to Wildfires
\label{subsec:Identification-of-Wildfire-Expos}}

Proposition~\ref{proposition:risk_aversion} can be used to calibrate
the portfolio shares held by investors given literature-driven choices
of the risk aversion parameter $\gamma$ with respect to wealth. Following
relationship \eqref{eq:relationship_arrow_pratt_risk_aversion}, the
portfolio share in the wildfire-exposed stock is:
\begin{equation}
q^{w}=\frac{\gamma^{w}}{\gamma}-\rho\frac{\sigma}{\sigma^{w}}q,\label{eq:portfolio_share}
\end{equation}
where $\gamma^{w}$ is the elasticity of state prices with respect
to the wildfire stock, $\rho$ is the correlation of the stock $S_{T}^{w}$
and the index $S_{T},$$\sigma$ is the standard deviation of the
diffusion of the index, $\sigma^{w}$ is the standard deviation of
the diffusion of the wildfire-exposed stock. Numerical simulations
suggest this relationship holds for other types of stochastic processes
than geometric brownian motions, including with stochastic volatility
and jumps. 

A literature provides mappings between portfolio shares and beliefs
including~\citeasnoun{eganrecovering}, \citeasnoun{egan2021drives},
and \possessivecite{heipertz2019transmission} proposition~5 on the
inversion of factor models in a Markowitz setting. This paper's approach
differs in two respects. First, options reveal an additional and granular
distribution $f^{*}()$ of state prices. Second, in previous work
portfolio shares do not reveal the physical distribution (beliefs
about future returns) but rather the risk neutral distribution, which
this paper's approach~\eqref{eq:portfolio_share} aims at breaking
down into marginal utilities and physical probabilities. An extension
of \eqref{eq:portfolio_share} allows for a joint recovery of $q_{j}^{w}$
for the set $j\in\mathcal{J}$ of wildfire-exposed stocks.

We estimate the share of the portfolio in a wildfire-exposed stock
$q_{j}^{w}$ on a range of possible measures of risk aversion w.r.t.
wealth $\gamma$. \citeasnoun{barro2006rare} suggests the coefficient
of relative risk aversion $\gamma$, has to be substantially above
one, cannot be below 3, and it suggests a value of 4. \citeasnoun{bansal2004risks}
match the observed equity premium using $\gamma=10$. This value is
the upper bound of \citeasnoun{mehra1985equity}. With the \citeasnoun{bansal2004risks}
approach, estimated portfolio shares are still very substantial and
above their weight in the index. A reassuring fact is that $q_{jt}^{w}\in[0,1]$
for 70\% of the firms of the treated sample. 87.7\% of treated firms
do not exhibit a pricing kernel puzzle ($q_{jt}^{w}>0$) and 82\%
of the treated firms have option-implied portfolio shares below 1.
For this sample the distribution of firm-level portfolio shares is:

\begin{center}
\begin{tabular}{lcccccc}
\toprule
& Min. &  1st Qu. &  Median &   Mean & 3rd Qu.  &  Max. \\
\midrule 
$q_{jt}^w$ & 0.002 & 0.194 & 0.382 & 0.425 & 0.657 & 0.971  \\
\bottomrule
\end{tabular}
\end{center}which suggests higher portfolio allocations than the market index
even with high values of relative risk aversion.

\section{Calibrating Risk Neutral Tail Risk\label{sec:Calibrating-Risk-Neutral}}

Parameterizing the risk neutral distribution enables an identification
of downward and upward jumps that can account for the shape of the
risk neutral distribution. We use here two simple models that feature
jumps to analyze the impact of wildfires. For important recent contributions
on the general topic, with models featuring more complex stochastic
structures, see, e.g. \citeasnoun{ait2021closed}, \citeasnoun{todorov2011econometric}.

\subsubsection*{Merton Model}

The first model we consider is the Merton model, which is a Black
and Scholes geometric brownian motion with Jumps (BSJ). Denote $s_{t}=\log(S_{t})$
the log stock price, whose dynamic can be written under the risk neutral
distribution as a set of two equations, one for the log stock, and
one for the time-varying stochastic volatility. The stock evolves
according to a process with a diffusion $B_{t}^{1}$ and a jump process
$Z_{t}^{s}$: 
\begin{equation}
ds_{t}=(r-\delta-\lambda^{s}\mu^{s}-\frac{1}{2}\sigma^{2})dt+\sigma dB_{t}^{1}+dZ_{t}^{s}\label{eq:ajd}
\end{equation}
where $r$ is the risk free rate, $\delta$ is the dividend yield,
$\lambda^{s}$ the average jump intensity for the stock, occurring
with Poisson probability $\lambda^{s}$. $\mu^{s}$ is the average
jump magnitude for the stock. $f^{s}$ denotes the pdf of stock jumps.
The jumps in the BSJ a have a log normal distribution with mean $\mu^{s}$
and standard deviation $\sigma^{s}$. 

This model parameterizes the risk neutral distribution obtained in
the previous sections as, at a given time $t$, a choice of $\Theta=(r,\delta,\sigma,\lambda^{s},\mu^{s},\sigma^{s})$
defines a risk neutral distribution $f^{*}(s_{T})$ for the log stock
at $T>t$ conditional on $s_{t}$. Our calibration goal is to estimate
the impact of wildfires on $(r,\delta,\sigma,\lambda^{s},\mu^{s},\sigma^{s})$
.

\subsubsection*{The Double Jump Model}

The risk neutral distribution results of Section~\ref{sec:wildfires_and_rnd}
suggest that wildfires might lead to an increasing probability of
\emph{both} upward and downward jumps. The parametric model just presented
only allows for one type of jump. The second model we consider is
\possessivecite{kou2004option} double exponential jump model. The
risk neutral dynamic of the stock is:
\begin{equation}
ds_{t}=(r-\delta-\frac{1}{2}\sigma^{2}-\lambda\zeta)dt+\sigma dB_{t}+d(\sum_{i=1}^{N(t)}(V_{i}-1)),\label{eq:kou_model}
\end{equation}
where $dB_{t}$ is a Brownian motion, $N(t)$ is a Poisson process
with intensity $\lambda$, and $V_{i}$ is the magnitude of the jumps,
distributed as $V_{i}=\exp(Y_{i})$ where $Y_{i}$ has a downward
exponential distribution with probability $1-p$ and mean $1/\eta_{1}$,
and an upward exponential distribution with probability $p$ and mean
$1/\eta_{2}$.

Here a choice of $\Theta=(r,\delta,\sigma,\lambda,\zeta,p,\eta_{1},\eta_{2})$
defines the risk neutral distribution $f^{*}(s_{T})$. Our calibration
goal is to estimate the impact of wildfires on $(r,\delta,\sigma,\lambda,\zeta,p,\eta_{1},\eta_{2})$
.

\subsection{Calibration of the Option-Implied Parameters}

The paper follows \citeasnoun{duffie2000transform} in pricing options
using the characteristic function. We estimate the structural parameters
as follows. The risk-free rate $r$ is calibrated using the LIBOR
at the maturity of the options.\footnote{The risk-free rate is assumed constant here. An extension would allow
for a stochastic rate following, for instance, a Cox-Ingersoll-Ross
(CIR) dynamic.} The dividend yield $\delta$ is matched to the historical dividend
yield. For a given vector of parameters, the option prices are converted
into Black and Scholes implied volatilities. We then find by a least-squares
approach the parameters for which the observed implied volatilities
match the ones predicted by the AJD option pricing model. 
\begin{equation}
\Theta\equiv\min\sum_{k=1}^{N_{K}}\sum_{\ell=1}^{N_{T}}\left\{ \widehat{IV_{k\ell}^{j}}-IV(K_{k},T_{\ell},t;S_{t},\Theta)\right\} ^{2}\label{eq:least_squares_calibration}
\end{equation}
where $k=1,\dots,N_{K}$ indexes observed strikes, and $t=1,2,\dots,N_{T}$
indexes observed maturities. $IV(\cdot)$ generates implied volatilities
using the parametric model~\ref{eq:ajd}. $\widehat{IV_{k\ell}^{jt}}$
are the observed implied volatilities for firm $j$ on day $t$, at
observed strike $K_{k}$ and maturity $T_{\ell}$.

\subsection{Identifying the Impact of Wildfires on Tail Risk}

Table~\ref{tab:calibration_treatment_group} presents the results
of the calibration of the two models. For each model, the calibration
is performed for the control group (first column) and for the treatment
group (second column). For each model, the mean squared error is also
reported, as the ratio of squared errors over the total variance.

Overall model calibration suggest that the most noticeable difference
between the control and the treatment group is in the jump magnitude:
in the jump diffusion model, the average magnitude of jumps goes from
$-0.308$ in the control group to $-1.10$ in the treatment group.
In the double exponential model the exponential parameter of downward
jumps increases from $0.908$ to $1.765$. 

These results suggest a pricing of tail risk consistent with the asymmetry
of the volatility surface observed for PG\&E on Figure~\ref{fig:pge_iv_surface}
and for the panel data set on Figure~\ref{fig:non_parametric}. Downward
jumps are systematically priced: in the double exponential jump model,
the probability of downward jumps increases 5 percentage points in
the treatment group, and the magnitude of downward jumps almost doubles.
Overall, results are consistent with wildfires causing a significant
increase in the pricing of asymmetric tail risk of the underlying
stock in equity options.

\section{Conclusion}

\label{sec:conclusion}

This paper develops a fast method for the identification of investors'
hedging of localized natural disaster risk, and the relationship between
such risk and investors' marginal utility of wealth. 45 million firm-,
maturity-, moneyness-level option prices over two decades reveal the
market's demand for climate risk hedging on the entire surface of
out-of-the-money options by strike and maturity at daily frequency.
The convexity of option prices yields a complete set of Arrow Debreu
prices for each state on the support of quoted prices, including those
states when the firm's establishments are exposed to physical risk.
Availability of granular data on physical risk exposure could spark
a literature at the intersection of option hedging and climate risk,
where financial market participants design delta hedging strategies\footnote{Classic references on delta hedging include \citeasnoun{hull2017optimal},
\citeasnoun{clewlow1997optimal}, \citeasnoun{hutchinson1994nonparametric}. } that take into account the exposure of firms to natural disasters,
including its impact on stochastic volatility and jumps.

Observed option prices suggest that investors hedge for natural disaster
risk, with larger demand for out of the money puts than what is suggested
by the small share of wildfire-exposed stocks in the market portfolio.
This finding is consistent with the hiatus described by \citeasnoun{cochrane2022portfolios}
between financial theory's guidance~\cite{markowitz1952utility,merton1969lifetime}
of holding the market portfolio and its practice by investors. This
direct evidence is also consistent with prior empirical work on the
lack of diversification of portfolios~\cite{goetzmann2005individual}.

Results could also be consistent with the findings of the behavioral
economics literature. \citeasnoun{rabin2001anomalies} presents
evidence that agents are risk averse for small gambles,  an anomaly
in the expected utility framework. 

While \citeasnoun{merton1976option} suggests that jumps may lead
to market incompleteness, the paper provides insights into the pricing
of parametric insurance by identifying of the risk neutral probabilities
of downward jumps for wildfire-exposed stock. Parametric insurance
delivers a fixed payoff in the event of a spatially-located natural
disaster. The Arrow Debreu prices obtained by calibrating the option-implied
risk neutral distribution pin down such price. Comparing such risk
neutral prices with the price of a corresponding set of weather derivatives\footnote{An extensive literature estimates the parametric relationship between
wildfire occurrence, natural variability and anthropogenic warming
\citeasnoun{zhuang2021quantifying}.} could enable the separate identification of weather surprises from
predictable events. 

\clearpage\pagebreak{}

\bibliographystyle{agsm}
\bibliography{options_wildfires}

\clearpage\pagebreak{}

\clearpage\pagebreak{}

\clearpage\pagebreak{}

\newgeometry{left=0.5in,bottom=0.5in,top=0.5in,right=0.5in}

\restoregeometry\clearpage\pagebreak\newgeometry{left=0.5in,bottom=0.5in,top=0.5in,right=0.5in}

\restoregeometry\clearpage\pagebreak{}

\newgeometry{left=0.5in,bottom=0.5in,top=0.5in,right=0.5in}
\restoregeometry\clearpage\pagebreak{}

\clearpage\pagebreak{}

\clearpage\pagebreak{}

\clearpage\pagebreak{}

\clearpage\pagebreak{}

\begin{table}
{\footnotesize{}\caption{The Impact of Wildfires on the Volatility Smile : Panel Data Evidence,
2000-2018}
\label{tab:panel_data_evidence} }\emph{\footnotesize{}These two tables
present the impact of a wildfire, on the days of the wildfire, on
the implied volatility skew for puts (upper panel) and for calls (lower
panel). The specification is a linear panel fixed effect regression,
as in (\ref{eq:linear_specification}) with double-clustered standard
errors. $S$ is the forward price. We take the square root of maturity,
as in \citeasnoun{dumas1998implied} and consistent with the empirical
term structure of implied volatilities.}{\footnotesize\par}

\footnotesize
\begin{centering}
{\footnotesize{}}%
\begin{tabular*}{1\textwidth}{@{\extracolsep{\fill}}@{\extracolsep{\fill}}lcccc}
\toprule 
 & \multicolumn{4}{c}{{\footnotesize{}Implied Volatility, Puts}}\tabularnewline
\cmidrule{2-5} \cmidrule{3-5} \cmidrule{4-5} \cmidrule{5-5} 
 & {\footnotesize{}(1) } & {\footnotesize{}(2) } & {\footnotesize{}(3) } & {\footnotesize{}(4) }\tabularnewline
\midrule 
{\footnotesize{}Constant } & {\footnotesize{}1.093{*}{*}{*} } &  &  & \tabularnewline
 & {\footnotesize{}(0.001) } &  &  & \tabularnewline
{\footnotesize{}Moneyness $K/S$ } & {\footnotesize{}$-$0.342{*}{*}{*} } & {\footnotesize{}$-$0.394{*}{*}{*} } & {\footnotesize{}$-$0.355{*}{*}{*} } & {\footnotesize{}$-$0.413{*}{*}{*} }\tabularnewline
 & {\footnotesize{}(0.001) } & {\footnotesize{}(0.022) } & {\footnotesize{}(0.023) } & {\footnotesize{}(0.032) }\tabularnewline
{\footnotesize{}Maturity $\sqrt{T}$ } & {\footnotesize{}$-$0.044{*}{*}{*} } & {\footnotesize{}$-$0.051{*}{*}{*} } & {\footnotesize{}$-$0.044{*}{*}{*} } & {\footnotesize{}$-$0.051{*}{*}{*} }\tabularnewline
 & {\footnotesize{}(0.000) } & {\footnotesize{}(0.001) } & {\footnotesize{}(0.002) } & {\footnotesize{}(0.002) }\tabularnewline
{\footnotesize{}$K/S$ $\times$ $\sqrt{T}$ } & {\footnotesize{}0.014{*}{*}{*} } & {\footnotesize{}0.014{*}{*}{*} } & {\footnotesize{}0.014{*}{*}{*} } & {\footnotesize{}0.013{*}{*}{*} }\tabularnewline
 & {\footnotesize{}(0.000) } & {\footnotesize{}(0.002) } & {\footnotesize{}(0.002) } & {\footnotesize{}(0.002) }\tabularnewline
{\footnotesize{}Treated } & {\footnotesize{}0.131{*}{*}{*} } & {\footnotesize{}0.151{*}{*} } & {\footnotesize{}0.124{*}{*} } & {\footnotesize{}0.145{*}{*} }\tabularnewline
 & {\footnotesize{}(0.006) } & {\footnotesize{}(0.063) } & {\footnotesize{}(0.046) } & {\footnotesize{}(0.070) }\tabularnewline
{\footnotesize{}Treated $\times$ $K/S$ } & {\footnotesize{}$-$0.197{*}{*}{*} } & {\footnotesize{}$-$0.176{*}{*} } & {\footnotesize{}$-$0.184{*}{*}{*} } & {\footnotesize{}$-$0.159{*}{*} }\tabularnewline
 & {\footnotesize{}(0.006) } & {\footnotesize{}(0.070) } & {\footnotesize{}(0.052) } & {\footnotesize{}(0.076) }\tabularnewline
{\footnotesize{}Treated $\times$ $\sqrt{T}$ } & {\footnotesize{}$-$0.015{*}{*}{*} } & {\footnotesize{}$-$0.015{*}{*} } & {\footnotesize{}$-$0.014{*}{*} } & {\footnotesize{}$-$0.014{*}{*} }\tabularnewline
 & {\footnotesize{}(0.001) } & {\footnotesize{}(0.005) } & {\footnotesize{}(0.004) } & {\footnotesize{}(0.006) }\tabularnewline
{\footnotesize{}Treated $\times$ $K/S$ $\times$ $\sqrt{T}$ } & {\footnotesize{}0.016{*}{*}{*} } & {\footnotesize{}0.016{*}{*} } & {\footnotesize{}0.016{*}{*}{*} } & {\footnotesize{}0.016{*}{*} }\tabularnewline
 & {\footnotesize{}(0.001) } & {\footnotesize{}(0.006) } & {\footnotesize{}(0.005) } & {\footnotesize{}(0.006) }\tabularnewline
\midrule 
{\footnotesize{}Firm fixed effects } &  & {\footnotesize{}Yes } &  & {\footnotesize{}Yes }\tabularnewline
{\footnotesize{}Day fixed effects } &  &  & {\footnotesize{}Yes } & {\footnotesize{}Yes }\tabularnewline
\midrule 
{\footnotesize{}$N$ } & {\footnotesize{}16,842,463 } & {\footnotesize{}16,842,463 } & {\footnotesize{}16,842,463 } & {\footnotesize{}16,842,463 }\tabularnewline
{\footnotesize{}$R^{2}$ } & {\footnotesize{}0.109 } & {\footnotesize{}0.309 } & {\footnotesize{}0.146 } & {\footnotesize{}0.352 }\tabularnewline
 &  &  &  & \tabularnewline
\midrule 
 & \multicolumn{4}{c}{{\footnotesize{}Implied Volatility, Calls}}\tabularnewline
\cmidrule{2-5} \cmidrule{3-5} \cmidrule{4-5} \cmidrule{5-5} 
 & {\footnotesize{}(1) } & {\footnotesize{}(2) } & {\footnotesize{}(3) } & {\footnotesize{}(4) }\tabularnewline
\midrule 
{\footnotesize{}Constant } & {\footnotesize{}0.509{*}{*}{*} } &  &  & \tabularnewline
 & {\footnotesize{}(0.001) } &  &  & \tabularnewline
{\footnotesize{}Moneyness $K/S$ } & {\footnotesize{}0.219{*}{*}{*} } & {\footnotesize{}0.066{*}{*} } & {\footnotesize{}0.206{*}{*}{*} } & {\footnotesize{}0.048 }\tabularnewline
 & {\footnotesize{}(0.001) } & {\footnotesize{}(0.023) } & {\footnotesize{}(0.020) } & {\footnotesize{}(0.031) }\tabularnewline
{\footnotesize{}Maturity $\sqrt{T}$ } & {\footnotesize{}$-$0.003{*}{*}{*} } & {\footnotesize{}$-$0.016{*}{*}{*} } & {\footnotesize{}$-$0.003{*}{*} } & {\footnotesize{}$-$0.017{*}{*}{*} }\tabularnewline
 & {\footnotesize{}(0.000) } & {\footnotesize{}(0.002) } & {\footnotesize{}(0.002) } & {\footnotesize{}(0.002) }\tabularnewline
{\footnotesize{}$K/S$ $\times$ $\sqrt{T}$ } & {\footnotesize{}$-$0.026{*}{*}{*} } & {\footnotesize{}$-$0.020{*}{*}{*} } & {\footnotesize{}$-$0.026{*}{*}{*} } & {\footnotesize{}$-$0.020{*}{*}{*} }\tabularnewline
 & {\footnotesize{}(0.000) } & {\footnotesize{}(0.002) } & {\footnotesize{}(0.001) } & {\footnotesize{}(0.002) }\tabularnewline
{\footnotesize{}Treated } & {\footnotesize{}0.139{*}{*}{*} } & {\footnotesize{}0.148{*} } & {\footnotesize{}0.134{*}{*} } & {\footnotesize{}0.148 }\tabularnewline
 & {\footnotesize{}(0.006) } & {\footnotesize{}(0.087) } & {\footnotesize{}(0.049) } & {\footnotesize{}(0.091) }\tabularnewline
{\footnotesize{}Treated $\times$ $K/S$ } & {\footnotesize{}$-$0.194{*}{*}{*} } & {\footnotesize{}$-$0.173{*} } & {\footnotesize{}$-$0.180{*}{*}{*} } & {\footnotesize{}$-$0.155 }\tabularnewline
 & {\footnotesize{}(0.006) } & {\footnotesize{}(0.092) } & {\footnotesize{}(0.047) } & {\footnotesize{}(0.095) }\tabularnewline
{\footnotesize{}Treated $\times$ $\sqrt{T}$ } & {\footnotesize{}$-$0.016{*}{*}{*} } & {\footnotesize{}$-$0.015{*}{*} } & {\footnotesize{}$-$0.015{*}{*}{*} } & {\footnotesize{}$-$0.013{*} }\tabularnewline
 & {\footnotesize{}(0.001) } & {\footnotesize{}(0.007) } & {\footnotesize{}(0.004) } & {\footnotesize{}(0.008) }\tabularnewline
{\footnotesize{}Treated $\times$ $K/S$ $\times$ $\sqrt{T}$ } & {\footnotesize{}0.017{*}{*}{*} } & {\footnotesize{}0.015{*}{*} } & {\footnotesize{}0.017{*}{*}{*} } & {\footnotesize{}0.015{*} }\tabularnewline
 & {\footnotesize{}(0.001) } & {\footnotesize{}(0.007) } & {\footnotesize{}(0.004) } & {\footnotesize{}(0.008) }\tabularnewline
\midrule 
{\footnotesize{}Firm fixed effects } &  & {\footnotesize{}Yes } &  & {\footnotesize{}Yes }\tabularnewline
{\footnotesize{}Day fixed effects } &  &  & {\footnotesize{}Yes } & {\footnotesize{}Yes }\tabularnewline
\midrule 
{\footnotesize{}$N$ } & {\footnotesize{}16,839,339 } & {\footnotesize{}16,839,339 } & {\footnotesize{}16,839,339 } & {\footnotesize{}16,839,339 }\tabularnewline
{\footnotesize{}$R^{2}$ } & {\footnotesize{}0.091 } & {\footnotesize{}0.280 } & {\footnotesize{}0.127 } & {\footnotesize{}0.321 }\tabularnewline
\bottomrule
\end{tabular*}{\footnotesize\par}
\par\end{centering}
\begin{centering}
{\footnotesize{}\bigskip{}
}{\footnotesize\par}
\par\end{centering}
\centering{}\emph{\footnotesize{}{*}{*}{*}: significant at 1\%, {*}{*}:
significant at 5\%, {*}: significant at 10\%.}{\footnotesize\par}
\end{table}

\clearpage\pagebreak{}

\begin{table}
\caption{By Industry: Control and Treatment Group}
\label{tab:by_industry}

\emph{These two tables present industry-level statistics on the share
of firm $\times$ date observations treated, i.e. exposed to a wildfire,
with either more than 10\% of establishments, employment, or sales
in the perimeter of a wildfire between 2000 and 2018. The first table
displays overall statistics for the 238 6-digit NAICS codes. 74.4\%
of such industries have at least one treated firm $\times$date observation,
and on average 1.5\% of observations are treated. The bottom table
presents the industries ranked in decreasing order of the share of
observations treated.}

\begin{center}

\subfloat[Industry Statistics]{

\footnotesize
\begin{tabular}{cccccccc}
\toprule
Share of Industries & \multicolumn{6}{c}{Fraction Treated} & Number of Industries\\
\cmidrule(lr){2-7}
 with Treated Firm(s) & Average & P10 & Median & P90 & P95 & P99 & 6-Digit NAICS \\
\midrule
0.744 & 0.015 &  0.000 & 0.012 & 0.035 & 0.043 & 0.064 & 238 \\
\bottomrule
\end{tabular}}

\bigskip{}

\subfloat[Share of Firm $\times$ Date Treated, by Industry]{\footnotesize
% latex table generated in R 4.1.2 by xtable 1.8-4 package
% Tue Jun 21 15:20:11 2022
\begin{tabular}{lp{7cm}cc}
  \toprule
NAICS Code & 2017 NAICS Title & Fraction Treated & Distinct Firm $\times$ Dates \\ 
  \midrule
621610 & Home Health Care Services & 0.08 & 251 \\ 
  332993 & Ammunition (except Small Arms) Manufacturing & 0.07 & 178 \\ 
  424720 & Petroleum and Petroleum Products Merchant Wholesalers (except Bulk Stations and Terminals) & 0.07 & 461 \\ 
  335999 & All Other Miscellaneous Electrical Equipment and Component Manufacturing & 0.06 & 281 \\ 
  722320 & Caterers & 0.05 & 300 \\ 
  211130 & Natural Gas Extraction & 0.05 & 351 \\ 
  517311 & Wired Telecommunications Carriers & 0.05 & 765 \\ 
  532490 & Other Commercial and Industrial Machinery and Equipment Rental and Leasing & 0.05 & 259 \\ 
  561920 & Convention and Trade Show Organizers & 0.05 & 522 \\ 
  333999 & All Other Miscellaneous General Purpose Machinery Manufacturing & 0.05 & 397 \\ 
  621512 & Diagnostic Imaging Centers & 0.04 & 404 \\ 
  447190 & Other Gasoline Stations & 0.04 & 412 \\ 
  451211 & Book Stores & 0.04 & 278 \\ 
  334515 & Instrument Manufacturing for Measuring and Testing Electricity and Electrical Signals & 0.04 & 285 \\ 
  424110 & Printing and Writing Paper Merchant Wholesalers & 0.04 & 285 \\ 
  221310 & Water Supply and Irrigation Systems & 0.04 & 1125 \\ 
  483112 & Deep Sea Passenger Transportation & 0.04 & 1078 \\ 
  316998 & All Other Leather Good and Allied Product Manufacturing & 0.04 & 470 \\ 
  445110 & Supermarkets and Other Grocery (except Convenience) Stores & 0.04 & 315 \\ 
  541614 & Process, Physical Distribution, and Logistics Consulting Services & 0.04 & 315 \\ 
   \bottomrule
\end{tabular}

}

\bigskip{}

\subfloat[Treatment Effects by Industry]{\footnotesize
\begin{tabular}{lcccc}
\toprule
& Share of Industries with             & \multicolumn{3}{c}{Statistically Significant} \\
\cmidrule(lr){3-5}
&  Steepening Volatility Smile &  at 90\% &  at 95\% &  at 99\% \\
\midrule
Out-of-the-money Puts & 0.657 & 0.600 & 0.571 & 0.572 \\
Out-of-the-money Calls & 0.257 & 0.243 & 0.243 & 0.243 \\
\bottomrule
\end{tabular}}

\end{center}
\end{table}

\clearpage\pagebreak{}
\begin{table}
\caption{Permanent Effects of a Wildfire on Options' Implied Volatility Smile}
\label{tab:permanent_effects}

\emph{\footnotesize{}This table estimates the permanent effects of
wildfires on option implied volatilities. "After First Wildfire"
is equal to 1 on any day after the first wildfire affecting the publicly
traded company. "After Last Wildfire" is equal to 1 on any day after
the first wildfire affecting the company. Other covariates are similar
to the linear specification of table~\ref{tab:panel_data_evidence}
and specification~(\ref{eq:linear_specification}). As in Table~\ref{tab:panel_data_evidence},
standard errors are double-clustered at the firm and day levels. All
four regressions include two-way fixed effects.}{\footnotesize\par}
\begin{centering}
{\footnotesize{}}%
\begin{tabular*}{1\textwidth}{@{\extracolsep{\fill}}@{\extracolsep{\fill}}lcccc}
\toprule 
 & \multicolumn{4}{c}{{\footnotesize{}Implied Volatility}}\tabularnewline
\cmidrule{2-5} \cmidrule{3-5} \cmidrule{4-5} \cmidrule{5-5} 
 & {\footnotesize{}Puts } & {\footnotesize{}Calls } & {\footnotesize{}Puts } & {\footnotesize{}Calls }\tabularnewline
\cmidrule{2-5} \cmidrule{3-5} \cmidrule{4-5} \cmidrule{5-5} 
 & {\footnotesize{}(1) } & {\footnotesize{}(2) } & {\footnotesize{}(3) } & {\footnotesize{}(4) }\tabularnewline
{\footnotesize{}Moneyness $K/S$ } & {\footnotesize{}-0.420{*}{*}{*} } & {\footnotesize{}0.080{*}{*} } & {\footnotesize{}-0.414{*}{*}{*} } & {\footnotesize{}0.067{*}{*} }\tabularnewline
 & {\footnotesize{}(0.031) } & {\footnotesize{}(0.031) } & {\footnotesize{}(0.030) } & {\footnotesize{}(0.031) }\tabularnewline
{\footnotesize{}Maturity $\sqrt{T}$ } & {\footnotesize{}-0.053{*}{*}{*} } & {\footnotesize{}-0.015{*}{*}{*} } & {\footnotesize{}-0.052{*}{*}{*} } & {\footnotesize{}-0.016{*}{*}{*} }\tabularnewline
 & {\footnotesize{}(0.002) } & {\footnotesize{}(0.002) } & {\footnotesize{}(0.002) } & {\footnotesize{}(0.002) }\tabularnewline
{\footnotesize{}Moneyness $K/S$ $\times$ Maturity $\sqrt{T}$ } & {\footnotesize{}0.014{*}{*}{*} } & {\footnotesize{}-0.022{*}{*}{*} } & {\footnotesize{}0.013{*}{*}{*} } & {\footnotesize{}-0.021{*}{*}{*} }\tabularnewline
 & {\footnotesize{}(0.002) } & {\footnotesize{}(0.002) } & {\footnotesize{}(0.002) } & {\footnotesize{}(0.002) }\tabularnewline
 &  &  &  & \tabularnewline
{\footnotesize{}After First Wildfire } & {\footnotesize{}-0.047 } & {\footnotesize{}0.126{*}{*} } &  & \tabularnewline
 & {\footnotesize{}(0.062) } & {\footnotesize{}(0.056) } &  & \tabularnewline
{\footnotesize{}After First Wildfire $\times$ Moneyness $K/S$ } & {\footnotesize{}0.017 } & {\footnotesize{}-0.152{*}{*} } &  & \tabularnewline
 & {\footnotesize{}(0.066) } & {\footnotesize{}(0.057) } &  & \tabularnewline
{\footnotesize{}After First Wildfire $\times$ Maturity $\sqrt{T}$ } & {\footnotesize{}0.005 } & {\footnotesize{}-0.009{*}{*} } &  & \tabularnewline
 & {\footnotesize{}(0.005) } & {\footnotesize{}(0.004) } &  & \tabularnewline
{\footnotesize{}After First Wildfire $\times$ Moneyness $K/S$ $\times$
Maturity $\sqrt{T}$ } & {\footnotesize{}-0.001 } & {\footnotesize{}0.013{*}{*} } &  & \tabularnewline
 & {\footnotesize{}(0.005) } & {\footnotesize{}(0.004) } &  & \tabularnewline
 &  &  &  & \tabularnewline
{\footnotesize{}After Last Wildfire } &  &  & {\footnotesize{}-0.016 } & {\footnotesize{}0.106 }\tabularnewline
 &  &  & {\footnotesize{}(0.075) } & {\footnotesize{}(0.066) }\tabularnewline
{\footnotesize{}After Last Wildfire $\times$ Moneyness $K/S$ } &  &  & {\footnotesize{}-0.013 } & {\footnotesize{}-0.132{*}{*} }\tabularnewline
 &  &  & {\footnotesize{}(0.081) } & {\footnotesize{}(0.066) }\tabularnewline
{\footnotesize{}After Last Wildfire $\times$ Maturity $\sqrt{T}$ } &  &  & {\footnotesize{}0.002 } & {\footnotesize{}-0.008{*} }\tabularnewline
 &  &  & {\footnotesize{}(0.006) } & {\footnotesize{}(0.005) }\tabularnewline
{\footnotesize{}After Last Wildfire $\times$ Moneyness $K/S$ $\times$
Maturity $\sqrt{T}$ } &  &  & {\footnotesize{}0.002 } & {\footnotesize{}0.011{*}{*} }\tabularnewline
 &  &  & {\footnotesize{}(0.007) } & {\footnotesize{}(0.005) }\tabularnewline
\midrule 
{\footnotesize{}Firm fixed effects } & {\footnotesize{}Yes } & {\footnotesize{}Yes } & {\footnotesize{}Yes } & {\footnotesize{}Yes }\tabularnewline
{\footnotesize{}Day fixed effects } & {\footnotesize{}Yes } & {\footnotesize{}Yes } & {\footnotesize{}Yes } & {\footnotesize{}Yes }\tabularnewline
\midrule 
{\footnotesize{}$N$ } & {\footnotesize{}16,842,463 } & {\footnotesize{}16,839,339 } & {\footnotesize{}16,842,463 } & {\footnotesize{}16,839,339 }\tabularnewline
{\footnotesize{}$R^{2}$ } & {\footnotesize{}0.352 } & {\footnotesize{}0.322 } & {\footnotesize{}0.352 } & {\footnotesize{}0.322 }\tabularnewline
\bottomrule
\end{tabular*}{\footnotesize\par}
\par\end{centering}
\begin{centering}
{\footnotesize{}\bigskip{}
}{\footnotesize\par}
\par\end{centering}
\centering{}\emph{\footnotesize{}{*}{*}{*}: significant at 1\%, {*}{*}:
significant at 5\%, {*}: significant at 10\%.}{\footnotesize\par}
\end{table}

\clearpage\pagebreak{}

\begin{table}
\caption{Implied Jumps and Stochastic Volatility for Wildfire-Exposed Stocks:
US Exchange-Traded Equity Options Exposed to Wildfires}
\label{tab:calibration_treatment_group}

\emph{\footnotesize{}This table presents the estimation of the stochastic
volatility and jump parameters implied by the equity options of the
publicly-listed firms exposed to wildfires. This is the treatment
group of Section~\ref{sec:panel_data_evidence}. We also present
the calibration for the control group. Estimation is performed by
minimizing the distance between the model-based implied volatilities
to the implied volatilities of the treatment group (resp. the control
group). The baseline Black and Scholes diffusion without jumps has
a mean squared error of 0.703 in the treatment group and 0.677 in
the control group.}{\footnotesize\par}
\begin{centering}
{\footnotesize{}}%
\begin{tabular*}{1\textwidth}{@{\extracolsep{\fill}}@{\extracolsep{\fill}}lcccc}
\toprule 
 & \multicolumn{2}{c}{\emph{\footnotesize{}Jump Diffusion}} & \multicolumn{2}{c}{\emph{\footnotesize{}Double Exponential Jumps}}\tabularnewline
 & \multicolumn{2}{c}{\emph{\footnotesize{}\citeasnoun{merton1976option}}} & \multicolumn{2}{c}{\emph{\footnotesize{}\citeasnoun{kou2004option}}}\tabularnewline
\cmidrule{2-5} \cmidrule{3-5} \cmidrule{4-5} \cmidrule{5-5} 
 & {\footnotesize{}Control} & {\footnotesize{}Treatment} & {\footnotesize{}Control} & {\footnotesize{}Treatment}\tabularnewline
\cmidrule{2-5} \cmidrule{3-5} \cmidrule{4-5} \cmidrule{5-5} 
{\footnotesize{}Volatility ($\sigma$)} & {\footnotesize{}0.064} & {\footnotesize{}0.003} & {\footnotesize{}0.001} & {\footnotesize{}0.003}\tabularnewline
 &  &  &  & \tabularnewline
{\footnotesize{}Jump intensity ($\lambda$)} & {\footnotesize{}0.148} & {\footnotesize{}0.146} & {\footnotesize{}0.721} & {\footnotesize{}0.428}\tabularnewline
{\footnotesize{}Mean jump magnitude ($\mu^{s}$)} & {\footnotesize{}$-$0.308} & {\footnotesize{}$-$1.013} & {\footnotesize{}-} & {\footnotesize{}-}\tabularnewline
{\footnotesize{}S.D. of jump magnitude ($\sigma^{s}$)} & {\footnotesize{}1.354} & {\footnotesize{}1.533} & {\footnotesize{}-} & {\footnotesize{}-}\tabularnewline
 &  &  &  & \tabularnewline
{\footnotesize{}Probability of an upward jump ($p$)} & {\footnotesize{}-} & {\footnotesize{}-} & {\footnotesize{}0.796} & {\footnotesize{}0.743}\tabularnewline
{\footnotesize{}Mean downward jump ($-1/\eta_{1}$)} & {\footnotesize{}-} & {\footnotesize{}-} & {\footnotesize{}$-$0.908} & {\footnotesize{}$-$1.765}\tabularnewline
{\footnotesize{}Mean upward jump ($1/\eta_{2}$)} & {\footnotesize{}-} & {\footnotesize{}-} & {\footnotesize{}0.378} & {\footnotesize{}0.383}\tabularnewline
 &  &  &  & \tabularnewline
{\footnotesize{}%Mean Squared Error                    & 0.240 & 0.258 & 0.270 & 0.241 \\ 
Mean Squared Error} & {\footnotesize{}0.056} & {\footnotesize{}0.066} & {\footnotesize{}0.072} & {\footnotesize{}0.058}\tabularnewline
\bottomrule
\end{tabular*}{\footnotesize\par}
\par\end{centering}
\emph{\footnotesize{}Parameters are calibrated by pricing options
using the characteristic function approach. The calibration minimizes
the sum of squared deviations between the model and the empirical
implied volatilities.}{\footnotesize\par}
\end{table}
\clearpage\pagebreak\setcounter{proposition}{0}

\section{Proof of Proposition 1\label{sec:Proof-of-Proposition}}
\begin{prop*}
\textbf{\emph{Sensitivity of State Prices $\zeta_{T}$ to the Terminal
Value of the Wildfire-Exposed Stock}}

When the parameters of the stocks' stochastic processes $\mu,\alpha,\beta,\sigma,\sigma^{w}$
are constant, when the investor has CRRA preferences $u(w)=\frac{w^{1-\gamma}}{1-\gamma}$,
a closed form ties the sensitivity of state prices w.r.t the wildfire-exposed
stock to the investor's risk aversion w.r.t. wealth: 
\begin{equation}
\gamma^{w}=(q^{w}+\rho\frac{\sigma}{\sigma^{w}}q)\gamma\label{eq:prop1}
\end{equation}
where $q^{w}$ the share of wealth invested in the wildfire-exposed
stock, $q$ is the share of wealth invested in the diversified portfolio;
$\rho$ is the correlation between the returns on the wildfire stock
and the diversified portfolio, $\rho=\beta\sigma/\sqrt{\beta^{2}\sigma^{2}+(\sigma^{w})^{2}}$;
$\sigma$ and $\sigma^{w}$ the volatilities of the stock and the
wildfire-exposed stock respectively. 
\end{prop*}
\begin{proof}
This framework corresponds to \possessivecite{merton1969lifetime}
continuous-time portfolio problem with two risky assets. Risk aversion
is the elasticity of the pricing kernel w.r.t. either wealth (for
$\gamma$) or the wildfire-exposed stock (for $\gamma^{w}$): 
\begin{equation}
\gamma=-\frac{d\log\zeta_{T}}{d\log W_{T}},\qquad\gamma^{w}=-\frac{d\log E(\zeta_{T}\vert S_{T}^{w})}{d\log S_{T}^{w}}
\end{equation}
In \citeasnoun{merton1969lifetime}, the portfolio share $q_{t}$
invested in the index and the share invested in the wildfire-exposed
stock $q_{t}^{w}$ are constant, $q_{t}=q$ and $q_{t}^{w}=q^{w}$.
The shares held in the wildfire-exposed stock and in the index are
obtained using the Hamilton-Jacobi-Bellman equation: 
\begin{equation}
q^{w}=\frac{\alpha+\beta\mu-r}{\gamma(\sigma^{w})^{2}},\qquad q=\frac{\mu-r}{\gamma\sigma^{2}}-q^{w}\beta
\end{equation}
where $\gamma=-\frac{V_{W}}{V_{WW}W}$ is the Arrow-Pratt relative
risk aversion with respect to wealth, $V_{W}=\frac{\partial V}{\partial W}$,
$V_{WW}=\frac{\partial^{2}V}{\partial W^{2}}$. Such risk aversion
is simply the scalar $\gamma$ when the investor's utility is CRRA.

Terminal wealth conditional on the value $S_{T}^{w}$ of the wildfire
stock is a function of the shares $q$, $q^{w}$ invested and the
paths of the Brownians:
\begin{equation}
W_{T}=W_{0}e^{(q(\mu-r)+q^{w}(\alpha+\beta\mu-r)+r-\frac{1}{2}q^{2}\sigma^{2})t}e^{(q+\beta q^{w})\sigma(B_{T}-B_{0})}e^{q^{w}\sigma^{w}(B_{T}^{w}-B_{0}^{w})}\label{eq:terminal_wealth}
\end{equation}
and $\zeta_{T}=u'(W_{T})=W_{T}^{-\gamma}$, hence the log of the pricing
kernel:
\begin{align}
\log\zeta_{T} & =-\gamma\log W_{0}-\gamma(q(\mu-r)+q^{w}(\alpha+\beta\mu-r)+r-\frac{1}{2}q^{2}\sigma^{2})t\nonumber \\
 & \quad\quad\quad\quad\quad\quad\quad-\gamma(q+\beta q^{w})\sigma(B_{T}-B_{0})+q^{w}\sigma^{w}(B_{T}^{w}-B_{0}^{w})\label{eq:log_state}
\end{align}
this can also be rewritten as a function of the values of the stocks.
To see this, first express the terminal values of $S_{T}$ and $S_{T}^{w}$:
\begin{align}
\log S_{T} & =\log S_{0}+(\mu-\frac{1}{2}\sigma^{2})t+\sigma(B_{T}-B_{0})\label{eq:terminal_value_index}\\
\log S_{T}^{w} & =\log S_{0}^{w}+(\alpha+\beta\mu-\frac{1}{2}\beta^{2}\sigma^{2}-\frac{1}{2}(\sigma^{w})^{2})t+\beta\sigma(B_{T}-B_{0})+\sigma^{w}(B_{T}^{w}-B_{0}^{w})\label{eq:terminal_value_wildfire_stock}
\end{align}
$(\log S_{T},\log S_{T}^{w})$ is bivariate normal with correlation
$\rho$ and with variances $\sigma T$ and $\sigma^{w}T$. The correlation
$\rho$ is a function of the beta of the stock and the standard deviations
$\rho=\beta\sigma/\sqrt{\beta^{2}\sigma^{2}+(\sigma^{w})^{2}}$. Hence
the classic bivariate conditional expectation formula applies. The
expected value of $\log S_{T}$ conditional on $S_{T}^{w}$ is $E(\log S_{T})+\rho\frac{\sigma}{\sigma^{w}}(\log S_{T}^{w}-E(\log S_{T}^{w}))$.
Hence $d\log E(\zeta_{T}\vert S_{T}^{w})/d\log S_{T}^{w}=q^{w}+\rho\frac{\sigma}{\sigma^{w}}q$. 
\end{proof}
\setcounter{figure}{0} 
\global\long\def\figurename{Appendix Figure}%
 
\global\long\def\thefigure{\Alph{figure}}%

\setcounter{table}{0} 
\global\long\def\tablename{Appendix Table}%
 
\global\long\def\thetable{\Alph{table}}%

\begin{figure}
\caption{Wildfire Exposure Over Time}
\label{fig:wildfire_exposure}

\emph{At daily frequency, these three panels present the log number
of establishments, the log number of employees, and the log dollar
sales of establishments of listed firms in wildfire perimeters.}

\medskip{}

\begin{center}
\begin{centering}
\subfloat[Establishments]{\includegraphics[scale=0.5]{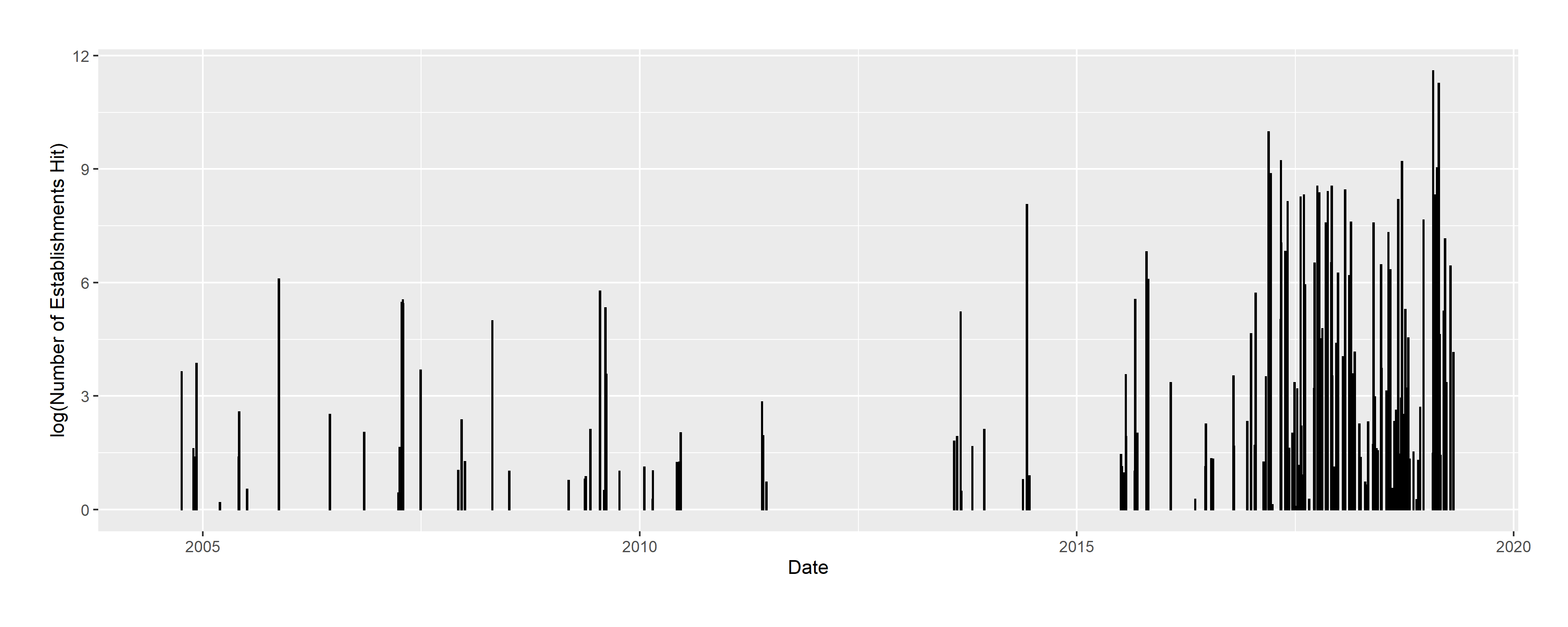}

}
\par\end{centering}
\begin{centering}
\subfloat[Employment]{\includegraphics[scale=0.5]{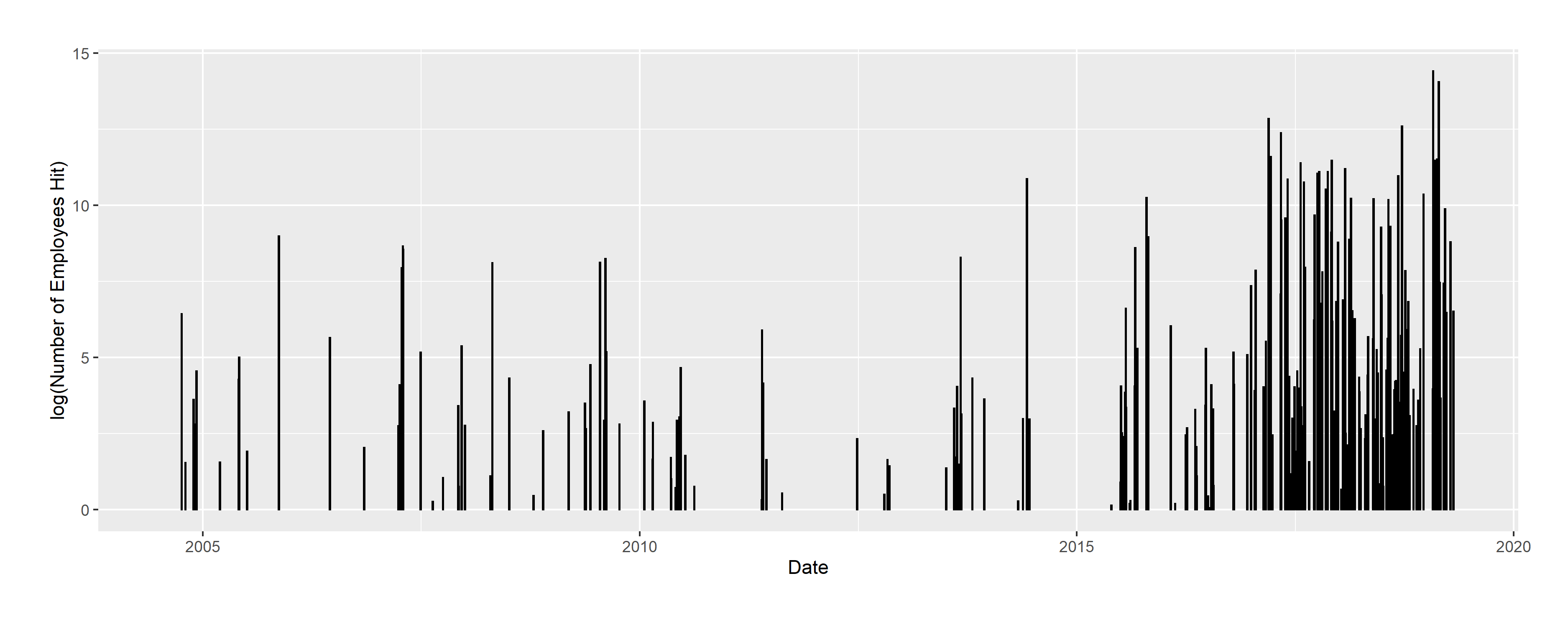}

}
\par\end{centering}
\begin{centering}
\subfloat[Sales]{\includegraphics[scale=0.5]{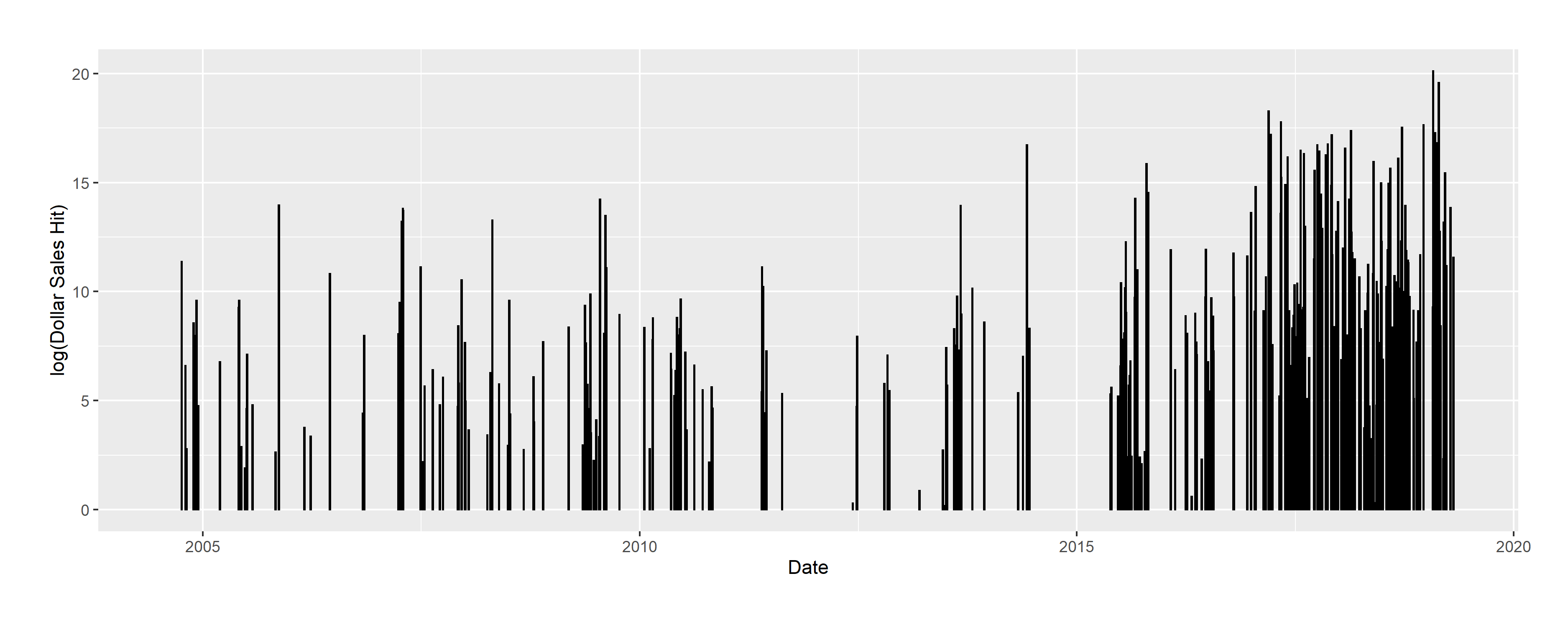}

}
\par\end{centering}
\end{center}
\end{figure}

\clearpage\pagebreak{}
\begin{table}
\caption{Industries Hit by Wildfires, Ranked by Number of Establishments Hit}
\label{tab:descriptive_statistics_industries_hit_by_wildfires}

\emph{For each 6-digit NAICS industry, this table presents the number
of establishments, the employment, in thousands, and the sales, in
thousands of \$, in wildfire perimeters. Wildfire perimeters are as
reported by the National Interagency Fire Center. Establishments,
employment, and sales as reported by Dun and Bradstreet's National
Establishment Timer Series (NETS). These numbers are cumulative over
the 681 days with a wildfire.}

\bigskip{}

{\scriptsize{}}%
\begin{tabular}{rlccc}
\toprule 
 &  & \multicolumn{3}{c}{{\scriptsize{}In Wildfire Perimeter At Any Point}}\tabularnewline
 &  & {\scriptsize{}Establishments } & {\scriptsize{}Employment } & {\scriptsize{}Sales}\tabularnewline
{\scriptsize{}NAICS } & {\scriptsize{}Description } & {\scriptsize{}('000) } & {\scriptsize{}('000) } & {\scriptsize{}('000) }\tabularnewline
\midrule 
{\scriptsize{}722513 } & {\scriptsize{}Limited-Service Restaurants } & {\scriptsize{}18.1 } & {\scriptsize{}410.8 } & {\scriptsize{}21,914.1}\tabularnewline
{\scriptsize{}522110 } & {\scriptsize{}Commercial Banking } & {\scriptsize{}15.6 } & {\scriptsize{}189.9 } & {\scriptsize{}51,623.9}\tabularnewline
{\scriptsize{}721110 } & {\scriptsize{}Hotels (except Casino Hotels) and Motels } & {\scriptsize{}14.7 } & {\scriptsize{}156.8 } & {\scriptsize{}17,780.6}\tabularnewline
{\scriptsize{}445110 } & {\scriptsize{}Supermarkets and Other Grocery (except Convenience)
Stores } & {\scriptsize{}14.5 } & {\scriptsize{}123.5 } & {\scriptsize{}29,724.8}\tabularnewline
{\scriptsize{}519130 } & {\scriptsize{}Internet Publishing and Broadcasting and Web Search
Portals } & {\scriptsize{}13.4 } & {\scriptsize{}75.3 } & {\scriptsize{}19,953.2}\tabularnewline
{\scriptsize{}523930 } & {\scriptsize{}Investment Advice } & {\scriptsize{}12.4 } & {\scriptsize{}17.2 } & {\scriptsize{}4,580.3}\tabularnewline
{\scriptsize{}333318 } & {\scriptsize{}Other Commercial and Service Industry Machinery Manufacturing } & {\scriptsize{}12.3 } & {\scriptsize{}15.1 } & {\scriptsize{}3,878.0}\tabularnewline
{\scriptsize{}561621 } & {\scriptsize{}Security Systems Services (except Locksmiths) } & {\scriptsize{}12.2 } & {\scriptsize{}14.5 } & {\scriptsize{}3,747.2}\tabularnewline
{\scriptsize{}324110 } & {\scriptsize{}Petroleum Refineries } & {\scriptsize{}11.4 } & {\scriptsize{}78.0 } & {\scriptsize{}93,234.3}\tabularnewline
{\scriptsize{}722511 } & {\scriptsize{}Full-Service Restaurants } & {\scriptsize{}10.1 } & {\scriptsize{}264.9 } & {\scriptsize{}13,746.9}\tabularnewline
{\scriptsize{}561499 } & {\scriptsize{}All Other Business Support Services } & {\scriptsize{}8.1 } & {\scriptsize{}12.4 } & {\scriptsize{}2,776.6}\tabularnewline
{\scriptsize{}611310 } & {\scriptsize{}Colleges, Universities, and Professional Schools } & {\scriptsize{}7.5 } & {\scriptsize{}11.5 } & {\scriptsize{}2,611.0}\tabularnewline
{\scriptsize{}452319 } & {\scriptsize{}All Other General Merchandise Stores } & {\scriptsize{}6.9 } & {\scriptsize{}359.4 } & {\scriptsize{}85,674.0}\tabularnewline
{\scriptsize{}532120 } & {\scriptsize{}Truck, Utility Trailer, and RV (Recreational Vehicle)
Rental and Leasing } & {\scriptsize{}5.8 } & {\scriptsize{}26.2 } & {\scriptsize{}5,279.1}\tabularnewline
{\scriptsize{}531210 } & {\scriptsize{}Offices of Real Estate Agents and Brokers } & {\scriptsize{}5.6 } & {\scriptsize{}120.1 } & {\scriptsize{}14,804.7}\tabularnewline
{\scriptsize{}486110 } & {\scriptsize{}Pipeline Transportation of Crude Oil } & {\scriptsize{}5.5 } & {\scriptsize{}31.4 } & {\scriptsize{}29,888.7}\tabularnewline
{\scriptsize{}446110 } & {\scriptsize{}Pharmacies and Drug Stores } & {\scriptsize{}4.6 } & {\scriptsize{}87.0 } & {\scriptsize{}31,673.4}\tabularnewline
{\scriptsize{}722515 } & {\scriptsize{}Snack and Nonalcoholic Beverage Bars } & {\scriptsize{}3.7 } & {\scriptsize{}61.0 } & {\scriptsize{}3,530.2}\tabularnewline
{\scriptsize{}515210 } & {\scriptsize{}Cable and Other Subscription Programming } & {\scriptsize{}3.4 } & {\scriptsize{}29.9 } & {\scriptsize{}10,010.1}\tabularnewline
{\scriptsize{}523920 } & {\scriptsize{}Portfolio Management } & {\scriptsize{}3.3 } & {\scriptsize{}69.7 } & {\scriptsize{}10,801.0}\tabularnewline
\bottomrule
\end{tabular}{\scriptsize\par}
\end{table}

\clearpage\pagebreak{}

\begin{table}
\caption{Trading of Short-Maturity Options and Wildfires: Panel Data Evidence}
\label{tab:trading_short_maturity}

\emph{These tables present the impact of a wildfire period, during
the wildfire, on the maturity of traded options. }

\begin{centering}
{\footnotesize{}}%
\begin{tabular*}{1\textwidth}{@{\extracolsep{\fill}}@{\extracolsep{\fill}}lccccc}
\toprule 
 & \multicolumn{5}{c}{{\footnotesize{}Puts}}\tabularnewline
 & \multicolumn{1}{c}{{\footnotesize{}$T<5$}} & \multicolumn{1}{c}{{\footnotesize{}$T<10$}} & \multicolumn{1}{c}{{\footnotesize{}$T<20$}} & \multicolumn{1}{c}{{\footnotesize{}$T<50$}} & \multicolumn{1}{c}{{\footnotesize{}$T<100$}}\tabularnewline
 & {\footnotesize{}(1) } & {\footnotesize{}(2) } & {\footnotesize{}(3) } & {\footnotesize{}(4) } & {\footnotesize{}(5) }\tabularnewline
\midrule 
{\footnotesize{}Treated } & {\footnotesize{}0.011{*}{*} } & {\footnotesize{}0.020{*}{*} } & {\footnotesize{}0.049{*}{*}{*} } & {\footnotesize{}0.027 } & {\footnotesize{}-0.003 }\tabularnewline
 & {\footnotesize{}(0.004) } & {\footnotesize{}(0.007) } & {\footnotesize{}(0.013) } & {\footnotesize{}(0.017) } & {\footnotesize{}(0.012) }\tabularnewline
{\footnotesize{}Moneyness $K/S$ } & {\footnotesize{}0.006{*}{*}{*} } & {\footnotesize{}0.008{*}{*}{*} } & {\footnotesize{}0.014{*}{*}{*} } & {\footnotesize{}0.032{*}{*}{*} } & {\footnotesize{}0.007 }\tabularnewline
 & {\footnotesize{}(0.001) } & {\footnotesize{}(0.002) } & {\footnotesize{}(0.003) } & {\footnotesize{}(0.007) } & {\footnotesize{}(0.005) }\tabularnewline
{\footnotesize{}Treated $\times$ Moneyness $K/S$ } & {\footnotesize{}-0.011{*}{*}{*} } & {\footnotesize{}-0.020{*}{*}{*} } & {\footnotesize{}-0.050{*}{*}{*} } & {\footnotesize{}-0.028{*} } & {\footnotesize{}0.004 }\tabularnewline
 & {\footnotesize{}(0.003) } & {\footnotesize{}(0.005) } & {\footnotesize{}(0.013) } & {\footnotesize{}(0.016) } & {\footnotesize{}(0.012) }\tabularnewline
\midrule 
{\footnotesize{}Firm fixed effects } & {\footnotesize{}Yes } & {\footnotesize{}Yes } & {\footnotesize{}Yes } & {\footnotesize{}Yes } & {\footnotesize{}Yes }\tabularnewline
{\footnotesize{}Day fixed effects } & {\footnotesize{}Yes } & {\footnotesize{}Yes } & {\footnotesize{}Yes } & {\footnotesize{}Yes } & {\footnotesize{}Yes }\tabularnewline
\midrule 
{\footnotesize{}$N$ } & {\footnotesize{}16,842,463 } & {\footnotesize{}16,842,463 } & {\footnotesize{}16,842,463 } & {\footnotesize{}16,842,463 } & {\footnotesize{}16,842,463 }\tabularnewline
{\footnotesize{}$R^{2}$ } & {\footnotesize{}0.059 } & {\footnotesize{}0.068 } & {\footnotesize{}0.076 } & {\footnotesize{}0.147 } & {\footnotesize{}0.083 }\tabularnewline
 &  &  &  &  & \tabularnewline
\midrule 
 & \multicolumn{5}{c}{{\footnotesize{}Calls}}\tabularnewline
 & \multicolumn{1}{c}{{\footnotesize{}$T<5$}} & \multicolumn{1}{c}{{\footnotesize{}$T<10$}} & \multicolumn{1}{c}{{\footnotesize{}$T<20$}} & \multicolumn{1}{c}{{\footnotesize{}$T<50$}} & \multicolumn{1}{c}{{\footnotesize{}$T<100$}}\tabularnewline
 & {\footnotesize{}(1) } & {\footnotesize{}(2) } & {\footnotesize{}(3) } & {\footnotesize{}(4) } & {\footnotesize{}(5) }\tabularnewline
\midrule 
{\footnotesize{}Treated } & {\footnotesize{}0.008{*} } & {\footnotesize{}0.021{*}{*} } & {\footnotesize{}0.022{*}{*} } & {\footnotesize{}0.029{*} } & {\footnotesize{}0.008 }\tabularnewline
 & {\footnotesize{}(0.005) } & {\footnotesize{}(0.007) } & {\footnotesize{}(0.011) } & {\footnotesize{}(0.017) } & {\footnotesize{}(0.014) }\tabularnewline
{\footnotesize{}Moneyness $K/S$ } & {\footnotesize{}0.012{*}{*}{*} } & {\footnotesize{}0.021{*}{*}{*} } & {\footnotesize{}0.034{*}{*}{*} } & {\footnotesize{}0.037{*}{*}{*} } & {\footnotesize{}0.044{*}{*}{*} }\tabularnewline
 & {\footnotesize{}(0.002) } & {\footnotesize{}(0.002) } & {\footnotesize{}(0.004) } & {\footnotesize{}(0.006) } & {\footnotesize{}(0.005) }\tabularnewline
{\footnotesize{}Treated $\times$ Moneyness $K/S$ } & {\footnotesize{}-0.010{*}{*} } & {\footnotesize{}-0.023{*}{*}{*} } & {\footnotesize{}-0.024{*}{*} } & {\footnotesize{}-0.031{*}{*} } & {\footnotesize{}-0.008 }\tabularnewline
 & {\footnotesize{}(0.004) } & {\footnotesize{}(0.006) } & {\footnotesize{}(0.010) } & {\footnotesize{}(0.015) } & {\footnotesize{}(0.013) }\tabularnewline
\midrule 
{\footnotesize{}Firm fixed effects } & {\footnotesize{}Yes } & {\footnotesize{}Yes } & {\footnotesize{}Yes } & {\footnotesize{}Yes } & {\footnotesize{}Yes }\tabularnewline
{\footnotesize{}Day fixed effects } & {\footnotesize{}Yes } & {\footnotesize{}Yes } & {\footnotesize{}Yes } & {\footnotesize{}Yes } & {\footnotesize{}Yes }\tabularnewline
\midrule 
{\footnotesize{}$N$ } & {\footnotesize{}16,839,339 } & {\footnotesize{}16,839,339 } & {\footnotesize{}16,839,339 } & {\footnotesize{}16,839,339 } & {\footnotesize{}16,839,339 }\tabularnewline
{\footnotesize{}$R^{2}$ } & {\footnotesize{}0.056 } & {\footnotesize{}0.066 } & {\footnotesize{}0.075 } & {\footnotesize{}0.151 } & {\footnotesize{}0.086 }\tabularnewline
\bottomrule
\end{tabular*}{\footnotesize\par}
\par\end{centering}
\begin{centering}
{\footnotesize{}\bigskip{}
}{\footnotesize\par}
\par\end{centering}
\centering{}\emph{\footnotesize{}{*}{*}{*}: significant at 1\%, {*}{*}:
significant at 5\%, {*}: significant at 10\%.}{\footnotesize\par}
\end{table}

\clearpage\pagebreak{}

\begin{table}
\caption{Wildfires and Arbitrage Opportunities in Option Prices}
\label{tab:wildfires_arbitrage}

\emph{These tables present the regression of the difference between
the observed option prices, as the midpoint between the bid and the
ask, and the arbitrage-free prices obtained using \citeasnoun{ait2003nonparametric}
approach described in Section~\ref{subsec:Estimation-Technique:-Arbitrage}.
Column (1) has the \$ difference noted $\Delta C$ as the lhs. Column
(2) has the absolute \$ difference $\Delta|C|$. Column (3) uses the
log difference $\Delta\log C$ as the lhs. The regression is performed
for each of the five quintiles of maturities. Performing the regression
for the entire sample or for a different set of cuts has no significant
impact on results.}

\footnotesize

\begin{center}

{\footnotesize{}}%
\begin{tabular*}{1\textwidth}{@{\extracolsep{\fill}}@{\extracolsep{\fill}}lccc}
\toprule 
 & {\footnotesize{}(1) } & {\footnotesize{}(2) } & {\footnotesize{}(3) }\tabularnewline
{\footnotesize{}Dependent variable: } & {\footnotesize{}$\Delta C$ } & {\footnotesize{}$\Delta|C|$ } & {\footnotesize{}$\Delta\log\vert C\vert$ }\tabularnewline
\midrule 
{\footnotesize{}Sample: } & \multicolumn{3}{c}{{\footnotesize{}Maturities $\leq17$ days, Q1}}\tabularnewline
\midrule 
{\footnotesize{}During a Wildfire } & {\footnotesize{}$-0.00$ } & {\footnotesize{}$0.01^{*}$ } & {\footnotesize{}$-0.49$ }\tabularnewline
 & {\footnotesize{}$(0.01)$ } & {\footnotesize{}$(0.01)$ } & {\footnotesize{}$(0.38)$ }\tabularnewline
\midrule 
{\footnotesize{}Observations } & {\footnotesize{}$121,263$ } & {\footnotesize{}$121,263$ } & {\footnotesize{}$121,263$ }\tabularnewline
{\footnotesize{}R$^{2}$ } & {\footnotesize{}$0.04$ } & {\footnotesize{}$0.14$ } & {\footnotesize{}$0.04$ }\tabularnewline
{\footnotesize{}Ticker fixed effects } & {\footnotesize{}$462$ } & {\footnotesize{}$462$ } & {\footnotesize{}$462$ }\tabularnewline
{\footnotesize{}Date fixed effects } & {\footnotesize{}$398$ } & {\footnotesize{}$398$ } & {\footnotesize{}$398$ }\tabularnewline
\midrule 
{\footnotesize{}Sample: } & \multicolumn{3}{c}{{\footnotesize{}Maturities $\in(17,38]$ days, Q2}}\tabularnewline
\midrule 
{\footnotesize{}During a Wildfire } & {\footnotesize{}$-0.00$ } & {\footnotesize{}$0.00$ } & {\footnotesize{}$0.07$ }\tabularnewline
 & {\footnotesize{}$(0.01)$ } & {\footnotesize{}$(0.00)$ } & {\footnotesize{}$(0.21)$ }\tabularnewline
\midrule 
{\footnotesize{}Observations } & {\footnotesize{}$134,323$ } & {\footnotesize{}$134,323$ } & {\footnotesize{}$134,323$ }\tabularnewline
{\footnotesize{}R$^{2}$ } & {\footnotesize{}$0.02$ } & {\footnotesize{}$0.12$ } & {\footnotesize{}$0.04$ }\tabularnewline
{\footnotesize{}Ticker fixed effects } & {\footnotesize{}$508$ } & {\footnotesize{}$508$ } & {\footnotesize{}$508$ }\tabularnewline
{\footnotesize{}Date fixed effects } & {\footnotesize{}$426$ } & {\footnotesize{}$426$ } & {\footnotesize{}$426$ }\tabularnewline
\midrule 
{\footnotesize{}Sample: } & \multicolumn{3}{c}{{\footnotesize{}Maturities $\in(38,92]$ days, Q3}}\tabularnewline
\midrule 
{\footnotesize{}During a Wildfire } & {\footnotesize{}$-0.01$ } & {\footnotesize{}$0.01$ } & {\footnotesize{}$-0.41^{*}$ }\tabularnewline
 & {\footnotesize{}$(0.01)$ } & {\footnotesize{}$(0.01)$ } & {\footnotesize{}$(0.22)$ }\tabularnewline
\midrule 
{\footnotesize{}Observations } & {\footnotesize{}$122,052$ } & {\footnotesize{}$122,052$ } & {\footnotesize{}$122,051$ }\tabularnewline
{\footnotesize{}R$^{2}$ } & {\footnotesize{}$0.02$ } & {\footnotesize{}$0.10$ } & {\footnotesize{}$0.03$ }\tabularnewline
{\footnotesize{}Ticker fixed effects } & {\footnotesize{}$536$ } & {\footnotesize{}$536$ } & {\footnotesize{}$536$ }\tabularnewline
{\footnotesize{}Date fixed effects } & {\footnotesize{}$472$ } & {\footnotesize{}$472$ } & {\footnotesize{}$472$ }\tabularnewline
\midrule 
{\footnotesize{}Sample: } & \multicolumn{3}{c}{{\footnotesize{}Maturities $\in(92,207]$ days, Q4}}\tabularnewline
\midrule 
{\footnotesize{}During a Wildfire } & {\footnotesize{}$0.00$ } & {\footnotesize{}$-0.00$ } & {\footnotesize{}$-0.19$ }\tabularnewline
 & {\footnotesize{}$(0.01)$ } & {\footnotesize{}$(0.01)$ } & {\footnotesize{}$(0.14)$ }\tabularnewline
\midrule 
{\footnotesize{}Observations } & {\footnotesize{}$128,410$ } & {\footnotesize{}$128,410$ } & {\footnotesize{}$128,404$ }\tabularnewline
{\footnotesize{}R$^{2}$ } & {\footnotesize{}$0.03$ } & {\footnotesize{}$0.04$ } & {\footnotesize{}$0.01$ }\tabularnewline
{\footnotesize{}Ticker fixed effects } & {\footnotesize{}$538$ } & {\footnotesize{}$538$ } & {\footnotesize{}$538$ }\tabularnewline
{\footnotesize{}Date fixed effects } & {\footnotesize{}$512$ } & {\footnotesize{}$512$ } & {\footnotesize{}$512$ }\tabularnewline
\midrule 
{\footnotesize{}Sample: } & \multicolumn{3}{c}{{\footnotesize{}Maturities $\geq207$ days, Q5}}\tabularnewline
\midrule 
{\footnotesize{}During a Wildfire } & {\footnotesize{}$-0.01$ } & {\footnotesize{}$0.01$ } & {\footnotesize{}$-0.01$ }\tabularnewline
 & {\footnotesize{}$(0.01)$ } & {\footnotesize{}$(0.01)$ } & {\footnotesize{}$(0.08)$ }\tabularnewline
\midrule 
{\footnotesize{}Observations } & {\footnotesize{}$126,964$ } & {\footnotesize{}$126,964$ } & {\footnotesize{}$126,964$ }\tabularnewline
{\footnotesize{}R$^{2}$ } & {\footnotesize{}$0.02$ } & {\footnotesize{}$0.02$ } & {\footnotesize{}$0.01$ }\tabularnewline
{\footnotesize{}Ticker fixed effects } & {\footnotesize{}$480$ } & {\footnotesize{}$480$ } & {\footnotesize{}$480$ }\tabularnewline
{\footnotesize{}Date fixed effects } & {\footnotesize{}$502$ } & {\footnotesize{}$502$ } & {\footnotesize{}$502$ }\tabularnewline
\bottomrule
\end{tabular*}{\footnotesize\par}

\end{center}
\begin{centering}
{\footnotesize{}\bigskip{}
}{\footnotesize\par}
\par\end{centering}
\centering{}\emph{\footnotesize{}{*}{*}{*}: significant at 1\%, {*}{*}:
significant at 5\%, {*}: significant at 10\%.}{\footnotesize{} }{\footnotesize\par}
\end{table}

\end{document}